\documentclass[aps,floatfix,pra,superscriptaddress,reprint,showpacs,10pt,preprintnumbers,longbibliography]{revtex4-2}
\usepackage[utf8]{inputenc}
\usepackage[pdftex]{graphicx}
\usepackage{float}
\usepackage{amssymb}
\usepackage{amsmath}
\usepackage{dsfont}
\usepackage{array}
\usepackage{bm}
\usepackage{subfigure}
\usepackage{mathrsfs}
\usepackage{pifont}
\usepackage{multirow}
\usepackage{upgreek}
\usepackage{xcolor}
\usepackage[pdftex,
            pdftitle={Measurement Error Mitigation in Quantum Computers Through Classical Bit-Flip Correction},
            pdfauthor={Lena Funcke,Tobias Hartung,Karl Jansen,Stefan Kuehn,Paolo Stornati,Xiaoyang Wang},
            bookmarks,
            colorlinks,            
            menucolor=black,            
            plainpages=false,
            linkcolor=myblue,
            citecolor=mymagenta,
            menucolor=black,
            urlcolor=myblue,
            pdfpagelabels,
            hypertexnames=false]{hyperref}
\usepackage{verbatim}
\usepackage{slashed}
\usepackage{cleveref}
\usepackage[braket,qm]{qcircuit}

\definecolor{mymagenta}{RGB}{200, 0, 100}
\definecolor{myblue}{RGB}{45, 48, 146}

\newcommand{\Id}{\ensuremath{\mathds{1}}}

\newcommand{\En}{\ensuremath{\mathbb{E}}}
\newcommand{\Vn}{\ensuremath{\mathbb{V}}}
\newcommand{\Cov}{\ensuremath{\mathrm{Cov}}}
\newcommand{\tr}{\ensuremath{\textrm{Tr}}}
\newcommand{\abs}[1]{\mbox{$ \left| #1 \right| $}}
\newcommand{\bs}{\vec}

\newcommand{\hist}{\ensuremath{\mathrm{hist}}}

\newcommand{\nshots}{\ensuremath{s}}

\DeclareMathAlphabet{\mathesstixfrak}{U}{esstixfrak}{m}{n}

\begin{document}
\title{Measurement Error Mitigation in Quantum Computers Through Classical Bit-Flip Correction}
\author{Lena Funcke}
\affiliation{Perimeter Institute for Theoretical Physics, 31 Caroline Street North, Waterloo, ON N2L 2Y5, Canada}
\author{Tobias Hartung}
\affiliation{Department of Mathematics, King's College London, Strand, London WC2R 2LS, United Kingdom}
\author{Karl Jansen}
\affiliation{NIC, DESY Zeuthen, Platanenallee 6, 15738 Zeuthen, Germany}
\author{Stefan K{\"u}hn}
\affiliation{Computation-Based Science and  Technology Research Center, The Cyprus  Institute, 20 Kavafi Street, 2121 Nicosia, Cyprus}
\author{Paolo Stornati}
\affiliation{NIC, DESY Zeuthen, Platanenallee 6, 15738 Zeuthen, Germany}
\affiliation{Institut für Physik, Humboldt-Universität zu Berlin, Zum Großen Windkanal 6, D-12489 Berlin, Germany}
\author{Xiaoyang Wang}
\affiliation{School of Physics, Peking University, 5 Yiheyuan Rd, Haidian District, Beijing 100871, China}
\date{\today}

\begin{abstract}
We develop a classical bit-flip correction method to mitigate measurement errors on quantum computers. This method can be applied to any operator, any number of qubits, and any realistic bit-flip probability. We first demonstrate the successful performance of this method by correcting the noisy measurements of the ground-state energy of the longitudinal Ising model. We then generalize our results to arbitrary operators and test our method both numerically and experimentally on IBM quantum hardware. As a result, our correction method reduces the measurement error on the quantum hardware by up to one order of magnitude. We finally discuss how to pre-process the method and extend it to other errors sources beyond measurement errors. For local Hamiltonians, the overhead costs are polynomial in the number of qubits, even if multi-qubit correlations are included.
\end{abstract}

\maketitle

\section{Introduction}
Quantum computers have the potential to outperform classical computers in a variety of tasks ranging from combinatorial optimization~\cite{Montanaro2016,Brandao2017} over cryptography~\cite{Gisin2002,Pirandola2020} to machine learning~\cite{Schuld2015,Biamonte2017}. In particular, the prospect of being able to efficiently simulate quantum systems makes them a promising tool for solving quantum many-body problems in physics and chemistry. Despite recent progress, a large scale, fault tolerant digital quantum computer is still not available, and current intermediate scale devices suffer from a considerable level of noise. Although this limits the depth of the circuits that can be executed faithfully, these noisy intermediate-scale quantum (NISQ) devices~\cite{Preskill:2018} are already able to exceed the capabilities of classical computes in certain cases~\cite{Arute2019}.

In the context of quantum many-body systems, a promising approach for exploiting the power of NISQ devices is variational quantum simulation (VQS), a class of hybrid quantum-classical algorithms for solving optimization problems~\cite{Peruzzo2014, McClean:2016}. These make use of a feedback loop between a classical computer and a quantum coprocessor; the latter is used to efficiently evaluate the cost function for a given set of variational parameters, which are optimized on a classical computer based on the measurement outcome obtained from the quantum coprocessor. In particular, it has been experimentally demonstrated that VQS allows for finding both the ground state and low-lying excitations of systems relevant for condensed matter and particle physics as well as quantum chemistry~\cite{OMalley:2016, Kandala:2017, Shen:2017, Colless:2018, Dumitrescu:2018, Hempel2018, Ganzhorn:2019, Kokail:2019, Hartung2019, Jansen2020}.

NISQ devices are susceptible to errors, which can only be partially mitigated using error mitigation procedures (see, e.g., Refs.~\cite{Kandala:2017,Li:2017,Temme:2017,McClean:2017,BonetMonroig:2018,Endo:2018,McArdle:2019,Endo:2019,Kandala:2019,McClean:2020,Otten:2019a,Otten:2019b,Sagastizabal:2019,Urbanek:2019,Crawford:2019,Baek:2019, Corcoles:2015,Sheldon:2016,Tannu:2019,YeterAydeniz2019,YeterAydeniz2020,berg2021modelfree}). In particular, the qubit measurement is among the most error-prone operations on NISQ devices, with error rates ranging from 8\% to 30\% for current hardware~\cite{Tannu:2019}. These errors arise from bit flips, i.e., from erroneously recording an outcome as 0 given it was actually 1, and vice versa.

The goal of this paper is to mitigate these types of measurement errors, in principle, for any operator, any number of qubits, and any bit-flip probability. We develop an efficient mitigation method that relies on cancellations of different erroneous measurement outcomes. This cancellation results from relative minus signs stemming from the default measurement basis of current hardware, $Z=\text{diag}(1,-1)$. The only input requirement for this approach is the knowledge of the different bit-flip probabilities during readout for each qubit.  Our method mainly focuses on measurement bit flips that are uncorrelated between the qubits for multi-qubit measurements, which is true in good approximation in many cases (see, e.g., Refs.~\cite{Kaufmann2017,Qiskit:2020,Mooney2021}). However, our method can also be extended to multi-qubit correlations and different error sources beyond measurement errors, as we discuss in the end of the paper.

Our paper is organized as follows. In Sec.~\ref{sec:mitenergy}, we demonstrate the performance of our mitigation method by correcting the noisy energy histograms for the longitudinal Ising (LI) model [the transversal Ising (TI) model is discussed in Appendix~\ref{sec:TI}]. For simplicity, we assume all bit-flip probabilities to be equal. In Sec.~\ref{sec:mitmeasure}, we generalize our method to different bit-flip probabilities and arbitrary operators. We now correct each bit flip directly at the measurement step, which allows us to mitigate the measurement errors of any expectation value of any operator. In Sec.~\ref{sec:results}, we demonstrate the experimental applicability of our method on IBM quantum hardware. In Sec.~\ref{sec:discussion}, we discuss our results and compare them to previous work. Moreover, we comment on the inclusion of multi-qubit correlations, provide an extension of our method to mitigate relaxation errors, work out a probabilistic implementation of our method, and finally discuss pre-processing and overhead costs. In Sec.~\ref{sec:conclusion}, we summarize our results.

\section{Mitigation of measurement errors for energy histograms\label{sec:mitenergy}}

Throughout this article, we focus on classical bit-flip errors (referred to as measurement or readout errors) and neglect any other sources of error, such as gate errors and decoherence. Thus, we assume that the quantum device prepares a pure state $\ket{\psi}$ for $N$ qubits, which we measure in the computational basis
\begin{align}
  \ket{\psi} = \sum_{i=0}^{2^N-1} c_{i} \ket{i}.
\end{align}
Here, $\ket{i}$ is a shorthand notation for the computational-basis state corresponding to a bit string for the binary representation of $i$ (e.g., for $N=4$ the state  $\ket{5}$ corresponds to $\ket{0101}$). A perfect, noise-free projective measurement would thus yield the bit string $q$ with probability $|c_{i}|^2$; however, bit flips during readout can lead to erroneously recording $j\neq i$ instead. Throughout the main body of this article, we make the assumption that each bit flips independently of the others, which is a good approximation on current quantum hardware~(see, e.g., Ref.~\cite{Qiskit:2020}). Eventually, we will discuss in Sec.~\ref{sec:discussion} how to relax this assumption and include multi-qubit correlations into our method.

Our goal is to obtain the expectation value $\bra{\psi}\mathcal{H}\ket{\psi}$ for a given Hamiltonian $\mathcal{H}$ from a quantum device. Without loss of generality, we assume that $\mathcal{H}$ is of the form
\begin{align}
  \mathcal{H} = \sum_k h_k U_k^*O_kU_k,
  \label{eq:PauliString}
\end{align}
where $O_k$ is a string of the Pauli matrices $\Id$ and $Z$ acting on $N$ qubits, and the unitary $U_k$  transforms this string to $U_k^*O_kU_k \in \{\Id,X,Y,Z\}^{\otimes N}$. Note that throughout the paper we denote the adjoint of operators with asterisks. Since in an experiment we can only measure the final state in the $Z$ basis, we cannot directly obtain $\bra{\psi}\mathcal{H}\ket{\psi}$. We have to determine instead the expectation values of individual Pauli strings $O_k$ by applying the post rotation $U_k$ to $\ket{\psi}$. Subsequently, we can correlate $O_k$ against the distribution of bit strings obtained from the measurement. Thus, we focus throughout the paper on Pauli strings of the form $\{\Id,Z\}^{\otimes N}$. Moreover, in the following we assume that each summand $U_k^*O_kU_k$ in Eq.~\eqref{eq:PauliString} is measured separately. For efficient implementations, multiple summands can also be measured simultaneously, which will be considered later (see Sec.~\ref{sec:variance-by-method}).

To obtain the distribution of bit strings, we have to execute the quantum circuit preparing $U_k\ket{\psi}$ a number of times and record the measurement outcome for each run. Throughout the paper, we refer to this number of repetitions as the number of shots $\nshots$.

\subsection{Prediction for the longitudinal Ising model}\label{sec:LI}

As a pedagogical introductory example that illustrates the basic idea of our method, let us briefly analyze the noisy energy histograms of the LI model with periodic boundary conditions. For this, we assume for simplicity that all bit-flip probabilities are equal, $p(\ket{0}\to\ket{1}) = p(\ket{1}\to\ket{0})=: p$, for all qubits. We will explain all technical details of this example in Appendix \ref{sec:illustration}, and we will also discuss the TI model in Appendix~\ref{sec:TI}. We will turn to the more general case in Sec.~\ref{sec:mitmeasure}, where we will discuss different bit-flip probabilities, arbitrary operators, and arbitrary (pure or mixed) states.

The Hamiltonian of the LI model reads
\begin{equation}
  \mathcal{H}_{\rm LI}=J\sum_{q=1}^N Z_qZ_{q+1}+h\sum_{q=1}^N Z_q,
  \label{eq:HlI}
\end{equation}
where we assume $J<0$ and $h>0$ and we identify $N+1$ with $1$. The true ground-state energy of the model is
\begin{equation}
  E_0 = E_{ZZ} + E_{Z} = NJ - Nh,
  \label{eq:E0lI}
\end{equation}
which is the sum of the individual ground-state energies for $h=0$ and $J=0$, which we call $E_{ZZ}$ and $E_{Z}$, respectively. 

Now we wish to determine the expectation $\En$ of the noisy ground-state energy $\tilde{E}_0$ measured on a quantum computer, where the tilde denotes a noisy outcome. We note that ``expectation'' here means the expectation with respect to the bit-flip probability $p$, which should not be confused with the quantum-mechanical expectation value of the Hamiltonian, $\bra{\psi}\mathcal{H}\ket{\psi}=E$. Thus, the expectation $\En \tilde{\mathcal{H}}$ is the expected value (as an operator to be measured subject to bit flips; see also Sec.~\ref{sec:mitmeasure}) for the noisy Hamiltonian $\tilde{\mathcal{H}}$, while $\En\bra{\psi}\tilde{\mathcal{H}}\ket{\psi}=\En \tilde{E}$ is the expected value for the noisy (quantum mechanical) expectation value $\bra{\psi}\tilde{\mathcal{H}}\ket{\psi}=\tilde{E}$.

To determine the noisy expectation of $E_0$ in Eq.~\eqref{eq:E0lI}, we will first discuss a single $Z_q$ operator, then a single $Z_qZ_{q+1}$ operator, and finally we will take the sum over all qubits to recover the LI model. Starting with a single $Z_q$ operator, we notice the following:
\begin{itemize}
\item If there are no bit flips and both possible measurement outcomes for the qubit are recorded correctly, i.e., $\ket{0}\xrightarrow{1-p}\ket{0}$, $\ket{1}\xrightarrow{1-p} \ket{1}$, we measure the true expectation value $\bra{\psi}Z\ket{\psi}$ with probability $(1-p)^2$.
\item If there are two bit flips and both measurement outcomes are recorded incorrectly, i.e., $\ket{0}\xrightarrow{p}\ket{1}$, $\ket{1}\xrightarrow{p} \ket{0}$, we measure the negative expectation value $-\bra{\psi}Z\ket{\psi}$ (due to $\bra{1}Z\ket{1}=-\bra{0}Z\ket{0}$) with probability $p^2$.
\item If there are single bit flips and one possible measurement outcome is recorded correctly, while the other one is recorded incorrectly, i.e., $\ket{0}\xrightarrow{p}\ket{1}$, $\ket{1}\xrightarrow{1-p} \ket{1}$ or $\ket{0}\xrightarrow{1-p}\ket{0}$, $\ket{1}\xrightarrow{p} \ket{0}$, we measure outcomes with opposite signs that cancel identically.
\end{itemize}
Thus, in total we get the expectation
\begin{align}
  \begin{split}
    \En \bra{\psi}\tilde{Z}\ket{\psi} &= (1-p)^2 \bra{\psi}Z\ket{\psi} + p^2 (-\bra{\psi}Z\ket{\psi}) \\
    &= (1-2p) \bra{\psi}Z\ket{\psi}.
  \end{split}
  \label{eq:E1q1}
\end{align}

\begin{figure}[t]
  \centering
  \includegraphics[width=1.0\columnwidth]{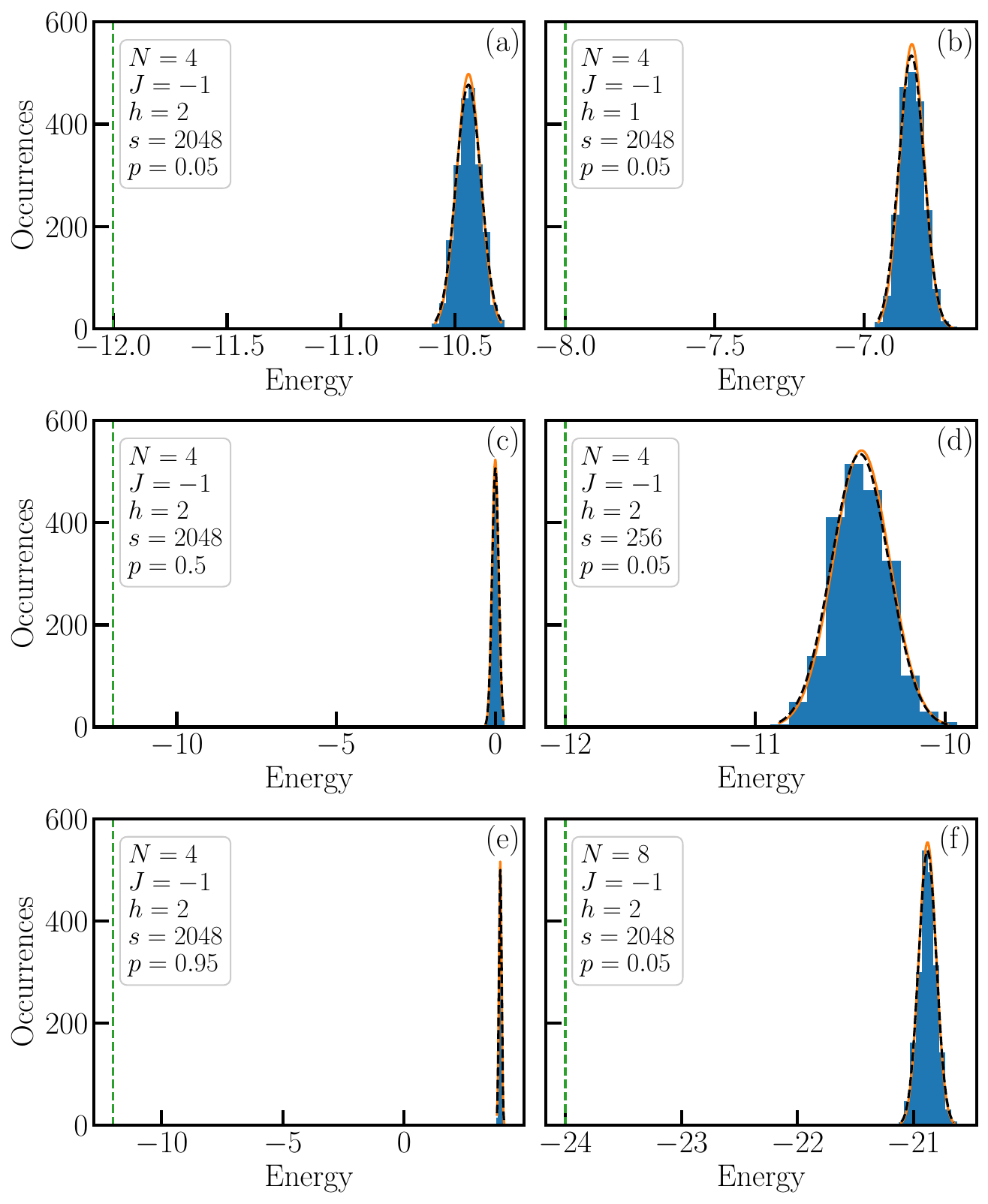}
  \caption{Energy histograms for the LI model. The vertical dashed green line indicates the true ground-state energy, the solid orange line the prediction from Eqs.~\eqref{eq:finalE} and~\eqref{eq:bit-string-variance}, and the dashed black line a fit to the data. The left column corresponds to $N=4$, $J=-1$, $h=2$, $\nshots=2048$ with (a)~$p=0.05$, (c)~$p=0.50$, and (e)~$p=0.95$. The right column shows varied $N$, $h$, and $\nshots$: (b)~$h=1$, (d)~$\nshots=256$, and (f)~$N=8$.}
  \label{fig:longIsing}
\end{figure}

For a single $Z_qZ_{q+1}$ operator, we get three different non-zero outcomes:
\begin{itemize}
\item the absence of any bit flip gives the true expectation value $\bra{\psi}Z_qZ_{q+1}\ket{\psi}$ with probability $(1-p)^2$, just as before,
\item total bit flips, $\ket{0}\xrightarrow{p}\ket{1}$ and $\ket{1}\xrightarrow{p} \ket{0}$ for both qubits, also give $\bra{\psi}Z_qZ_{q+1}\ket{\psi}$ (due to $\bra{00}Z_1Z_2\ket{00}=\bra{11}Z_1Z_2\ket{11}$) with probability $p^2$, unlike before,
\item total bit flips for \textit{one} qubit but no bit flip for the \textit{other} qubit gives the negative expectation value $-\bra{\psi}Z_qZ_{q+1}\ket{\psi}$ with a combined probability of $p(1-p)+(1-p)p=2p(1-p)$.
\end{itemize}
All other possible outcomes cancel identically, similar to the third case discussed previously for the $\bra{\psi}Z\ket{\psi}$ case. In total, this yields
\begin{align}
  \begin{split}
    \En \bra{\psi}\tilde{Z}_q\tilde{Z}_{q+1}\ket{\psi} =\, &(1-p)^2 \bra{\psi}Z_qZ_{q+1}\ket{\psi}\\ 
    &+ p^2 \bra{\psi}Z_qZ_{q+1}\ket{\psi} \\
    &+ 2p(1-p) (-\bra{\psi}Z_qZ_{q+1}\ket{\psi})\\
    =\, &(1-2p)^2 \bra{\psi}Z_qZ_{q+1}\ket{\psi}.
  \end{split}
  \label{eq:E2flip1}
\end{align}
A more detailed derivation of these results can be found in Appendix~\ref{sec:illustration} and Sec.~\ref{sec:multi_Z_section_same_p}.

Finally, to derive the noisy expectation of the full ground-state energy $E_0$ in Eq.~\eqref{eq:E0lI}, we can sum Eqs.~\eqref{eq:E1q1} and \eqref{eq:E2flip1} over the $N$ different qubits. Thus, the final result for the LI model reads
\begin{align}
  \begin{split}
    \En \tilde{E}_0 = (1 - 2p)E_Z +(1-2p)^2E_{ZZ}.
  \end{split}
  \label{eq:finalE}
\end{align}
Our method allows us to predict the variance of the noisy energy histograms as well, as we will explain in detail in Sec.~\ref{sec:variance-by-method} and Appendix~\ref{app:variances}. Based on these results, Fig.~\ref{fig:longIsing} shows the resulting energy histograms for the ground state of $\mathcal{H}_{\rm LI}$ with different choices of the parameters $N$, $J$, $h$, $\nshots$, and $p$, where we measure the ground state 2048 times for each parameter combination. The noise model, with the mean energy from Eq.~\eqref{eq:finalE} and the variance from Eq.~\eqref{eq:bit-string-variance}, agrees with the data for all the parameters. Indeed, our prediction (solid orange line in Fig.~\ref{fig:longIsing}) perfectly matches the fitted data of the histogram (dashed black line). This allows to retrieve the true ground state energy $E_0$ (dashed green line) using Eq.~\eqref{eq:finalE}.

\section{Mitigation of measurement errors for arbitrary operators\label{sec:mitmeasure}}

In this section, we generalize our previous results to \textit{arbitrary} operators acting on $Q$ different qubits $q=1,...,Q\leq N$, where $N$ is the total number of qubits in the system (including the ones the operators are not acting on). We also generalize our previous results to allow for different bit-flip probabilities, $p(\ket{0}\to\ket{1}) \neq p(\ket{1}\to\ket{0})$, which can also differ among the qubits.

These generalizations are greatly aided by a change in point of view. Whereas previously, we treated the bit-flip error as part of the measurement process, i.e., we projectively measured the state $\ket\psi$ onto a basis bit string and randomly flipped the bits of this bit string, we now consider the bit flip as part of the operator. In other words, the measurement process no longer includes the bit flips and instead we consider \textit{random} operators to be measured. While this point of view is conceptually very different, we will demonstrate that these random operators yield a distribution of measurements that precisely coincides with the distribution of measurements for a non-random operator subject to bit flips.

Our analysis will be split into four parts. First, we will consider a single $Z$ operator acting on a single qubit, while allowing for different bit-flip probabilities, $p(\ket{0}\to\ket{1}) \neq p(\ket{1}\to\ket{0})$, in Sec.~\ref{sec:single_Z_section}. In particular, we will compute the operator's expectation as a random operator subject to classical bit flips during measurement. This computation will be the stepping stone to subsequently construct the expectations for noisy measurements of $Z_Q\otimes\cdots\otimes Z_1$ operators with $Q>1$ in Sec.~\ref{sec:multi_Z_section}. This construction is inductive with respect to $Q$ and will allow us to construct a classical bit-flip correction procedure for the noisy measurement of $Z_Q\otimes\cdots\otimes Z_1$. It is important to note that the classical bit-flip correction procedure can be pre-processed (replacing the operator to be measured; see Sec.~\ref{sec:pre-processing}) as well as post-processed (measuring the necessary information first and then extracting the bit-flip corrected expectation values from the measured data).

In Sec.~\ref{sec:multi_Z_section_same_p}, we will consider the special case of equal bit-flip probabilities for all qubits, to compare the results directly to Sec.~\ref{sec:mitenergy}. In Sec.~\ref{sec:general_operators}, we will generalize the classical bit-flip correction procedure to arbitrary operators that are measured from bit-string distributions of the state $\ket\psi$. We note that Sec.~\ref{sec:general_operators} denotes a change in measurement paradigm compared to the previous sections, which affects the variance of the histogram means. We will discuss the different measurement paradigms in detail in Sec.~\ref{sec:variance-by-method} and return to the TI model for an explicit illustration. The derivation of the corresponding variances is provided in Appendix~\ref{app:variances}.

\subsection{Measurement of a single $Z$ operator\label{sec:single_Z_section}}

\subsubsection{Prediction for the noisy expectation value}
For $Q=1$ and arbitrary $N$, the noise-free operator $Z_q$ gets replaced by the random noisy operator $\tilde{Z}_q$, which can take the values

\begin{itemize}
  \item $Z_q$ with probability $(1-p_{q,0})(1-p_{q,1})$,
  \item $-\Id_q$ with probability $p_{q,0}(1-p_{q,1})$,
  \item $\Id_q$ with probability $(1-p_{q,0})p_{q,1}$,
  \item or $-Z_q$ with probability $p_{q,0}p_{q,1}$.
\end{itemize}

Here, $p_{q,b}$ is the probability of flipping the qubit $q$ given that it is in the state $b=\ket{0}$ or $\ket{1}$. For example, $p_{3,0}$ is the probability of flipping $\ket{0}\to\ket{1}$ for qubit~$3$.

Then, we obtain the noisy expectation $\En \tilde{Z}_q$ for the random operator $\tilde{Z}_q$,
\begin{align}
  \En \tilde{Z}_q=(1-p_{q,0}-p_{q,1}) Z_q+(p_{q,1}-p_{q,0})\Id_q,
  \label{eq:expvalue}
\end{align}
which reduces to Eq.~\eqref{eq:E1q1} for $p_{q,0}=p_{q,1}=:p$. As before, ``expectation'' here means the expectation with respect to the bit-flip probabilities, which should not be confused with the quantum mechanical expectation value $\bra{\psi}O\ket{\psi}$ of the operator $O$. The expectation $\En \tilde{O}$ is the expected value (as an operator) for the noisy operator $\tilde{O}$, while $\En\bra{\psi}\tilde{O}\ket{\psi}$ is the expected value for the noisy (quantum mechanical) expectation value $\bra{\psi}\tilde{O}\ket{\psi}$ of the operator $\tilde{O}$.

\subsubsection{Density matrix description and visualization of measurement noise}
For the single-qubit case, it is instructive to express our results in terms of density matrices. Starting from an arbitrary single-qubit density operator
\begin{align}
\rho = (\Id +\bs{r}\cdot\bs{\sigma})/2,
\end{align}
where $\bs{r}$ is a real vector with $\|\bs{r}\|\leq 1$ and $\bs{\sigma}$ is the vector containing the Pauli matrices, any quantum channel acting on the state $\rho$ is an affine linear map,
\begin{align}
  \bs{r}\mapsto \bs{r}\,'= M \bs{r}+\bs{c},
\end{align}
where $M$ is a $3\times 3$ real matrix and $\bs{c}$ is a constant real vector~\cite{Nielsen2000}. In particular, a noise-free projective measurement in the computational basis corresponds to a unital channel with $M = \text{diag}(0,0,1)$ and $\bs{c} = 0$. For an arbitrary pure single-qubit state, $\ket{\psi} = \alpha \ket{0} + \beta \ket{1}$, with density operator
\begin{align}
  \rho=  \begin{pmatrix}
    |\alpha|^2 & \alpha\beta^* \\ \beta\alpha^*&|\beta|^2
  \end{pmatrix},
\end{align}
such a  projective measurement yields the classical mixture $\rho_c = \text{diag}(|\alpha|^2, |\beta|^2)$.

In case of a noisy measurement, the bit flips change the classical state that one obtains after the measurement. As discussed above, (i)~with probability $(1-p_0)(1-p_1)$ we obtain the original state, (ii)~with probability $p_0(1-p_1)$ the $\ket{0}$ flips to a $\ket{1}$, (iii)~with probability $(1-p_0)p_1$ the $\ket{1}$ flips to a $\ket{0}$, and (iv)~with probability $p_0p_1$ both measurement outcomes flip. The resulting classical state can be expressed as a convex linear combination of the different outcomes,
\begin{widetext}
  \begin{align}
    \begin{split}
      \tilde{\rho}_c=&  \begin{pmatrix}
        |\alpha|^2 & 0 \\ 0&|\beta|^2
      \end{pmatrix}(1-p_0)(1-p_1)
      + \begin{pmatrix}
        0 & 0 \\ 0&1
      \end{pmatrix}
      p_0(1-p_1)
      + \begin{pmatrix}
        1 & 0 \\ 0&0
      \end{pmatrix}
      p_1(1-p_0)
     +  \begin{pmatrix}
        |\beta|^2 & 0 \\ 0&|\alpha|^2
      \end{pmatrix}
      p_1 \: p_0\\
      =&  \begin{pmatrix}
        (1-p_0-p_1)|\alpha|^2+p_1 & 0 \\ 0&(1-p_0-p_1)|\beta|^2+p_0
      \end{pmatrix}.
    \end{split}
    \label{eq:rho_convex_linear_combination}
  \end{align}
\end{widetext}
The expectation value of the $Z_q$ operator then reads
\begin{align}
  \begin{aligned}
    \langle \tilde{Z}_q \rangle & = \text{Tr}(\tilde{\rho}_c Z_q)                       \\
                        & =(1-p_0-p_1)(|\alpha |^2-|\beta |^2)+p_1-p_0,
  \end{aligned}
  \label{eq:rhoZ}
\end{align}
which is equivalent to computing the quantum expectation value of Eq.~\eqref{eq:expvalue}, $\text{Tr}(\rho \,\En \tilde{Z}_q)$. 

\begin{figure}[b]
  \centering
  \includegraphics[width=1.0\columnwidth]{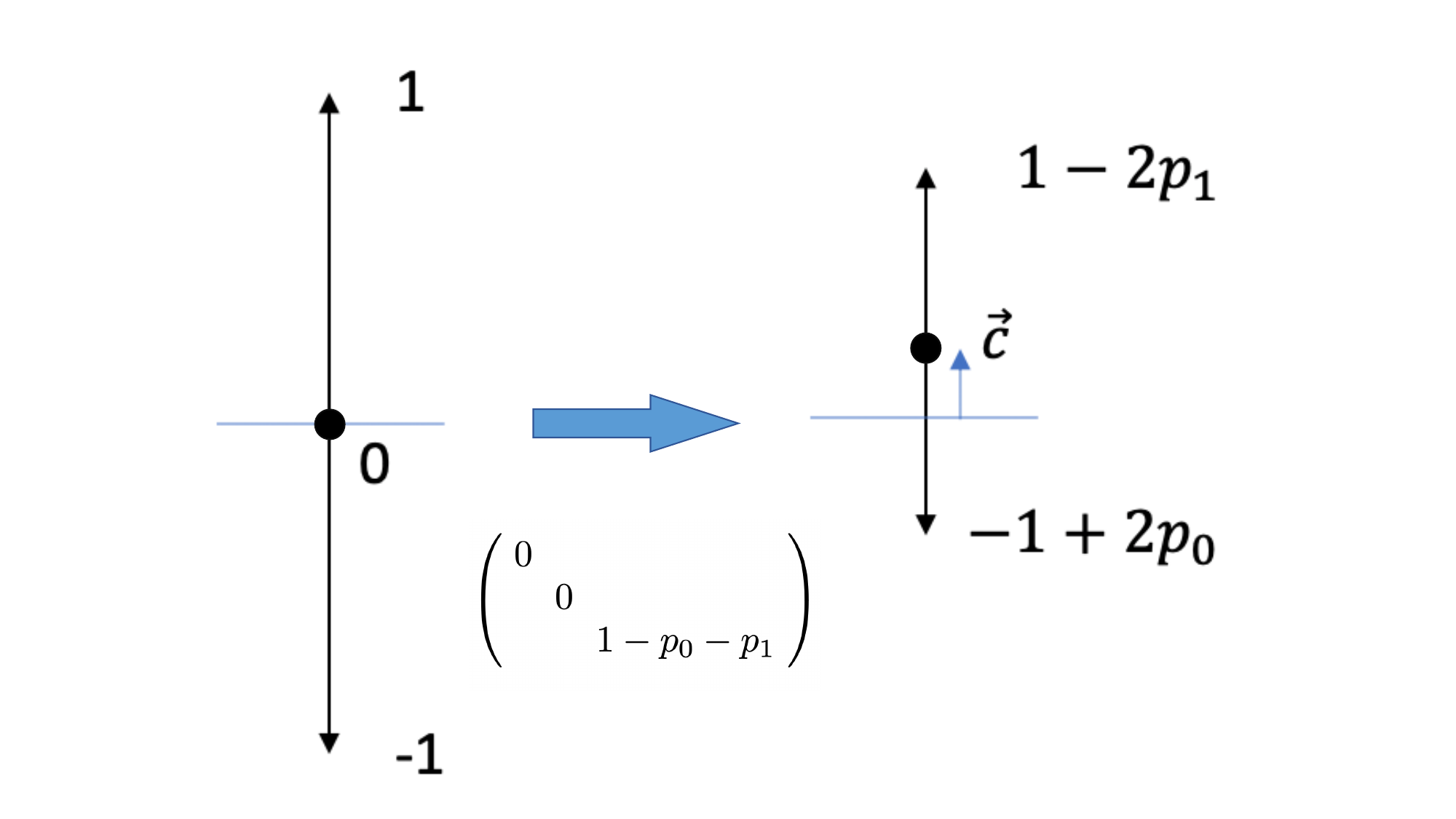}
  \caption{Left panel: Possible range of Bloch vectors of the classical states $\rho_c$ obtained from a noise-free projective measurement in the computational basis. Right panel: Deformed range of Bloch vectors corresponding to the classical state $\tilde{\rho}_c$ resulting from a measurement in the presence of measurement noise.}
  \label{fig:Bloch2}
\end{figure}

Moreover, we see that Eq.~\eqref{eq:rho_convex_linear_combination} arises from the original density operator $\rho$ by applying the quantum channel
\begin{align}
  \tilde{M}=  \begin{pmatrix}
    0 &  & \\ &0&\\&&1-p_0-p_1
  \end{pmatrix}
  ,\,\,
  \tilde{\bs{c}} =  \begin{pmatrix}
    0 \\ 0\\p_1-p_0
  \end{pmatrix}.
\end{align}
From the equation above, it is apparent that the channel is no longer unital. For $p_0=p_1$ all quantum states $\rho$ in the equatorial plane of the Bloch sphere, corresponding to $r_z=0$, are unaffected. The closer the state is to the polar region of the sphere, the more pronounced is the effect of the measurement errors. Compared to the classical state $\rho_c$ obtained from a noise-free projective measurement, the Bloch vector corresponding to $\tilde{\rho}_c$ is shortened because of $\tilde{M}$, and translated along the $z$ axis by $\tilde{\bs{c}}$ (see Fig.~\ref{fig:Bloch2}). Moreover, for $p_0+p_1=1$, the channel maps any state to the same point inside the Bloch sphere. As a result, our mitigation method is not applicable to that special case, which will be further discussed in the next section.

\subsection{Measurement of $Z_Q\otimes\cdots\otimes Z_1$ operators}\label{sec:multi_Z_section}

Going beyond $Q=1$, we can now compute the noisy expectations for arbitrary operators $Z_Q\otimes\cdots\otimes Z_1$ with $Q>1$ and arbitrary $N$. For this, we assume that the expectations of the individual operators can be measured \textit{independently} of each other. In this case, the noisy expectation of the tensor product $\tilde{Z}_Q\otimes\cdots\otimes \tilde{Z}_1$ equals the tensor product of the individual noisy expectations,
\begin{align}
  \En\left(\tilde{Z}_Q\otimes\cdots\otimes \tilde{Z}_1\right)= \En \tilde{Z}_Q\otimes\cdots\otimes \En \tilde{Z}_1.
  \label{eq:En}
\end{align}
Equation~\eqref{eq:En} can be proven by considering two different noisy operators $\tilde{O}_1$ and $\tilde{O}_2$ acting on different qubits, and defining their conditional expectations $\En^{\tilde{O}_1}\tilde{O}_1=:\Omega_1$ and $\En^{\tilde{O}_2} \tilde{O}_2=:\Omega_2$. The term ``conditional'' here means that the expectations are only taken with respect to the qubits on which the operators are acting, leaving the other qubits untouched. Now, if we assume $\tilde{O}_1$ takes the values $\chi_\alpha$ with probabilities $p_\alpha$, for example $\tilde{O}_1=\tilde{Z}_q$ could take $\chi_\alpha\in \{Z_q,-\Id_q,\Id_q,-Z_q\}$ as above, then we observe
\begin{align}\label{eq:independence-operators-different-qubits}
  \begin{split}
    \En\left(\tilde{O}_1\otimes \tilde{O}_2\right)&=\sum_\alpha p_\alpha \,\En^{\tilde{O}_2}\left(\chi_\alpha\otimes \tilde{O}_2\right)\\
    &=\sum_\alpha p_\alpha \, \chi_\alpha\otimes\Omega_2=\Omega_1\otimes\Omega_2,
  \end{split}
\end{align}
which directly yields Eq.~\eqref{eq:En}.

Our final goal is to \textit{reconstruct} the noise-free quantum mechanical expectation value $\bra{\psi} O\ket{\psi}$ of an arbitrary operator $O=O_Q\otimes\cdots\otimes O_1 \in \{\Id,Z\}^{\otimes Q}$ from its noisy measurement. To this end, we need to find a matrix $\omega^{-1}$ that multiplies the noisy expectations $\En \bra{\psi} \tilde{O}\ket{\psi}$ and yields the noise-free expectation values $\bra{\psi} O\ket{\psi}$,
\begin{align}\label{eq:p_q_b_hist_extraction}
  \bra\psi O\ket\psi=\sum_{\tilde{O}\in\{\Id,Z\}^{\otimes Q}}\omega^{-1}_{O, \tilde{O}}\En \bra\psi \tilde{O}\ket\psi.
\end{align}
For this, we first express the noisy expectation of $\tilde{Z}_Q\otimes\cdots\otimes \tilde{Z}_1$ in Eq.~\eqref{eq:En} in terms of the noise-free operators $O_Q\otimes\cdots\otimes O_1$. Using Eq.~\eqref{eq:expvalue}, we find
\begin{align}
  \begin{split}
  &\; \En\left(\tilde{Z}_Q\otimes\cdots\otimes \tilde{Z}_1\right)\\
  &=\sum_{O\in\{\Id,Z\}^{\otimes Q}}\gamma(O_Q)O_Q\otimes\cdots\otimes \gamma(O_1)O_1,
  \end{split}
  \label{eq:p_q_b_EZ}
\end{align}
where the coefficients $\gamma$ in front of the noise-free operators are defined as
\begin{align}
  \gamma(O_q):=
  \begin{cases}
    1-p_{q,0}-p_{q,1}\ \ \qquad & \textrm{for}\ \ O_q=Z_q,   \\
    p_{q,1}-p_{q,0}\qquad       & \textrm{for}\ \ O_q=\Id_q.
  \end{cases}
\end{align}
In order to construct the value of $\En\left(\tilde{Z}_Q\otimes\cdots\otimes \tilde{Z}_1\right)$ in Eq.~\eqref{eq:p_q_b_EZ} inductively, it is advantageous to choose the ``lexicographic order'' $\preceq$ for both the noise-free operators $O\in\{\Id,Z\}^{\otimes Q}$ and the noisy operators $\tilde{O}\in\{\Id,Z\}^{\otimes Q}$,
\begin{align}
  \begin{split}
    &\Id_3\otimes\Id_2\otimes\Id_1\preceq \Id_3\otimes\Id_2\otimes Z_1\\
    \preceq &\Id_3\otimes Z_2\otimes\Id_1\preceq \Id_3\otimes Z_2\otimes Z_1\\
    \preceq &Z_3\otimes\Id_2\otimes\Id_1\preceq Z_3\otimes\Id_2\otimes Z_1\\
    \preceq &Z_3\otimes Z_2\otimes\Id_1\preceq Z_3\otimes Z_2\otimes Z_1
    \preceq\ldots
  \end{split}
\end{align}
This choice implies $O_Q\otimes\cdots\otimes O_1\preceq Z_Q\otimes\cdots\otimes Z_1$ and will later ensure that the matrix $\omega$ in Eq.~\eqref{eq:p_q_b_hist_extraction} is a lower triangular matrix, which is invertible as long as none of its diagonal entries vanish. To determine the matrix $\omega$, we need to generalize Eq.~\eqref{eq:p_q_b_EZ} to arbitrary noisy operators,
\begin{align}
  \begin{split}
    &\; \En \left( \tilde{O}_Q \otimes\cdots\otimes \tilde{O}_1\right)\\
    &=\sum_{O\in\{\Id,Z\}^{\otimes Q}}\Gamma(O_Q|\tilde{O}_Q)O_Q\otimes\cdots\otimes \Gamma(O_1| \tilde{O}_1)O_1,
  \end{split}
  \label{eq:p_q_b_EO}
\end{align}
where the coefficients $\Gamma$ in front of the noise-free operators are now defined as
\begin{align}
  \Gamma(O_q|\tilde{O}_q)= &
  \begin{cases}
    \gamma(O_q)\ \quad & \textrm{for}\ \  \tilde{O}_q=\tilde{Z}_q                     \\
    1\qquad            & \textrm{for}\ \ O_q=\Id_q\ \land \ \tilde{O}_q=\tilde{\Id}_q \\
    0\qquad            & \textrm{for}\ \ O_q=Z_q\ \land \ \tilde{O}_q=\tilde{\Id}_q.
  \end{cases}
  \label{eq:Gamma}
\end{align}
Using this definition, we can now define the matrix $\omega$ as
\begin{align}
  \begin{split}
    \omega\left(O|\tilde{O}\right) &:=\prod_{q=1}^Q\Gamma(O_q|\tilde{O}_q),\\
    \omega &:=\left(\omega\left(O|\tilde{O}\right)\right)_{\tilde{O},O\in\{\Id,Z\}^{\otimes Q}}.
  \end{split}
  \label{eq:omega}
\end{align}
It is important to note that $\tilde{O}\prec O$ implies $\omega\left(O|\tilde{O}\right)=0$. In other words, $\omega$ is a lower triangular matrix and therefore is invertible as long as none of its diagonal entries vanish.
The diagonal entries are $\prod_{q=1}^Q\Gamma\left(O_q|\tilde{O}_q\right)$ and thus can only vanish if one of the $\gamma(Z_q)$ vanishes, i.e., $\omega$ is invertible as long as $\forall q:\ p_{q,0}+p_{q,1}\ne1$. If that is the case, then we obtain the bit-flip corrected operators
\begin{align}
  \left(O\right)_{O\in\{\Id,Z\}^{\otimes Q}}=\omega^{-1}\left(\En \tilde{O}\right)_{\tilde{O}\in\{\Id,Z\}^{\otimes Q}}.
  \label{eq:corr_exp}
\end{align}
In particular, for $O=Z_2\otimes Z_1$, we obtain
\begin{align}
  \begin{split}
    Z_2\otimes Z_1=\,&\frac{1}{\gamma(Z_2)\gamma(Z_1)}\En\left(\tilde{Z}_2\otimes \tilde{Z}_1\right)\\
    &-\frac{\gamma(\Id_1)}{\gamma(Z_2)\gamma(Z_1)}\En\left(\tilde{Z}_2\right)\otimes\Id_1\\
    &-\frac{\gamma(\Id_2)}{\gamma(Z_2)\gamma(Z_1)} \Id_2\otimes\En\left(\tilde{Z}_1\right)\\
    &+\frac{\gamma(\Id_2)\gamma(\Id_1)}{\gamma(Z_2)\gamma(Z_1)} \Id_2\otimes\Id_1.
  \end{split}
\end{align}

\begin{figure}[htp!]
  \centering
  \includegraphics[width=1.0\columnwidth]{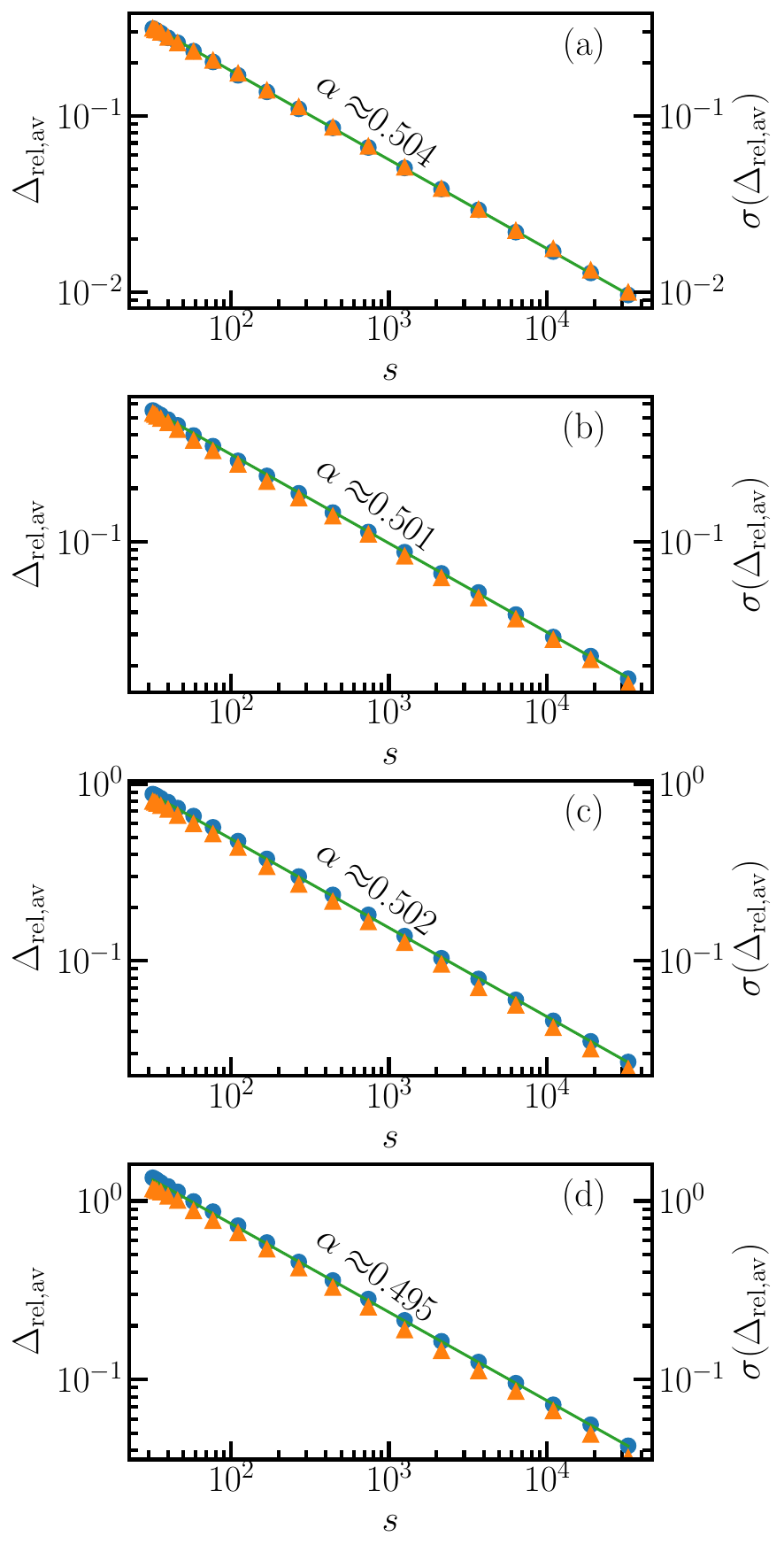}
  \caption{Mean value $\Delta_\mathrm{rel,av}$ (blue dots, left $y$-axis) and standard deviations $\sigma(\Delta_\mathrm{rel,av})$ (orange triangles, right $y$-axis) of the relative error for the bit-flip corrected expectation values of $\bra\psi \tilde{Z}_Q\otimes\cdots\otimes \tilde{Z}_1\ket\psi$, as retrieved from histogram data using Eq.~\eqref{eq:p_q_b_hist_extraction}, compared to the ``true'' bit-flip free expectation values of $\bra\psi Z_Q\otimes\cdots\otimes Z_1\ket\psi$, see Eq.~\eqref{eq:relerr}. Shown are the four different operators (a) $Z_1$, (b) $Z_2\otimes Z_1$, (c) $Z_3\otimes Z_2\otimes Z_1$, and (d) $Z_4\otimes Z_3\otimes Z_2\otimes Z_1$. The average relative errors are fitted with a power law in the number of shots $\nshots$, $y(\nshots)\propto \nshots^\alpha$ (green lines), the slopes obtained are indicated in the different panels. The standard deviations of the relative errors are extracted from $2^{12} = 4096$ random states $\ket\psi$ and random bit-flip probabilities $p_{q,b}$.}
  \label{fig:p_q_b}
\end{figure}

In Fig.~\ref{fig:p_q_b}, we show the relative error for the bit-flip corrected expectation value of $\bra\psi \tilde{Z}_Q\otimes\cdots\otimes \tilde{Z}_1\ket\psi$, as retrieved from histogram data using Eq.~\eqref{eq:p_q_b_hist_extraction}, compared to the bit-flip free expectation value $\bra\psi Z_Q\otimes\cdots\otimes Z_1\ket\psi$:
\begin{equation}
  \Delta_\mathrm{rel} = \frac{\abs{\bra\psi \tilde{Z}_Q\otimes\cdots\otimes \tilde{Z}_1\ket\psi - \bra\psi Z_Q\otimes\cdots\otimes Z_1\ket\psi}}{\abs{\bra\psi Z_Q\otimes\cdots\otimes Z_1\ket\psi}}.
  \label{eq:relerr}
\end{equation}
We also plot the standard deviation of this relative error, alternatively to plotting the error bars. Figure~\ref{fig:p_q_b} also contains a fit $y(\nshots)= C\nshots^{-\alpha}$ of the relative error in Eq.~\eqref{eq:relerr}, where $\nshots$ is again the number of shots, i.e., the number of $\bra\psi \tilde{Z}_Q\otimes\cdots\otimes \tilde{Z}_1\ket\psi$ evaluations needed to produce the histogram. In particular, the fit indicates Monte-Carlo type convergence $\alpha\approx 1/2$ for $Q\in\{1,2,3,4\}$. Figure~\ref{fig:p_q_b} has been generated using $2^{12} = 4096$ random states $\ket\psi$ satisfying $\abs{\bra\psi Z_Q\otimes\cdots\otimes Z_1\ket\psi}\ge.25$ to avoid dividing by small numbers when computing relative errors. For each $\ket\psi$ we randomly chose the bit-flip probabilities $p_{q,b}$ uniformly in $(0.05,0.25)$.

\subsection{Measurement of $Z_Q\otimes\cdots\otimes Z_1$ operators assuming equal bit-flip probabilities}\label{sec:multi_Z_section_same_p}

To compare the results of the previous two subsections with the results obtained in Sec.~\ref{sec:mitenergy}, we now set all bit-flip probabilities $p_{q,b}=p$ to be equal. For the case $Q=1$, the expectation $\En \tilde{Z}_q$ in Eq.~\eqref{eq:expvalue} reduces to
\begin{align}
  \En \tilde{Z}_q = (1-2p)Z_q,
\end{align}
in agreement with Eq.~\eqref{eq:E1q1}. For $Q>1$, the expectation in Eq.~\eqref{eq:En} reduces to
\begin{align}
  \En(\tilde{Z}_Q\otimes\cdots\otimes \tilde{Z}_1)=(1-2p)^Q\ Z_Q\otimes\cdots\otimes Z_1,
  \label{eq:Eequalp}
\end{align}
which yields Eq.~\eqref{eq:E2flip1} for $Q=2$. This implies that the matrix $\omega$ in Eq.~\eqref{eq:omega} becomes diagonal with
\begin{align}
  \begin{split}
    \En(\tilde{O}_Q\otimes\cdots\otimes  \tilde{O}_1)=\ &(1-2p)^{\#Z(O)}\\ 
    &\times O_Q\otimes\cdots\otimes O_1,
  \end{split}
  \label{eq:ExpArbitOp}
\end{align}
where $\#Z(O)$ is the number of terms $O_q=Z_q$ in the tensor product $O=O_N\otimes\cdots\otimes O_1$. In particular, $\omega$ is invertible as long as $p\ne 1/2$. We again observe in Eqs.~\eqref{eq:Eequalp} and \eqref{eq:ExpArbitOp} that the noisy expectations of arbitrary operators can be related to the true operators in a surprisingly simple way, which requires no knowledge of the quantum hardware apart from the different bit-flip probabilities of the qubits.

\subsection{Measurement of general operators $\mathcal{H}$ from bit-string distributions of $\ket{\psi}$}\label{sec:general_operators}

\subsubsection{Prediction for the noisy expectation value}

Our analysis of the bit-flip error above assumed that we measure general operators $\mathcal{H}$ by expressing them as linear combinations of operators $U^*OU$ with $O\in\{\Id,Z\}^{\otimes N}$ on an $N$-qubit machine, and by measuring each $O$ independently ($U$ being the transformation into the $Z$ basis). For example, if we are interested in measuring $\mathcal{H}_{ZZ}=J\sum_{i=1}^NZ_iZ_{i+1}$ with $N=3$ qubits, then we generate independent histograms for $\bra{\psi} \Id_3\otimes Z_2\otimes Z_1\ket{\psi}$, $\bra{\psi} Z_3\otimes Z_2\otimes \Id_1\ket{\psi}$, and $\bra{\psi} Z_3\otimes \Id_2\otimes Z_1\ket{\psi}$, we extract their expectation values, and we recover $\bra{\psi} \mathcal{H}_{ZZ}\ket{\psi}$ accordingly. Alternatively, we can measure the \textit{distribution} of $\ket{\psi}$ and obtain a single histogram in terms of the computational basis $\{\ket{j};\ j\in\mathbb{N}_{0,<2^N}\}$. Hence, if the probability of measuring $\ket j$ is $\mathesstixfrak{p}_j$, then we can recover $\bra\psi\mathcal{H}_{ZZ}\ket\psi$ from $\sum_j \mathesstixfrak{p}_j\bra j\mathcal{H}\ket j$. While both approaches yield the same expectation value, the variance obtained for both approaches will in general be different, as we will discuss further below.

Moreover, for a general Hamiltonian $\mathcal{H}$ the full expectation value $\bra\psi\mathcal{H}\ket\psi$ cannot always be recovered from a single histogram via $\sum_j \mathesstixfrak{p}_j\bra j\mathcal{H}\ket j$. For example, if we are interested in measuring the TI Hamiltonian $\mathcal{H}_{\rm TI}=J\sum_{i=1}^NZ_iZ_{i+1} + h\sum_{i=1}^NX_i$, we cannot directly recover the full expectation value $\bra\psi\mathcal{H}_{\rm TI}\ket\psi$ from measuring the distribution of $\ket{\psi}$, because the terms in the Hamiltonian do not all commute. However, as we discussed below Eq.~\eqref{eq:PauliString}, an efficient implementation on the quantum hardware can be achieved by splitting the Hamiltonian into two sums of Pauli strings $U_k^* O_k U_k\in \{\Id,X,Y,Z\}^{\otimes N}$, where multiple summands of the Hamiltonian are measured simultaneously. For example, both $\mathcal{H}_{ZZ}=J\sum_{i=1}^NZ_iZ_{i+1}$ and $\mathcal{H}_X=h\sum_{i=1}^NX_i$ can be measured using bit-string distributions. Here, $\mathcal{H}_{ZZ}$ can be measured directly by using the bit-string distribution of the state $\ket\psi$ and $\mathcal{H}_X$ can be measured by using $h\sum_{i=1}^NZ_i$ and the bit-string distribution of the state $H^{\otimes N}\ket\psi$, i.e., after applying a Hadamard gate $H$ on each qubit. Hence, using the bit-string distribution, we can measure all the $ZZ$ terms and all the $X$ terms in the TI Hamiltonian simultaneously. In other words, we are only required to measure two bit-string distributions instead of measuring each of the $2N$ Pauli-terms separately. This allows for an efficient implementation on the quantum hardware.

If we measure the distribution of $\ket{\psi}$, the measurements of $\bra\psi U^*OU\ket\psi$ comprising $\bra\psi\mathcal{H}\ket\psi$ are no longer independent. This has an impact on the variance of measurement histograms, as we will discuss in Sec.~\ref{sec:variance-by-method}. However, it has no impact on the expectation subject to bit flips, since linearity of the expectation value implies
\begin{align}
  \begin{split}
    \En\bra\psi\tilde{\mathcal{H}}\ket\psi=&\En\bra\psi\sum_\alpha \lambda_\alpha U_\alpha^*\tilde{O}_\alpha U_\alpha\ket\psi\\
    =&\bra\psi\sum_\alpha \lambda_\alpha U_\alpha^*\left(\En \tilde{O}_\alpha\right) U_\alpha\ket\psi,
  \end{split}
  \label{eq:linearity}
\end{align}
which is precisely the expression we would obtain from summing the independently measured operators $\tilde{O}_\alpha$.

\subsubsection{Prediction for the bit-flip corrected operator}

In order to correct for bit flips in this setting, we need to keep in mind that the general case requires measurements of all operators $O\preceq O_\alpha$ (with respect to the lexicographic order $\preceq$ on $\{\Id,Z\}^{\otimes N}$) for all operators $O_\alpha$ in $\mathcal{H}=\sum_\alpha \lambda_\alpha U_\alpha^* O_\alpha U_\alpha$. Hence, the histogram for $\bra\psi\tilde{\mathcal{H}}\ket\psi$ does not contain sufficient information. However, we can use the classical bit-flip correction method as discussed above to find coefficients $\omega_{\alpha,O}$ such that
\begin{align}
  O_\alpha = \sum_{O\preceq O_\alpha}\omega_{\alpha,O}\En \tilde{O}
\end{align}
holds. Inserting this into $\mathcal{H}$, we can express $\mathcal{H}$ as
\begin{align}
  \mathcal{H}=\sum_\alpha \lambda_\alpha U_\alpha^*\sum_{O\preceq O_\alpha}\omega_{\alpha,O}\En \tilde{O} U_\alpha.
\end{align}
In other words, we can replace the operator $\mathcal{H}$ by the bit-flip corrected noisy operator
\begin{align}\label{eq:bit_flip_corrected_noisy_operator}
  \tilde{\mathcal{H}}_\mathrm{bfc}:=\sum_\alpha \lambda_\alpha U_\alpha^*\sum_{O\preceq O_\alpha}\omega_{\alpha,O}\tilde{O} U_\alpha
\end{align}
and obtain
\begin{align}
  \En\bra\psi\tilde{\mathcal{H}}_\mathrm{bfc}\ket\psi=\bra\psi\mathcal{H}\ket\psi.
\end{align}

\subsubsection{Prediction for equal bit-flip probabilities}

To compare our results to Secs.~\ref{sec:mitenergy} and \ref{sec:multi_Z_section_same_p}, let us assume that the bit-flip probabilities $p_{q,b}$ satisfy $p_{q,0}=p_{q,1}=p_q$, i.e., there is no difference between $p(\ket0\to\ket1)$ and $p(\ket1\to\ket0)$ for each qubit, but this value might depend on the individual qubit. Then we obtain $\omega_{\alpha,O}=0$ unless $O=O_\alpha=O_{\alpha,N}\otimes\cdots\otimes O_{\alpha,1}$, for which we find
\begin{equation}
  \omega_{\alpha,O_\alpha}=:\omega_\alpha = \prod_q \frac{1}{(1-2p_q)},
\end{equation}
where $q$ ranges over all qubits satisfying $O_{\alpha,q}=Z_q$. For $p_{q,b}=p$, this result agrees with Eqs.~\eqref{eq:E1q1}, \eqref{eq:E2flip1}, and \eqref{eq:Eequalp}.

Thus, the bit-flip corrected noisy operator
\begin{align}
  \tilde{\mathcal{H}}_\mathrm{bfc}:=\sum_\alpha \lambda_\alpha\omega_{\alpha}U_\alpha^*\tilde{O}_\alpha U_\alpha
\end{align}
has the same Pauli-sum structure as the original operator $\mathcal{H}$, changing only the coefficients. This is completely analogous to the independent measurement case. In both cases, if we have $p_{q,0}=p_{q,1}$, then we can correct for bit flips without additional cost to the quantum device.

\subsection{Impact of measurement choices}\label{sec:variance-by-method}

In general, we will extract the quantum-mechanical expectation of an operator by running the circuit preparing $\ket{\psi}$ followed by a projective measurement in the computational basis a number of times. As before, we refer to these repetitions as the number of shots, $\nshots$. Of course, these shots are still subject to statistical fluctuations. Hence, if we generate $N_\hist$ histograms with $\nshots$ shots each, we can generate a histogram from the means extracted from each histogram. This will yield results as in Fig.~\ref{fig:longIsing} and Fig.~\ref{fig:transIsing}. Using bit-flip corrected operators as in Eq.~\eqref{eq:bit_flip_corrected_noisy_operator}, we can shift the expected mean to coincide with the quantum mechanical expectation of the operator we wish to measure. However, the variance of histogram means is then highly dependent on the measurement paradigm.

For illustration, let us consider the TI model $\mathcal{H}_{\rm TI}=J\sum_{j=1}^N Z_jZ_{j+1}+h\sum_{j=1}^N X_j$, which we will measure on the ground state $\ket\psi$. The first step is to compute the bit-flip corrected noisy Hamiltonian $\tilde{\mathcal{H}}_{\rm TI,bfc}$. For simplicity, we will assume that all bit-flip probabilities $p_{q,b}$ coincide with some value $p$. This yields
\begin{align}
  \tilde{\mathcal{H}}_{\rm TI,bfc} = J_p\sum_{j=1}^N \tilde{Z}_j\tilde{Z}_{j+1}+h_p\sum_{j=1}^N \tilde{X}_j
  \label{eq:HTIbfc}
\end{align}
with $J_p:=J(1-2p)^{-2}$ and $h_p:=h(1-2p)^{-1}$. Of course, this process changes the variances. In particular, since Fig.~\ref{fig:longIsing} and Fig.~\ref{fig:transIsing} show histograms without the bit-flip correction, the prediction of variances in Fig.~\ref{fig:longIsing} (and Fig.~\ref{fig:transIsing} in Appendix~\ref{sec:illustration}) uses $J$ and $h$ instead of $J_p$ and $h_p$.

At this point, we need to decide upon the precise way of measuring the Hamiltonian. Essentially, we have a spectrum of possibilities which contains three interesting cases:

\begin{itemize}
  \item \emph{Method 1:} measure each $\tilde{Z}_j\tilde{Z}_{j+1}$ and $\tilde{X}_j$ in Eq.~\eqref{eq:HTIbfc} independently
  \item \emph{Method 2:} measure the entire Hamiltonian $\tilde{\mathcal{H}}_{\rm TI,bfc}$ in Eq.~\eqref{eq:HTIbfc} from distributions of $\ket\psi$ measurements
  \item \emph{Method 3:} measure $\tilde{\mathcal{H}}_{ZZ}:=J_p\sum_{j=1}^N \tilde{Z}_j\tilde{Z}_{j+1}$ and $\tilde{\mathcal{H}}_X:=h_p\sum_{j=1}^N \tilde{X}_j$ independently from distributions of $\ket\psi$ measurements
\end{itemize}

Methods 1 and 2 are the two extremes discussed in Secs.~\ref{sec:single_Z_section}--\ref{sec:multi_Z_section_same_p} and Sec.~\ref{sec:general_operators}, respectively. Note that Method 2 would require us to perform global projective measurements in the eigenbasis of the Hamiltonian, and therefore it is in general not applicable on real hardware devices. Nevertheless, the results allow us to quantify the effect of the bit-flip variance for the idealized setting where the quantum mechanical contribution to the variance vanishes (up to statistical fluctuations due to a finite number of shots). Method 3 is a reasonable compromise, and it is precisely the method we used for Fig.~\ref{fig:longIsing} and Fig.~\ref{fig:transIsing}. Method 3 is also an example that is closely related to implementations of quantum algorithms that are optimized for the number of calls to the quantum device, i.e., implementations in which only parts of an operator can be measured simultaneously, and both Methods 1 and 2 are impractical to various degrees. 

The variance of histogram means has two contributions: bit-flip variance and quantum-mechanical variance. These contributions for each of the three methods are shown in Fig.~\ref{fig:variance_contributions}. The derivation of these variances can be found in Appendix~\ref{app:variances}; in particular, Fig.~\ref{fig:variance_contributions} shows Eq.~\eqref{eq:var-m1}, Eq.~\eqref{eq:var-m2}, and Eq.~\eqref{eq:var-m3}.  To remove the dependence on the number of shots per histogram, all variances are multiplied by the number of shots $\nshots$, i.e., all values in Fig.~\ref{fig:variance_contributions} correspond to the normalization $\nshots=1$.

\begin{figure}[t]
  \centering
  \includegraphics[width=1.0\columnwidth]{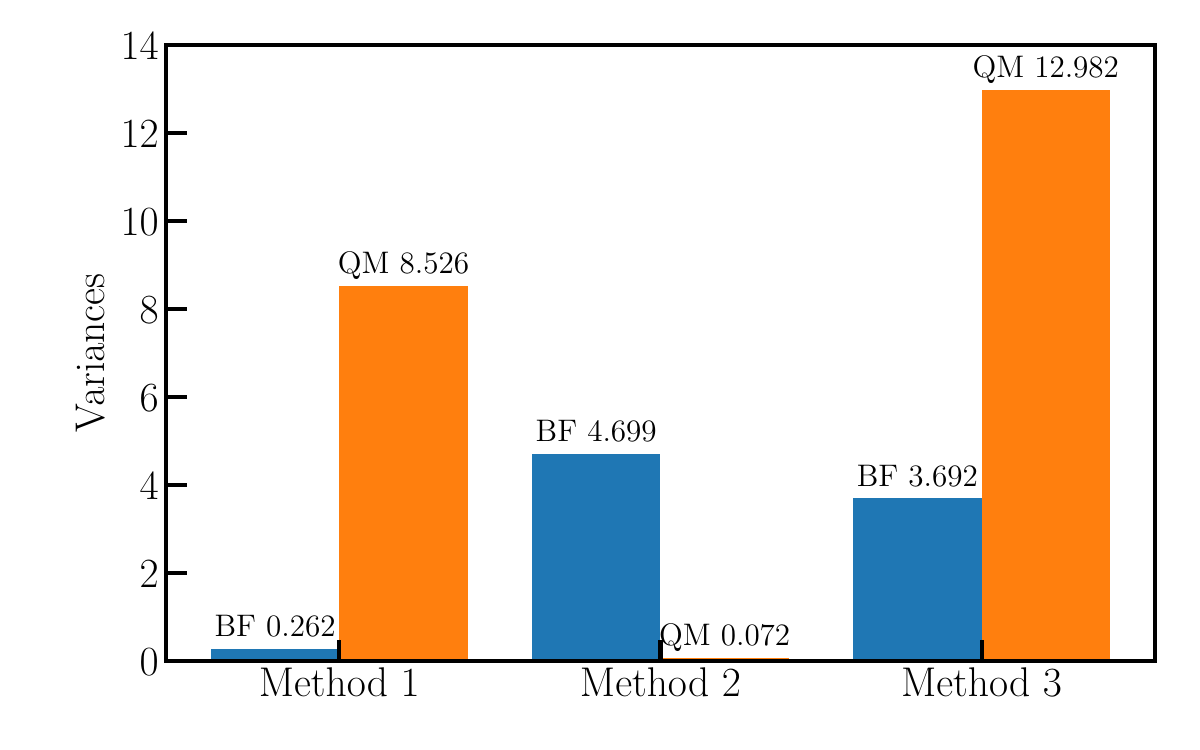}
  \caption{Contributions to the variance of histogram means for the bit-flip corrected TI Hamiltonian in Eq.~\eqref{eq:HTIbfc} evaluated on the ground state of the ``true'' TI Hamiltonian in Eq.~\eqref{eq:HtI}. The different bars correspond to the bit-flip (BF, blue) and quantum-mechanical (QM, orange) variance contributions for the three different measurement methods. We used the parameters $N=4$, $J=-1$, $h=2$, and $p_{q,b}=p=0.05$. All values are normalized by setting $\nshots=1$.}
  \label{fig:variance_contributions}
\end{figure}

It is interesting to note that not only does the full variance vary in magnitude, but also the relative contribution from bit flips and quantum mechanics is vastly different between the three methods.

If we compare the two extremes---Method 1 and Method 2---we notice that for Method 1 the bit-flip induced variance is small compared to the quantum-mechanical variance, whereas for Method 2 the situation is reversed. Generically, this pattern is to be expected. Method 1 is likely to produce a much smaller bit-flip contribution since all summands are measured independently. Meanwhile, measuring with Method 2 introduces $O(4^N)$ covariance terms, which vanish in Method~1 due to independent measurements of summands.
Moreover, concerning Method~2, we note that the quantum-mechanical variance vanishes upon evaluation on an eigenstate of the operator. In Fig.~\ref{fig:variance_contributions}, we evaluated the bit-flip corrected TI Hamiltonian $\tilde{\mathcal{H}}_{\rm TI,bfc} = J_p\sum_{j=1}^N \tilde{Z}_j\tilde{Z}_{j+1}+h_p\sum_{j=1}^N \tilde{X}_j$ with equal bit-flip probabilities $p_{q,b}=p=0.05$ on the ground state of the ``true'' TI Hamiltonian $\mathcal{H}_{\rm TI} = J\sum_{j=1}^N Z_jZ_{j+1}+h\sum_{j=1}^N X_j$. For small values of $p$, we can interpret the bit-flip correction as a small perturbation to the original operator. Hence, the ground state of $\mathcal{H}_{\rm TI}$ is close to an eigenstate of $\tilde{\mathcal{H}}_{\rm TI,bfc}$, and thus the quantum-mechanical contribution to the variance is small.

For intermediate methods, such as Method 3, it is generally difficult to predict the different contributions to the variance using similar arguments as above. Depending on the practical limitation of any given implementation, it will be imperative to balance the different contributions to the variance with the number of quantum device calls. For example, for the TI model, fewer quantum device calls per evaluation of the Hamiltonian introduce more covariance terms. In turn, this requires more quantum device calls to obtain the necessary statistical power if we aim to extract a histogram mean with a required level of precision. Thus, this balancing act is highly problem-specific. However, considering Method 3 for the TI model, it clearly shows that great care has to be taken when constructing an intermediate method if the aim is to reduce the overall variance on a given budget of quantum device calls.

\section{Experimental results\label{sec:results}}
To demonstrate the experimental applicability of our measurement error mitigation method, we generate data on IBM quantum hardware using the Qiskit software development kit (SDK)~\cite{Qiskit}. To assess the performance of our correction procedure, we first simulate the quantum hardware classically using the noise models for the different backends provided by Qiskit, before we proceed to the actual hardware.

\subsection{Single-qubit case\label{sec:results_single_qubit}}
To begin with, let us focus on the simplest case of a single qubit. In a first step, we determine the bit-flip probabilities of the qubit. The probability $p_{0}$ can be easily obtained by measuring the initial state $\ket{0}$ and recording the number of $1$ outcomes, while $p_{1}$ requires preparing the state $\ket{1}$ through applying a single $X$ gate to the initial $\ket{0}$ state and recording the number of 0 outcomes. In order to account for statistical fluctuations, we repeat this procedure several times and average over the bit-flip probabilities obtained for each run (see Appendix~\ref{app:calibration} for details).

After obtaining the bit-flip probabilities, we measure $\bra{\psi}Z\ket{\psi}$ for a randomly chosen $\ket{\psi}$. Starting from the initial state $\ket{0}$, we can prepare any state on the Bloch sphere by first applying a rotation gate around the $x$-axis followed by a rotation around the $z$-axis. Hence, we choose the circuit
\begin{align*}
  \Qcircuit @C=1em @R=.7em {
  \lstick{\ket{0}} & \gate{R_x(\theta_0)} & \gate{R_z(\theta_1)} & \meter      & \qw \\
  \lstick{c}       & \cw                  & \cw                  & \cw\cwx[-1] & \cw
  }
\end{align*}
in our experiments, where the angles $\theta_0$, $\theta_1$ are both drawn uniformly from the interval $[0,2\uppi]$. Our measurement outcomes allow us to determine the noisy expectation value of $Z$, $\En(\tilde{Z})$. Subsequently, we can apply our correction procedure using Eq.~\eqref{eq:expvalue}. To acquire statistics for $\En(\tilde{Z})$, we repeat the process for 1050 randomly chosen $\ket{\psi}$ and monitor the mean and the standard deviation of the absolute error,
\begin{align}
  \Delta_\mathrm{1,abs}=|\bra{\psi}\tilde{Z}\ket{\psi}_\text{measured} - \bra{\psi}Z\ket{\psi}_\text{exact}|,
  \label{eq:absolute_error_single_qubit}
\end{align}
for both the noisy expectation value and the corrected expectation value. Moreover, each individual measurement for fixed values of $\theta_0$ and $\theta_1$ requires running the circuit multiple times to get the probability distribution of basis states in $\ket{\psi}$. Thus we also explore the dependence of our results on the number of shots $s$.

\subsubsection{Classical simulation of quantum hardware}
\begin{figure*}
	\centering
	\includegraphics[width=1.0\textwidth]{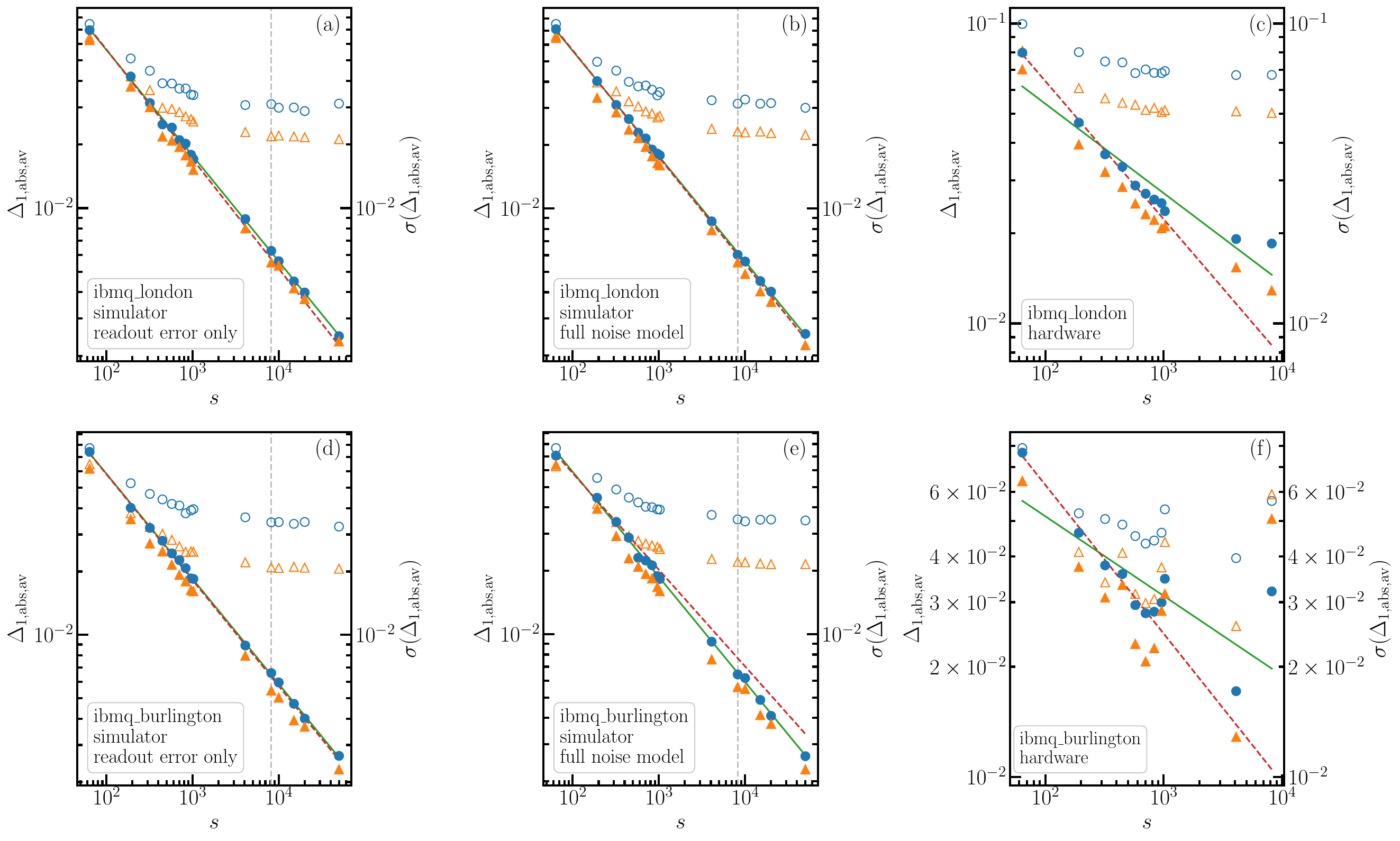}
	\caption{Mean value $\Delta_\mathrm{1,abs,av}$ (blue dots, left $y$-axes) and standard deviation $\sigma(\Delta_\mathrm{1,abs,av})$ (orange triangles, right $y$-axes) of the absolute error in Eq.~\eqref{eq:absolute_error_single_qubit} after applying the correction procedure (filled symbols) and without it (open symbols) as a function of the number of shots $s$. The panels in the upper row correspond to data obtained for ibmq\_london, the panels in the lower row to data obtained for ibmq\_burlington. The different columns correspond to a classical simulation of the quantum device taking into account only readout error (first column), the full hardware noise model (second column) and data obtained on actual quantum hardware (third column). The solid green lines corresponds to a power law fit to all our data points for the mean absolute error, the red dashed lines to fit including the lowest four number of shots. The vertical gray dashed lines in panels (a), (b) and (d), (e) indicate the maximum number of shots that can be executed on the actual hardware.}
	\label{fig:ibmq_burlington_london_single_qubit_vs_shots_all}
\end{figure*}

To benchmark the performance of our correction procedure, we first simulate ibmq\_london \cite{London2020} and ibmq\_burlington \cite{Burlington2020} classically. The Qiskit SDK provides a noise model for each of the respective chips comprising various sources of error, including readout errors during the measurement process, which can be switched on and off individually. To begin with, we simulate the quantum hardware incorporating the measurement errors only, and subsequently we use the full noise model to see the effect of the various other errors. Our results for the mean and the standard deviation of the absolute error as a function of $\nshots$ are shown in Figs.~\ref{fig:ibmq_burlington_london_single_qubit_vs_shots_all}(a), \ref{fig:ibmq_burlington_london_single_qubit_vs_shots_all}(b) and Figs.~\ref{fig:ibmq_burlington_london_single_qubit_vs_shots_all}(d), \ref{fig:ibmq_burlington_london_single_qubit_vs_shots_all}(e).

First, let us comment on our results before applying the mitigation procedure. For the case in which the only noise present on the device is the readout error, corresponding to Figs.~\ref{fig:ibmq_burlington_london_single_qubit_vs_shots_all}(a) and~\ref{fig:ibmq_burlington_london_single_qubit_vs_shots_all}(d), we can derive a relation between the bit-flip probabilities and the saturation values for the mean value of the absolute error in Eq.~\eqref{eq:absolute_error_single_qubit} (open blue dots). As we are interested in the plateau value of the average absolute error in the limit $s\to\infty$, we neglect in the following the statistical uncertainty due to a finite number of shots. Starting from Eq.~\eqref{eq:expvalue}, we can express the absolute error as
\begin{align}
    \left|\mathbb{E}\langle \tilde{Z}\rangle - \langle Z\rangle\right| = \left|-(p_0+p_1)\langle Z\rangle + p_1-p_0\right|.
\end{align}
For our choice of ansatz $\ket{\psi(\theta_1,\theta_0)} = R_z(\theta_1)R_x(\theta_0)\ket{0}$, a short analytic calculation yields that $ \langle Z\rangle = \cos^2(\theta_0/2) - \sin^2(\theta_0/2)$. Note that this expression is independent of $\theta_1$. Inserting this result into the equation above, we find
\begin{align}
    \left|\mathbb{E}\langle \tilde{Z}\rangle - \langle Z\rangle\right| &= \left|-2p_0\cos^2(\theta_0/2)  + 2p_1 \sin^2(\theta_0/2) \right|.
\end{align}
Averaging this quantity over the angles $\theta_0$ and $\theta_1$, we find the following expression for the average absolute error:
\begin{align}
\begin{split}
   & \overline{\left|\mathbb{E}\langle \tilde{Z}\rangle - \langle Z\rangle\right|}\\
   & = \frac{1}{4\pi^2}\int_{0}^{2\pi}\int_{0}^{2\pi} \left|-2p_0\cos^2(\theta_0/2)\right.  \\
    &\left. \hspace{7.5em}+ 2p_1 \sin^2(\theta_0/2) \right|d\theta_0 d\theta_1\\
     & = \frac{2(p_0+p_1)}{\pi}\left(2\arccos\sqrt{\frac{p_1}{p_0+p_1}}+2\frac{\sqrt{p_0p_1}}{p_0+p_1}-\frac\pi2\right)\\
  & \hspace{7.5em} +\frac{4p_1\left(\frac\pi2-2\arccos\sqrt{\frac{p_1}{p_0+p_1}}\right)}{\pi}.
  \end{split}
  \label{eq:noisy_prediction}
\end{align}
For our numerical simulations using the noise model that keeps the readout error only, the bit-flip probabilities from the Qiskit for the ibmq\_london device are $p_0=0$ and $p_1=0.0317$. Thus, we can insert these values into Eq.~\eqref{eq:noisy_prediction}, which results in a value of $ \overline{\left|\mathbb{E}\langle \tilde{Z}\rangle - \langle Z\rangle\right|} = p_1$, compatible with the plateau value of the absolute error (open blue dots) in Fig.~\ref{fig:ibmq_burlington_london_single_qubit_vs_shots_all}(a). For ibmq\_burlington, the noise model has the values $p_0 = 0.015$ and $p_1=0.034$, which yields $ \overline{\left|\mathbb{E}\langle \tilde{Z}\rangle - \langle Z\rangle\right|} \approx 0.0336$, in agreement with our results in Fig.~\ref{fig:ibmq_burlington_london_single_qubit_vs_shots_all}(d).

Now, let us comment on our results after applying the mitigation procedure. Focusing on the case with readout error only in Figs.~\ref{fig:ibmq_burlington_london_single_qubit_vs_shots_all}(a) and \ref{fig:ibmq_burlington_london_single_qubit_vs_shots_all}(d), we see that correcting our results according to Eq.~\eqref{eq:expvalue} clearly reduces the mean and the standard deviation of the absolute error in both cases. Without correction, the mean (standard deviation) of the absolute error converges to a value around 0.03 (0.02), and increasing $\nshots$ beyond 1024 does not significantly improve the results. In particular, this stagnation already happens for values of $\nshots$ below the maximum one possible on real hardware, hence showing that the readout error severely limits the accuracy that can be achieved. On the contrary, the corrected results show a significant improvement and a power-law decay of these quantities with $\nshots$. In particular, in the ideal, completely noise-free case, performing a projective measurement on $\ket{\psi}$ is nothing but sampling from a probability distribution, thus one would expect the mean error to decay as  $\propto \nshots^{-1/2}$. To check for that behavior, we can fit the same functional form as in Sec.~\ref{sec:multi_Z_section} to our data; the resulting exponents are shown in Tab.~\ref{tab:ibmq_burlington_london_single_qubit_vs_shots_simulator}. Indeed, we recover $\alpha = 1/2$, thus demonstrating that our correction procedure essentially allows us to recover the noise-free case.

Taking into account the full noise model in our simulations, which contains for instance gate errors and decoherence, we obtain the results in Figs.~\ref{fig:ibmq_burlington_london_single_qubit_vs_shots_all}(b) and \ref{fig:ibmq_burlington_london_single_qubit_vs_shots_all}(e). Compared to the case with readout errors only, the picture is very similar, which shows that the dominant error contribution for the single-qubit case is coming from the readout procedure. The mean and the standard deviation of the absolute error without any correction only approach marginally higher values than previously. Again, we observe a significant reduction of the mean and the standard deviation of the absolute error after applying the correction procedure, and a power-law decay with $\nshots$. Fitting a power law to our data yields once more exponents around $1/2$ (see Tab.~\ref{tab:ibmq_burlington_london_single_qubit_vs_shots_simulator}).

\begin{table}[t]
  \begin{tabular}{|l|c|c|}
    \hline
    readout error only & first 4 points & full range \\ \hline \hline
    ibmq\_london       & 0.519          & 0.501      \\ \hline
    ibmq\_burlington   & 0.503          & 0.499      \\ \hline \hline
    full noise model   & first 4 points & full range \\ \hline \hline
    ibmq\_london       & 0.508          & 0.500      \\ \hline
    ibmq\_burlington   & 0.459          & 0.503      \\ \hline
  \end{tabular}
  \caption{Exponents $\alpha$ obtained from fitting the power law $C\nshots^{-\alpha}$ to our simulator data for the mean absolute error in Figs.~\ref{fig:ibmq_burlington_london_single_qubit_vs_shots_all}(a), \ref{fig:ibmq_burlington_london_single_qubit_vs_shots_all}(b), and \ref{fig:ibmq_burlington_london_single_qubit_vs_shots_all}(d), \ref{fig:ibmq_burlington_london_single_qubit_vs_shots_all}(e) after applying the correction.}
  \label{tab:ibmq_burlington_london_single_qubit_vs_shots_simulator}
\end{table}

\subsubsection{Quantum hardware}
Our experiments can be readily carried out on quantum hardware, and we repeat the same simulations on ibmq\_london and ibmq\_burlington. The only difference with respect to the classical simulation is that $\nshots$ on those two devices is limited to a maximum number of 8192. Figures~\ref{fig:ibmq_burlington_london_single_qubit_vs_shots_all}(c) and \ref{fig:ibmq_burlington_london_single_qubit_vs_shots_all}(d) show our results obtained on real devices.

Comparing our data for the chip imbq\_london in Fig.~\ref{fig:ibmq_burlington_london_single_qubit_vs_shots_all}(c) to the classical simulation of the quantum hardware in Fig.~\ref{fig:ibmq_burlington_london_single_qubit_vs_shots_all}(a) and Fig.~\ref{fig:ibmq_burlington_london_single_qubit_vs_shots_all}(b), we observe qualitative agreement for $\nshots\leq 1024$. Compared to the classical simulation of the quantum device, the mean value and the standard deviation of the absolute error are in general larger on the hardware. Correcting for the readout error yields again a significant improvement and reduces the mean and the standard deviation of absolute error considerably. As before, we can fit our data to a power law. While for a small number of shots about $\nshots<500$ we observe again an exponent of about $1/2$, for a larger number of measurements the curve for the corrected result starts to flatten out and the exponent obtained for fitting the entire range is considerably smaller than $1/2$ (see Tab.~\ref{tab:ibmq_burlington_london_single_qubit_vs_shots_hardware} for details). Since increasing $\nshots$ should decrease the inherent statistical fluctuations of the projective measurements, and readout errors can be managed with our scheme, this might be an indication that in addition to readout errors, also other sources of noise play a significant role. Their effects cannot be corrected with our procedure and thus dominate from a certain point on.

Looking at the results for imbq\_burlington in Fig.~\ref{fig:ibmq_burlington_london_single_qubit_vs_shots_all}(f) and comparing them with the classical simulation of the quantum hardware in Fig.~\ref{fig:ibmq_burlington_london_single_qubit_vs_shots_all}(d) and Fig.~\ref{fig:ibmq_burlington_london_single_qubit_vs_shots_all}(e), we see that the discrepancies in this case are more severe and the data is less consistent. Applying our mitigation to the data again yields an improvement, which is less pronounced than in the case of imbq\_london. For a small number of shots, the mean of the absolute error after correction shows again roughly a power-law decay. The exponent obtained from a fit to our data in that range is smaller compared to the one from our data from ibmq\_london (see Tab.~\ref{tab:ibmq_burlington_london_single_qubit_vs_shots_hardware} for details). From $s= 1024$ on, the uncorrected data is already less consistent. Making use of our mitigation scheme still yields an improvement, however, the corrected results scatter similarly to the original ones and do not follow the same power law as for a small number of shots, as a fit to our data reveals.  This suggests that noise other than the one resulting from the measurement has a considerable contribution.

\subsection{Two-qubit case\label{sec:results_two_qubit}}
Since our bit-flip correction procedure is not limited to the single-qubit setup, we can straightforwardly apply it to multiple qubits. To assess the performance for that case, we repeat the same procedure that we did previously but now for a circuit encompassing two qubits. Since we assume the bit-flip probabilities $p_{q,b}$ (with $q=1,2$, $b=0,1$) of the qubits to be independent of each other, we apply the same procedure that we used to obtain the bit-flip probabilities in the single-qubit case, but this time for each qubit individually.

Subsequently we prepare a two-qubit state using the following circuit
\begin{align*}
  \Qcircuit @C=1em @R=.7em {
  \lstick{\ket{0}} & \gate{R_x(\theta_0)} & \gate{R_z(\theta_1)} & \ctrl{1} & \meter      & \qw \\
  \lstick{\ket{0}} & \gate{R_x(\theta_2)} & \gate{R_z(\theta_3)} & \targ    & \meter      & \qw \\
  \lstick{c}       & \cw                  & \cw                  & \cw      & \cw\cwx[-1] & \cw
  }
\end{align*}
where the angles $\theta_0,\dots,\theta_3$ are again random numbers drawn uniformly from $[0,2\uppi]$, and the final CNOT gate allows for creating entanglement between the two qubits. Analogous to the single-qubit case, we first simulate the quantum hardware classically before we eventually carry out our experiments on a real quantum device. In both cases we measure the noisy expectation value of $Z_2\otimes Z_1$, $\En(\tilde{Z}_2\otimes \tilde{Z}_1)$, and apply Eq.~\eqref{eq:p_q_b_hist_extraction} to correct for noise caused by readout errors. Again, we repeat the procedure for 1050 randomly chosen sets of angles and compute the mean and the standard deviation of the absolute error,
\begin{align}
  \Delta_\mathrm{2,abs} = |\bra{\psi}\tilde{Z}_2\otimes \tilde{Z}_1\ket{\psi}_\text{measured} - \bra{\psi}Z_2\otimes Z_1\ket{\psi}_\text{exact}|,
  \label{eq:absolute_error_two_qubit}
\end{align}
as a function of the number of shots, $\nshots$, with and without applying the mitigation scheme.

\begin{table}[t]
  \begin{tabular}{|l|c|c|}
    \hline
    chip name        & first 4 points & full range \\ \hline \hline
    ibmq\_london     & 0.460          & 0.298      \\ \hline
    ibmq\_burlington & 0.405          & 0.217      \\ \hline
  \end{tabular}
  \caption{Exponents $\alpha$ obtained from fitting the function $C\nshots^{-\alpha}$ to our hardware data for the mean absolute error in Figs.~\ref{fig:ibmq_burlington_london_single_qubit_vs_shots_all}(c) and ~\ref{fig:ibmq_burlington_london_single_qubit_vs_shots_all}(f) after applying the correction.}
  \label{tab:ibmq_burlington_london_single_qubit_vs_shots_hardware}
\end{table}

\subsubsection{Classical simulation of quantum hardware}
\begin{figure*}
  \centering
  \includegraphics[width=1.0\textwidth]{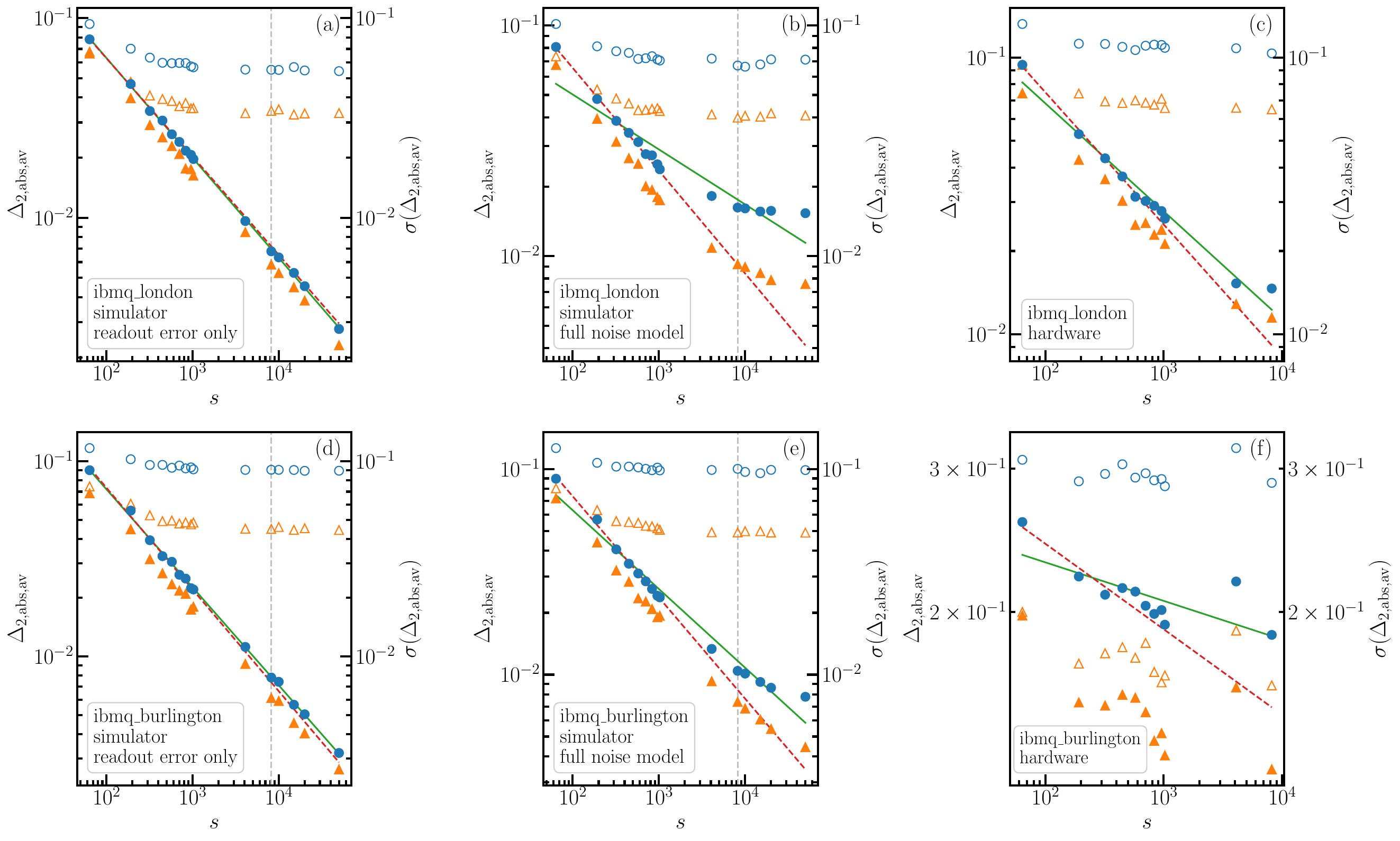}
  \caption{Mean value $\Delta_\mathrm{2,abs,av}$ (blue dots, left $y$-axes) and standard deviation $\sigma(\Delta_\mathrm{2,abs,av})$ (orange triangles, right $y$-axes) of the absolute error in Eq.~\eqref{eq:absolute_error_two_qubit} after applying the correction procedure (filled symbols) and without it (open symbols) as a function of the number of shots $s$. The panels in the upper row correspond to data obtained for ibmq\_london, the panels in the lower row to data obtained for ibmq\_burlington. The different columns correspond to a classical simulation of the quantum device taking into account only readout error (first column), the full hardware noise model (second column) and data obtained on actual quantum hardware (third column). The solid green lines corresponds to a power law fit to all our data points for the mean absolute error, the red dashed lines to fit including the lowest four number of shots. The vertical gray dashed lines in panels (a), (b) and (d), (e) indicate the maximum number of shots that can be executed on the actual hardware.}
  \label{fig:ibmq_burlington_london_two_qubit_vs_shots_all}
\end{figure*}

As for the single-qubit case, we use the Qiskit SDK to classically simulate the chips imq\_london and ibmq\_burlington first with readout error only and subsequently using the full noise model. Figures~\ref{fig:ibmq_burlington_london_two_qubit_vs_shots_all}(a), \ref{fig:ibmq_burlington_london_two_qubit_vs_shots_all}(b) and \ref{fig:ibmq_burlington_london_two_qubit_vs_shots_all}(d), \ref{fig:ibmq_burlington_london_two_qubit_vs_shots_all}(e) show our results for both cases.

Looking at Fig.~\ref{fig:ibmq_burlington_london_two_qubit_vs_shots_all}(a) and Fig.~\ref{fig:ibmq_burlington_london_two_qubit_vs_shots_all}(d), we see that the two-qubit case with just readout error behaves like the single-qubit case. Without applying any correction, the mean and the standard deviation of the absolute error initially decrease with increasing $\nshots$, before eventually converging to fixed values which are slightly higher than for the single-qubit case (compare Fig.~\ref{fig:ibmq_burlington_london_single_qubit_vs_shots_all}(a) with Fig.~\ref{fig:ibmq_burlington_london_two_qubit_vs_shots_all}(a) and Fig.~\ref{fig:ibmq_burlington_london_single_qubit_vs_shots_all}(b) with Fig.~\ref{fig:ibmq_burlington_london_two_qubit_vs_shots_all}(b)). Applying the correction procedure, we can significantly decrease the values and observe again a power-law decay with an exponent of $1/2$ over the entire range of $\nshots$ we study, as a fit to our corrected data reveals (see also Tab.~\ref{tab:ibmq_burlington_london_two_qubit_vs_shots_simulator}).

Repeating the same simulations, but this time with the full noise model, yields the results in Fig.~\ref{fig:ibmq_burlington_london_two_qubit_vs_shots_all}(b) and Fig.~\ref{fig:ibmq_burlington_london_two_qubit_vs_shots_all}(e). Comparing this to the case with readout error only, we see a more pronounced effect than in the single-qubit case. Applying the correction reduces the mean and the standard deviation of the absolute error still considerably, nevertheless one can observe that data after correction converge to a fixed value with increasing $\nshots$. In particular, the power law decay with $\alpha = 1/2$ is only present for a small number of shots. Considering the entire range of $\nshots$ we study, the classical simulation of ibmq\_london predicts that the data are not very compatible with a power law. In contrast, our simulation data for ibmq\_burlington are still reasonably well described by a power law, however with an exponent of $0.38$ and thus considerably smaller than $1/2$ (see Tab.~\ref{tab:ibmq_burlington_london_two_qubit_vs_shots_simulator} for details). Most notably, from a comparison between the results for classically simulating two qubits using the full noise model to the single-qubit case in Fig.~\ref{fig:ibmq_burlington_london_single_qubit_vs_shots_all}(b) and Figs.~\ref{fig:ibmq_burlington_london_single_qubit_vs_shots_all}(e), we see that noise has a substantially larger effect in the two-qubit case. This can be partially explained by the CNOT gate in the circuit, as the error rates for two-qubit gates are in general much larger than for single-qubit rotations.

\begin{table}[t]
  \begin{tabular}{|l|c|c|}
    \hline
    readout error only & first 4 points & full range \\ \hline \hline
    ibmq\_london       & 0.492          & 0.501      \\ \hline
    ibmq\_burlington   & 0.522          & 0.503      \\ \hline \hline
    full noise model   & first 4 points & full range \\ \hline \hline
    ibmq\_london       & 0.446          & 0.238      \\ \hline
    ibmq\_burlington   & 0.492          & 0.383      \\ \hline
  \end{tabular}
  \caption{Exponents $\alpha$ obtained from fitting the power law $C\nshots^{-\alpha}$ to our simulator data for the mean absolute error in Figs.~\ref{fig:ibmq_burlington_london_two_qubit_vs_shots_all}(a), \ref{fig:ibmq_burlington_london_two_qubit_vs_shots_all}(b) and \ref{fig:ibmq_burlington_london_two_qubit_vs_shots_all}(d), \ref{fig:ibmq_burlington_london_two_qubit_vs_shots_all}(e).}
  \label{tab:ibmq_burlington_london_two_qubit_vs_shots_simulator}
\end{table}

\subsubsection{Quantum hardware}
For the two-qubit case, we can carry out the simulations on real quantum hardware as well. Using again imbq\_london and ibmq\_burlington, we obtain the data depicted in Fig.~\ref{fig:ibmq_burlington_london_two_qubit_vs_shots_all}(c) and Fig.~\ref{fig:ibmq_burlington_london_two_qubit_vs_shots_all}(f).

Our results for ibmq\_london in Fig.~\ref{fig:ibmq_burlington_london_two_qubit_vs_shots_all}(c) show qualitative agreement with the classical simulation. Once more, we see that the  mean and the standard deviation of the absolute error obtained on the hardware converge to higher values than the ones obtained from the simulation (compare Fig.~\ref{fig:ibmq_burlington_london_two_qubit_vs_shots_all}(b) and Fig.~\ref{fig:ibmq_burlington_london_two_qubit_vs_shots_all}(c)). Correcting our data according to Eq.~\eqref{eq:p_q_b_hist_extraction}, the mean of the absolute error and its standard deviation are significantly reduced. Comparing the reduction to the single-qubit case in Fig.~\ref{fig:ibmq_burlington_london_single_qubit_vs_shots_all}, we observe that for the two-qubit case, the improvement is even larger. In particular, for our largest number of shots $\nshots=8192$, the mean and the standard deviation of the absolute error are reduced by approximately one order of magnitude. The corrected data is again well described by a power law. Fitting the first 4 data points, we obtain an exponent of $0.48$. Using the entire range of $\nshots$ for the fit, the exponent only decreases moderately to $0.39$ (see also Tab.~\ref{tab:ibmq_burlington_london_two_qubit_vs_shots_hardware}), thus showing that the readout error has still a significant contribution to the overall error.

Turning to our results for ibmq\_burlington in Fig.~\ref{fig:ibmq_burlington_london_two_qubit_vs_shots_all}(f), we see that the data for this chip is significantly worse. For one, the mean value (standard deviation) of the absolute error without applying any correction procedure is roughly a factor 3 (2) larger than the one obtained on ibmq\_london. Applying the correction procedure still yields an improvement, however, this time it is a lot smaller than for ibmq\_london, as a comparison between Figs.~\ref{fig:ibmq_burlington_london_two_qubit_vs_shots_all}(c) and~\ref{fig:ibmq_burlington_london_two_qubit_vs_shots_all}(f) shows. While for a small number of shots, the mean value of the absolute error after correction still shows a power law decay, albeit with an exponent a lot smaller than $1/2$, for a large number of shots this trend stops, as fits to our data reveal (see also Tab.~\ref{tab:ibmq_burlington_london_two_qubit_vs_shots_hardware}). This behavior indicates that for ibmq\_burlington, the readout error is not the dominant one, but rather other errors have a significant contribution that cannot be corrected for using our scheme.

\section{Discussion\label{sec:discussion}}
After demonstrating the applicability of our mitigation method to real quantum hardware, we discuss our results here in greater detail. We comment on the relation to previous works on error mitigation, and we address how our scheme allows for extensions beyond those. In particular, we discuss the inclusion of multi-qubit correlations and the generalization to other types of errors such as relaxation. Moreover, we address some questions regarding the practical implementation, such as the overhead costs introduced, pre-processing versus post-processing, and the possibility of doing probabilistic error mitigation.

\begin{table}[t]
  \begin{tabular}{|l|c|c|}
    \hline
    chip name        & first 4 points & full range \\ \hline \hline
    ibmq\_london     & 0.478          & 0.390      \\ \hline
    ibmq\_burlington & 0.105          & 0.047      \\ \hline
  \end{tabular}
  \caption{Exponents $\alpha$ obtained from fitting the power law $C\nshots^{-\alpha}$ to our hardware data for the mean absolute error in Fig.~\ref{fig:ibmq_burlington_london_two_qubit_vs_shots_all}(c) and \ref{fig:ibmq_burlington_london_two_qubit_vs_shots_all}(f).}
  \label{tab:ibmq_burlington_london_two_qubit_vs_shots_hardware}
\end{table}

\subsection{Comparison to previous work}

One way to mitigate measurement errors that has been put forward in the literature~(see, e.g., Ref.~\cite{Qiskit}) is to construct a linear map, which relates the observed measurement outcomes for each computational basis state to the state that was actually prepared. To this end, one prepares all computational basis states $\ket{i}$, $i=0,\dots,2^{N}-1$, on the quantum device and records the probabilities $p_{ji}$ of obtaining the computational basis state $\ket{j}$ after a projective measurement. The linear map $\omega = (p_{ji})_{i,j=0}^{2^N-1}$ now relates the observed probability distribution of basis states $\tilde{P}$ in a noisy measurement to the ideal distribution $P$ as $\tilde{P} = \omega P$. Thus, one can in principle obtain the exact solution $P$ from the observed results by inverting $\omega$ and post-processing $\tilde{P}$. Obviously, the method scales exponentially with $N$ in terms of the number of measurements and memory requirements. In addition, $\omega$ can be singular and a direct inversion might not be possible. Even if $\omega^{-1}$ exists, it is not guaranteed to be stochastic, such that the result obtained might not be a valid probability distribution.

To overcome these shortcomings, it has been proposed to mitigate measurement errors by expressing the error-corrected result in terms of a sum of noisy outcomes and combinations of bit-flip probabilities~\cite{Kandala:2017,YeterAydeniz2019,YeterAydeniz2020}. 
This was first studied by Kandala et al.~\cite{Kandala:2017} for the case of single-qubit $Z$-operator measurements. An extension of this method has been provided by Yeter-Aydeniz et al.~\cite{YeterAydeniz2019,YeterAydeniz2020} for multi-qubit $Z_Q\otimes\cdots\otimes Z_1$-operators with expectation values measured from bit-string distributions. Our approach provides an alternative proof for some of the results in Refs.~\cite{YeterAydeniz2019,YeterAydeniz2020}, which offers an implementation beyond bit-string distributions and allows for several further generalizations, which are discussed in the following subsections. In particular, our results can be extended to multi-qubit correlation errors (see Sec.~\ref{sec:correlations}), relaxation errors (see Sec.~\ref{sec:relaxation}), and probabilistic mitigation schemes (see Sec.~\ref{sec:probabilistic}). Moreover, while previous results all rely on post-processing of the error mitigation scheme, our method allows for pre-processing as well. Thus, it can be readily integrated into hybrid quantum-classical algorithms, such as the variational quantum eigensolver (VQE) (see Sec.~\ref{sec:pre-processing}).

\subsection{Inclusion of multi-qubit correlations}\label{sec:correlations}

In our paper, we have assumed for simplicity that there are no multi-qubit correlations in multi-qubit $Z_Q\otimes\cdots\otimes Z_1$-operator measurements. This is because most of the physically relevant Hamiltonians only contain local interaction terms (see the discussion in Sec.~\ref{sec:overhead}). As such, the number of qubits for each multi-qubit $Z_Q\otimes\cdots\otimes Z_1$-operator measurement is relatively small and the correlations are negligible. 

However, as more qubits are measured simultaneously, multi-qubit correlations can become significant and have to be taken into account. This has not been incorporated into the above-mentioned mitigation schemes~\cite{Kandala:2017,YeterAydeniz2019,YeterAydeniz2020} and, to our knowledge, it has only been addressed with methods that are exponentially costly with the number of qubits~\cite{Geller2020,Geller2020a}. 

Our measurement mitigation scheme can easily take multi-qubit correlations into account, because the fundamental step of our approach is the replacement of the operator to be measured with a probability distribution of operators. Adding multi-qubit correlations into this probability distribution is straight-forward and only requires multi-qubit calibration results similar to the single-qubit calibrations discussed in Appendix~C. For example, while the single-qubit calibrations required measuring $p(\ket{j}|\ket{k})$ for $j,k\in\{0,1\}$, the two-qubit calibrations would require measuring $p(\ket{j}|\ket{k})$ for $j,k\in\{00,01,10,11\}$. Since we are interested in $n$-local Hamiltonians with at most $n$-qubit interactions (see also the discussion in Sec.~\ref{sec:overhead}), the calibration cost scales polynomially in the number of qubits $N$, as no more than $n$ qubits are measured simultaneously. Indeed, the multi-qubit calibration method requires the calibration of ${N\choose n}$ $n$-qubit systems with fixed $n$, which requires $\mathcal{O}(N^n)$ calibrations. Thus, incorporating multi-qubit correlations into our mitigation scheme is straight-forward and requires relatively small overhead costs compared to previous approaches.

\subsection{Extension to relaxation errors}\label{sec:relaxation}

Our method of replacing noisy operators with random operators that model the noise behavior, as presented in Sec.~\ref{sec:mitmeasure}, can in principle be generalized to other types of errors on noisy quantum computers. For example, if we wish to measure the operator $Z$ and consider the relaxation error $T_1$ (decay of $\ket{1}$ to $\ket{0}$)~\cite{Qiskit}, then we have a probability distribution of measuring $Z$ (not yet decayed) and $\Id$ (decayed). Hence, the measurement outcome subject to the $T_1$ error is described by $\tilde{Z} = p(t)Z + [1-p(t)]\Id$, where $p(t)=\exp\left(-\frac{t}{T_1}\right)$ is the probability that $\ket{1}$ has not yet decayed. With our scheme of replacing noisy with random operators, this $T_1$ error can be corrected as $Z = \frac{1}{p(t)}\tilde{Z} - \frac{1-p(t)}{p(t)}\Id$. As such, our approach is generalizable to other types of errors beyond measurement errors, and it needs to be adapted accordingly in order to correctly incorporate the parameters underlying the specific type of error.

\subsection{Probabilistic implementation of the measurement error mitigation scheme}\label{sec:probabilistic}

The probabilistic description of the noisy operator naturally lends itself to a probabilistic implementation of the mitigation scheme. While deterministic mitigation schemes require the measurement of all mitigation terms, a probabilistic protocol allows for partial error mitigation if the full mitigation is too costly. For example, if the corrected operator $Z\otimes Z$ is given by $\alpha_1\tilde{Z}\otimes \tilde{Z}+\alpha_2\tilde{Z}\otimes \Id+\alpha_3\Id\otimes \tilde{Z}+\alpha_4\Id\otimes \Id=A(p_1s_1\tilde{Z}\otimes \tilde{Z}+p_2s_2\tilde{Z}\otimes \Id+p_3s_3\Id\otimes \tilde{Z}+p_4s_4\Id\otimes \Id)$ with $A:=\sum_j\abs{\alpha_j}$, $s_j:=\mathrm{sgn}(\alpha_j)$, $p_j:=\frac{\abs{\alpha_j}}{A}$, and $\sum_j p_j=1$, then we may randomly draw the operator to measure; namely, $As_1\tilde{Z}\otimes \tilde{Z}$ with probability $p_1$, $As_2\tilde{Z}\otimes \Id$ with probability $p_2$, et cetera. This can be used on a single-operator measurement, commuting sets of operators, as well as on the level of drawing random Hamiltonians. It can also be generalized to other types of errors, such as the inclusion of multi-qubit correlations.

\subsection{Pre-processing the mitigation scheme}\label{sec:pre-processing}

All of the previously known measurement mitigation schemes rely on post-processing, that is, first measuring without taking the error into account and afterwards manipulating the obtained data (see, e.g., Refs.~\cite{Qiskit,Kandala:2017,YeterAydeniz2019,YeterAydeniz2020,Geller2020,Geller2020a}). However, this is not always possible using ``black box subroutines'', such as VQE routines provided by SDKs. Such routines typically ask for the Hamiltonian to be passed as an argument, and they will return the optimized parameter set. They do not allow for user-supplied error mitigation methods to be incorporated. In contrast, the approach presented in this work allows for pre-processing. Hence, rather than passing on the Hamiltonian we are interested in solving, we can pass on the bit-flip corrected Hamiltonian instead. The user is therefore able to manually insert an error-mitigation scheme into the ``black-box subroutine''.

\subsection{Moderate overhead costs}\label{sec:overhead}

For local Hamiltonians, the computational cost of our mitigation routine scales polynomially with the number of qubits. 
For non-local Hamiltonians, our mitigation routine does not add any computational cost with respect to the measurement itself, as the measurement of the expectation value already exhibits exponential complexity. We will explain and exemplify both of these cases in the following.

For non-local Hamiltonians, let us consider the example of a generic operator acting on $N$ qubits. This operators is a linear combination of all $4^N$ $N$-qubit Pauli matrices, i.e., a tensor product of $N$ $2\times 2$ Pauli matrices, which include the $2\times 2$ identity. Our mitigation method now replaces each tensor product by a sum of up to $2^N$ operators, which already need to be measured for the full Hamiltonian measurement. Thus, the replacement only changes the coefficients of these operators but does not incur any overhead on the quantum device. Moreover, Hamiltonians with non-local interactions are likely to incur exponential complexity already in the evaluation of the expectation value, thus making them unfeasible to measure, let alone error-correct. 

For $n$-local Hamiltonians, the individual Pauli terms do not act on all $N$ qubits but on a given number $Q\leq n \leq N$ of qubits, which is independent of the total number $N$ of qubits. For example, the Ising model, the Heisenberg model, and the Schwinger model (after integrating out the gauge field, see, e.g., Refs.~\cite{Banuls2013,Banuls2016a,Banuls2016b,Funcke:2019zna}) exhibit at most two-qubit interaction terms and thus have $Q\leq n = 2$. For our mitigation method, we now need to replace each tensor product of the $Q$ non-identity Pauli matrices by up to $2^Q$ operators. For each $Q\leq n$, there are polynomially many of these terms, $N \choose Q$, and we can estimate the total number using the upper bound $\sum_{Q\le n} {N \choose Q} \le (n+1) N^n$. For each of these terms, the error correction matrix $\omega$ is of magnitude $2^Q$, i.e., bounded by a constant of magnitude $2^n$. Each of the matrices $\omega$ are triangular and thus can be inverted with a computational cost of $\mathcal{O}(4^n)$, which is still constant.  The entire operation requires fewer than $(n+1) N^n$ times $\mathcal{O}(1)$ operations, i.e., the computational complexity is $\mathcal{O}(N^n)$ and thus scales polynomially in the number of qubits $N$. Note that for $n$-local interactions between \textit{adjacent} qubits, the computational complexity gets reduced even further, because we then only have $nN$ instead of $\sum_{Q=1}^n {N\choose Q}$ different terms.

For the Ising, Heisenberg, and Schwinger models mentioned above, we can explicitly estimate the number of terms required for our error-mitigation method. For each qubit $q$, we have at most three single-qubit Paulis ($X_q$, $Y_q$, $Z_q$), as well as three two-qubit Paulis ($X_qX_{q+1}$, $Y_qY_{q+1}$, $Z_qZ_{q+1}$). Hence, for $N$ qubits, our mitigation method requires the inversion of $3N$ matrices $\omega$ (size $2\times2$) for the single-qubit Paulis and $3N$ matrices $\omega$ (size $4\times 4$) for the two-qubit Paulis. In other words, the overall complexity of the error mitigation for $N$ qubits is bounded by $3N$ triangular matrix inversions of size $2\times 2$ plus $3N$ triangular matrix inversions of size $4\times 4$. 

In some cases, the number of additional terms that have to be measured on the quantum device can be reduced even further. For example, the LI model discussed in Sec.~\ref{sec:LI} only contains the Paulis $Z_qZ_{q+1}$ and $Z_q$ for each qubit $q$. While error-correcting the $Z_qZ_{q+1}$ terms, we automatically error-correct the $Z_q$ terms as well. Hence, the overall complexity for the error mitigation of the LI model for $N$ qubits is equivalent to $N$ triangular matrix inversions of size $4\times 4$. Thus, the LI model incurs no overhead cost on the quantum device.

\section{Conclusions\label{sec:conclusion}}

In this paper, we proposed a classical bit-flip correction method to mitigate measurement errors on noisy quantum computers. This method relies on cancellations of different erroneous measurement outcomes and requires knowledge of the different bit-flip probabilities during readout for each qubit. We tested the performance of this method by correcting the noisy energy histograms of the longitudinal and transversal Ising models. Moreover, we demonstrated that the method can be applied to any operator, any number of qubits, and any realistic bit-flip probability. For the single-qubit case, we also provided a density matrix description and a visualization scheme of the measurement noise. Finally, we tested our method both numerically and experimentally for the IBM quantum devices ibmq\_london and ibmq\_burlington for both a single qubit and two qubits. We observe that our method is able to improve the data significantly for both cases and to reduce the error by up to one order of magnitude.

Our method of replacing noisy operators with random operators that model the noise behavior, as we presented in Sec.~\ref{sec:mitmeasure}, is generally applicable to arbitrary observables and could also be applied to other error sources, such as relaxation errors. As stated in Sec.~\ref{sec:discussion}, the computational cost of our mitigation routine is moderate (i.e., polynomial) for local Hamiltonians, even if multi-qubit correlations are included. For non-local Hamiltonians, our mitigation routine does not add any computational cost with respect to the measurement itself, as the measurement of the expectation value already exhibits exponential complexity.

In addition to the moderate overhead cost, another advantage of our mitigation scheme is that it can be readily integrated into hybrid quantum-classical algorithms, as for example the Quantum Approximate Optimization Algorithm~\cite{Farhi2014} and VQS. After initially measuring the bit-flip probabilities, one can simply correct the values obtained for the cost function from the quantum device, before passing them on to a classical algorithm for optimizing the variational parameters. Moreover, in contrast to previous mitigation schemes, our method also allows for pre-processing. Thus, the user can manually insert the bit-flip corrected Hamiltonian into ``black-box subroutines'' such as VQS routines provided by SDKs, which allows for on-the-fly error mitigation.

Finally, our method is completely platform-independent and lends itself not only to superconducting qubits, but also to other architectures such as trapped ions. 
As long as the measurement errors are constant to a certain degree and not excessively large, they can be reliably corrected for with our procedure. 
These advantages make our mitigation method promising for various applications on NISQ devices but also beyond.

\section*{Note added in proof}

Recently, we became aware of a related mitigation method~\cite{Nation:2021kye}, which also assumes uncorrelated measurement errors and thus replaces the exponential scaling of common methods by a polynomial one. The implementation in the current paper goes further by constructing operators that correct for the measurement error on the operator level. While the current paper demonstrates the performance of the method with two-qubit experiments, Ref.~\cite{Nation:2021kye} provides a demonstration on up to 42 qubits.

\begin{acknowledgments}
The authors thank Xu Feng, Georgios Polykratis, Philipp Stratmann, and Tom Weber for pointing out the probabilistic implementation of the mitigation scheme. Research at Perimeter Institute is supported in part by the Government of Canada through the Department of Innovation, Science and Industry Canada and by the Province of Ontario through the Ministry of Colleges and Universities.
S.K.\ acknowledges financial support from the Cyprus Research and Innovation Foundation under project "Future-proofing Scientific Applications for the Supercomputers of Tomorrow (FAST)", contract no.\ COMPLEMENTARY/0916/0048. P.S.\ thanks the Helmholtz Einstein International Berlin Research School in Data Science (HEIBRiDS) for funding. We acknowledge the use of IBM Quantum services for this work. The views expressed are those of the authors, and do not reflect the official policy or position of IBM or the IBM Quantum team.
\end{acknowledgments}

\appendix

\section{Illustration of mitigation method for simplified energy histograms}\label{sec:illustration}

In this appendix, we provide a pedagogic explanation of our mitigation method. We describe in Sec.~\ref{sec:bindistr} how the measured energy histograms can be described by a binomial distribution in certain cases. We then discover in Sec.~\ref{sec:p=0.5} that the mean energy of the distribution vanishes if all qubits have equal bit-flip probabilities, $p=0.5$. We finally discuss the deviation $\Delta \tilde{E}$ between the measured mean energy and the noiseless true energy for the specific cases of non-interacting Hamiltonians (Sec.~\ref{sec:non-int}), interacting Hamiltonians (Sec.~\ref{sec:int}), and the example of the TI model (Sec.~\ref{sec:TI}).

\subsection{Binomial distribution of measurements\label{sec:bindistr}}
To start with, let us focus on a diagonal Hamiltonian $\mathcal{H}$ with eigenstate $\ket{\psi}$. To evaluate the corresponding energy, $E=\bra{\psi}\mathcal{H}\ket{\psi}$, on a quantum device, we have to (i) run the quantum circuit preparing the state $\ket{\psi}$, (ii) projectively measure the energy in the (computational) basis, and (iii) record the outcomes. Performing $\nshots$ shots, we record $k$ correct results with $\tilde{E} = E$ but $\nshots-k$ incorrect results with $\tilde{E}\neq E$, where the tilde denotes a noisy outcome. For simplicity, we assume that (i) a wrong measurement originates from a \textit{single} bit flip with probability $p$, and (ii) each bit flip yields the \textit{same} deviation from $E$. We will see later that these assumptions will need to be modified in the presence of multi-qubit interactions, for example, for the LI and TI models.

The probability of getting $k$ correct measurement results is given by the probability mass function
\begin{equation}
  f(k,\nshots,1-p) = \binom{\nshots}{k}\ (1-p)^k\ p^{\nshots-k},
  \label{eq:binomial}
\end{equation}
where $p$ is the probability of \textit{incorrectly} measuring the energy.
The resulting noisy energy histograms can be described in terms of the number $k$ of correct measurements,
\begin{align}
  \begin{aligned}
    \tilde{E}(k) &= E + (\nshots-k)\Delta \tilde{E} \\
    &=\begin{cases}
      E                     & \text{for}\ k=\nshots, \\
      E + \nshots\Delta \tilde{E} & \text{for}\ k=0,
    \end{cases}
  \end{aligned}
  \label{eq:E(k)}
\end{align}
where $\Delta \tilde{E}$ is the deviation from $E$ per bit flip. In terms of the bit-flip probability $p$, the resulting noisy expectation $\En$ of the measured energy $\tilde{E}$ reads
\begin{align}
  \begin{aligned}
  \En \tilde{E} &=E + \nshots p\Delta \tilde{E} \\
  &=
  \begin{cases}
    E                     & \text{for}\ p=0, \\
    E + \nshots\Delta \tilde{E} & \text{for}\ p=1.
    \end{cases}
  \end{aligned}
  \label{eq:E(p)}
\end{align}
We note that ``expectation'' here means the expectation with respect to the probability $p$, which should not be confused with the quantum-mechanical expectation value of the Hamiltonian, $\bra{\psi}\mathcal{H}\ket{\psi}=E$. Thus, the expectation $\En \tilde{\mathcal{H}}$ is the expected value (as an operator to be measured subject to bit flips; see also Sec.~\ref{sec:mitmeasure}) for the noisy Hamiltonian $\tilde{\mathcal{H}}$, while $\En\bra{\psi}\tilde{\mathcal{H}}\ket{\psi}=\En \tilde{E}$ is the expected value for the noisy (quantum-mechanical) expectation value $\bra{\psi}\tilde{\mathcal{H}}\ket{\psi}=\tilde{E}$.

For a large number of shots $\nshots$, the noisy energy histograms can be described by a normal distribution with mean energy $\En \tilde{E}$ given by Eq.~\eqref{eq:E(p)}. The only free parameter of this measurement noise model is $\Delta \tilde{E}$, since $\nshots$ and $p$ are known input parameters.

\subsection{Mean energy vanishes for $p=0.5$\label{sec:p=0.5}}

The first step towards eliminating the free parameter $\Delta \tilde{E}$ is to study the dependence of this parameter on the bit-flip probability $p$, for example for $p=0.5$. Let us consider the noise-free Hamiltonian $\mathcal{H}$ acting on the state $\ket{\psi} = c_0 \ket{0} + c_1 \ket{1}$ and yielding the energy
\begin{equation}
  \bra{\psi} \mathcal{H} \ket{\psi} = (c_0^* \bra{0} + c_1^* \bra{1})\ \mathcal{H}\ (c_0 \ket{0} + c_1 \ket{1})=E.
  \label{eq:H}
\end{equation}
The noisy measurement of this energy on the quantum hardware is performed along the basis $Z=\text{diag}(1,-1)$. We note that this noisy measurement yields $\En \tilde{E}=0$ for $p=0.5$, due to the opposite signs of the terms resulting from bit-flipping the terms in Eq.~\eqref{eq:H}.

Let us demonstrate this for a simple example, $\mathcal{H}_X=X=HZH$, where $H$ is the Hadamard gate, and study the possible outcomes of the energy measurements in the single-qubit case:

\begin{itemize}
\item The absence of any bit flip gives the true energy of the noise-free Hamiltonian:
\begin{align}
\begin{split}
\bra{\psi} \mathcal{H}_X \ket{\psi} & = \bra{\psi} HZH \ket{\psi}                            \\
& = [c_0^* (\bra{0}+\bra{1}) + c_1^* (\bra{0}-\bra{1})]\ \\
& \quad \ Z\ [c_0 (\ket{0}+\ket{1}) + c_1 (\ket{0} -\ket{1})]         \\
& = |c_0+c_1|^2 - |c_0-c_1|^2 = E.
\end{split}
\end{align}        
\item The bit flip $\ket{0}\to\ket{1}$, $\ket{1}\to \ket{1}$ changes one sign: $\bra{\psi} \tilde{\mathcal{H}}_X \ket{\psi} = -|c_0+c_1|^2 - |c_0-c_1|^2$.
\item The bit flip $\ket{0}\to\ket{0}$, $\ket{1}\to \ket{0}$ changes the other sign: $\bra{\psi} \tilde{\mathcal{H}}_X \ket{\psi} =  |c_0+c_1|^2 + |c_0-c_1|^2$.
\item The bit flip $\ket{0}\to\ket{1}$, $\ket{1}\to \ket{0}$ changes both signs: $\bra{\psi} \tilde{\mathcal{H}}_X \ket{\psi} =  -|c_0+c_1|^2 + |c_0-c_1|^2$ and thus yields the outcome $-E$.
\end{itemize}
For $p=0.5$, each of these four possible outcomes has the same probability $p^2=0.25$, and thus cancellation yields $\En \tilde{E} =0$. This result holds true for any operator and any number of qubits, as we showed in Eq.~\eqref{eq:ExpArbitOp}.

\subsection{$\Delta \tilde{E}$ for non-interacting Hamiltonians}\label{sec:non-int}

The next step towards eliminating the free parameter $\Delta \tilde{E}$ is to examine the four possible measurement outcomes from the previous section for \textit{any} bit-flip probability $p$. We observe that the second and third outcomes have \textit{opposite} signs and \textit{equal} probability and thus cancel, given the above assumption that $p(\ket{0}\to\ket{1}) = p(\ket{1}\to\ket{0})$. For N qubits, one can similarly show that among the $4^N$ possible measurement outcomes, \textit{all} outcomes cancel apart from the ones corresponding to no bit flip and all bit flips. This justifies our previous assumption that we either measure a correct energy with probability $1-p$ or an incorrect energy with probability $p$. Crucially, the latter probability is \textit{not} given by $p^{2N}$ as one might expect at first glance. Thus, each incorrect measurement yields the same deviation from the correct energy of $-2E$ with the same probability $p$. This can be seen by evaluating the probabilities of the four different outcomes above:

\begin{itemize}
  \item The absence of any bit flip, $\ket{0}\xrightarrow{1-p}\ket{0}$, $\ket{1}\xrightarrow{1-p} \ket{1}$, gives $\bra{\psi} \mathcal{H}_X \ket{\psi}=E$ with probability $(1-p)^2$.
  \item The ``mixed'' bit flips $\ket{0}\xrightarrow{p}\ket{1}$, $\ket{1}\xrightarrow{1-p} \ket{1}$ and $\ket{0}\xrightarrow{1-p}\ket{0}$, $\ket{1}\xrightarrow{p} \ket{0}$ give $\bra{\psi} \tilde{\mathcal{H}}_X \ket{\psi}=0$ with a combined probability of $2p(1-p)$.
  \item The ``total'' bit flip $\ket{0}\xrightarrow{p}\ket{1}$, $\ket{1}\xrightarrow{p} \ket{0}$ gives $\bra{\psi} \tilde{\mathcal{H}}_X \ket{\psi}=-E$ with probability $p^2$.
\end{itemize}
Note that the cancellation $\bra{\psi} \tilde{\mathcal{H}} \ket{\psi}=0$ for the ``mixed'' bit flips seems to require $|\mathcal{H}|0\rangle| = |\mathcal{H}|1\rangle|$ at first glance. While this is not true in general, the measurement of $\bra{\psi} \tilde{\mathcal{H}} \ket{\psi}$ in the $Z$ basis reduces to measuring Pauli strings composed of $\Id$ and $Z$ matrices, which are unitary. Thus, after the appropriate post-rotation, the condition $|\mathcal{H}|0\rangle| = |\mathcal{H}|1\rangle|$ changes to $|\mathcal{Z}|0\rangle| = |\mathcal{Z}|1\rangle|$ and $|\mathcal{\Id}|0\rangle| = |\mathcal{\Id}|1\rangle|$, which is trivially fulfilled.

This yields the simple relation for the mean energy
\begin{align}
  \begin{split}
    \En \tilde{E} &= (1-p)^2 E + p^2 (-E) \\
    &= (1-2p) E.
  \end{split}
  \label{eq:E1q}
\end{align}
Combining Eqs.~\eqref{eq:E(p)} and \eqref{eq:E1q} we find for the parameter $\Delta \tilde{E}$
\begin{equation}
  \En \tilde{E} =E + \nshots p\Delta \tilde{E} \quad \leftrightarrow\quad \Delta \tilde{E} = - \frac{2E}{\nshots},
  \label{eq:delE1q}
\end{equation}
where $\Delta \tilde{E}$ is normalized by the number of shots $\nshots$, i.e., the number of evaluations of the energy~\eqref{eq:H} required to produce the energy histogram. For $p=1$, the first three possible measurement outcomes have zero probability, independently of any cancellations, and only the last outcome with $\bra{\psi} \tilde{\mathcal{H}}_X \ket{\psi} = -E$ contributes.

As we will discuss in the next subsection, Eq.~\eqref{eq:delE1q} \textit{only} applies to non-interacting Hamiltonians, i.e., without any multi-qubit interaction terms. For example, for the Hamiltonians $\mathcal{H}_X=h\sum_{i=1}^N X_i$ or $\mathcal{H}_Z=h\sum_{i=1}^N Z_i$ with the ground-state energy $E_0 = -Nh$, we would get $\Delta \tilde{E}_0 = 2Nh/\nshots$ when measuring the ground-state energy. Thus, after measuring the noisy expectation value of any (trivial) non-interacting Hamiltonian on a quantum computer, Eq.~\eqref{eq:delE1q} allows us to predict the corresponding true energy.

\subsection{$\Delta \tilde{E}$ for interacting Hamiltonians}\label{sec:int}

For two-qubit interaction terms in the Hamiltonian, e.g., for $\mathcal{H}_{ZZ}=J\sum_{i=1}^N Z_iZ_{i+1}$, our previous considerations need to be modified in two ways: first, we observe that the \textit{one-qubit} bit flips from the previous subsection give the same contribution to the mean energy as before, but now with a probability of $2p(1-p)$ instead of $p^2$. This is because the one-qubit ``total'' bit flips yield $ \bra{\psi} \tilde{\mathcal{H}}_{ZZ} \ket{\psi} = -E$. Here, ``one-qubit ``total'' bit flip'' means that one of the two qubits experiences a bit flip during readout ($\ket{0}\to\ket{1}$, $\ket{1}\to \ket{0}$), while the other qubit has no bit flip ($\ket{0}\to\ket{0}$, $\ket{1}\to \ket{1}$).
Second, the mean energy receives small $\mathcal{O}(p^2)$ corrections since the parameter $\Delta \tilde{E}$ becomes $p$-dependent for the interacting case. These $\mathcal{O}(p^2)$ corrections come from the \textit{two-qubit} bit flips and have the \textit{opposite} sign of the $\mathcal{O}(p)$ terms, because the two minus signs from the measurement bases $Z_1$ and $Z_2$ cancel. Indeed, the two-qubit ``total'' bit flips, i.e., $\ket{0}\to\ket{1}$ and $\ket{1}\to \ket{0}$ for both qubits, yield $\bra{\psi} \tilde{\mathcal{H}}_{ZZ} \ket{\psi} = E$ with probability $p^2$.

Let us demonstrate the latter for the simple two-qubit Hamiltonian $\mathcal{H}_{ZZ}=Z_1Z_2$, which gives
\begin{align}
  \begin{split}
    &\bra{\psi} \mathcal{H}_{11} \ket{\psi} = \bra{\psi} Z_1Z_2 \ket{\psi} \\
    &= [c_0^*\bra{00}+c_1^*\bra{01}+c_2^*\bra{10}+c_3^*\bra{11}]Z_1Z_2\\
    &\quad\,\, [c_0\ket{00}+c_1\ket{01}+c_2\ket{10}+c_3\ket{11}] \\
    &= |c_0|^2-|c_1|^2-|c_2|^2+|c_3|^2 = E
  \end{split}
\end{align}
without any bit flip. For two-qubit bit flips with $p=1$, we obtain the same result and thus recover the true energy $E$,
\begin{align}
  \begin{split}
    &\bra{\psi} \tilde{\mathcal{H}}_{ZZ} \ket{\psi}= \bra{\psi} \tilde{Z}_1 \tilde{Z}_2 \ket{\psi}\\
    &=[c_0^*\bra{11}+c_1^*\bra{10}+c_2^*\bra{01}+c_3^*\bra{00}]Z_1Z_2\\
    &\quad\,\, [c_1\ket{11}+c_2\ket{10}+c_3\ket{01}+c_4\ket{00}] \\
    &=|c_0|^2-|c_1|^2-|c_2|^2+|c_3|^2 = E,
  \end{split}
  \label{eq:2flip}
\end{align}
since the two minus signs from the $Z$-matrices cancel, i.e., $\bra{00}Z_1Z_2\ket{00}=\bra{11}Z_1Z_2\ket{11}$.

The contributions from ``mixed'' bit flips, such as all basis states $\ket{b_1b_0}$ flipping to $\ket{11}$, cancel for any $p$ due to opposite signs and equal probabilities, just as in the non-interacting case. Therefore, the ``total'' two-qubit bit flips as discussed in Eq.~\eqref{eq:2flip} have a probability of $p^2$ instead of $p^{4N}$. This yields for the total mean energy
\begin{align}
  \begin{split}
    \En \tilde{E} &= (1-p)^2 E + 2p(1-p) (-E) + p^2 E \\
    &= E - 4pE + 4p^2E= (1-2p)^2 E.
  \end{split}
  \label{eq:E2flip}
\end{align}
Thus, the parameter $\Delta \tilde{E}$ now has two contributions,
\begin{equation}
  \En \tilde{E} = E + \nshots p\Delta \tilde{E} \ \Leftrightarrow\  \Delta \tilde{E} = - \frac{4E}{\nshots}(1-p).
  \label{eq:delE2flip}
\end{equation}
Equations~\eqref{eq:E2flip} and \eqref{eq:delE2flip} imply that the two-qubit interacting Hamiltonian yields the correct energy $\En \tilde{E} = E$ for \textit{both} $p=0$ and $p=1$, in contrast to the non-interacting case where $p=1$ gave $\En \tilde{E} = -E$ (see Eq.~\eqref{eq:E1q}). Moreover, $\En \tilde{E} =0$ is still given for $p=0.5$.

\subsection{Prediction for the transversal Ising model}\label{sec:TI}

Next, we apply our results to the ground-state energy of the TI model with the Hamiltonian
\begin{equation}
  \mathcal{H}_{\rm TI}=J\sum_{i=1}^N Z_iZ_{i+1}+h\sum_{i=1}^N X_i,
  \label{eq:HtI}
\end{equation}
where we again assume $J<0$ and $h>0$ and periodic boundary conditions. The true ground-state energy can be derived as~\cite{Lieb1961,Pfeuty1970,Kitaev2001,Vidal2003a,Latorre2004}
\begin{align}
  \begin{split}
    E_0 =&\, - \frac{1}{2}\sum_k \gamma\, (\alpha^2+4\beta^2) \\
    =&\, - \frac{1}{2}\sum_k \gamma \left[4h^2 + 4J^2 - 8Jh\cos\left(\frac{2\pi k}{N}\right)\right],
  \end{split}
  \label{eq:Etrans}
\end{align}
where the sum runs from $k =-\left(\frac{N-1}{2}\right)$ to $\left(\frac{N-1}{2}\right)$ and the constants $\alpha$, $\beta$, and $\gamma$ are defined as
\begin{align}
  \begin{split}
    \alpha &= 2h- 2J\cos\left(\frac{2\pi k}{N}\right),\\
    \beta &= J\sin\left(\frac{2\pi k}{N}\right),\\
    \gamma &= \frac{\textrm{sign}(\alpha)}{\alpha}\, \sqrt{ \frac{\alpha^2}{\alpha^2+4\beta^2}}\, .
  \end{split}
\end{align}
\begin{figure}[t]
  \centering
  \includegraphics[width=1.0\columnwidth]{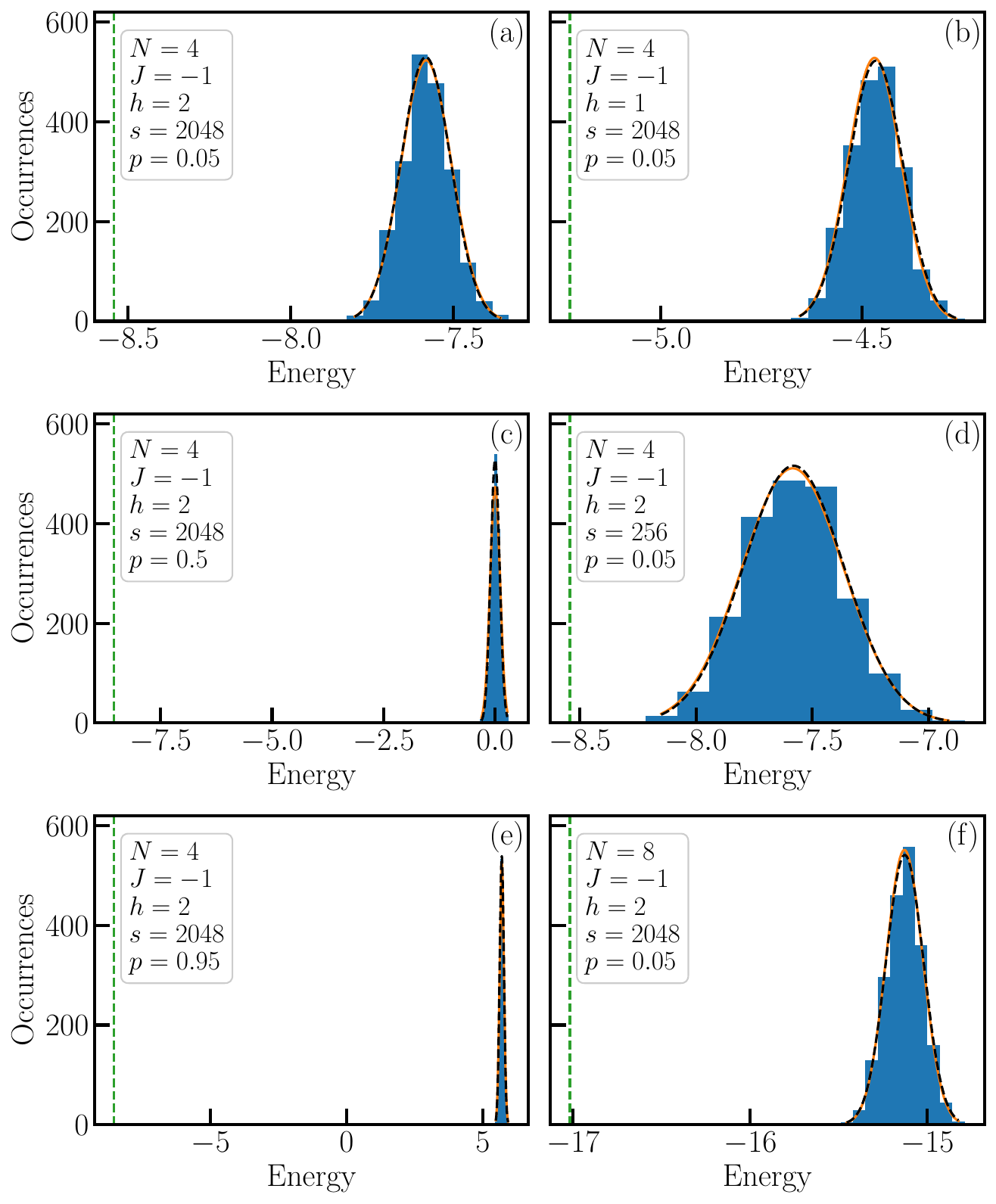}
  \caption{Energy histograms for the TI model. The vertical dashed green line indicates the true ground-state energy, the solid orange line the prediction and the dashed black line a fit to the data. The left column corresponds to $N=4$, $J=-1$, $h=2$, $\nshots=2048$ with (a) $p=0.05$, (c) $p=0.50$, and (e) $p=0.95$. The right column shows varied $N$, $h$, and $\nshots$: (b) $h=1$, (d) $\nshots=256$, and (f) $N=8$.}
  \label{fig:transIsing}
\end{figure}
Just as for the LI model~\eqref{eq:finalE}, the mean energy of the noisy ground-state energy histograms receives three different contributions,
\begin{align}
  \En \tilde{E}_0 & = (1-p)^2 E_1 + 2p(1-p)E_2 + p^2 E_3. \label{eq:finalEtrans}
\end{align}
The probabilities of the three different terms in Eqs.~\eqref{eq:finalE} and \eqref{eq:finalEtrans} are the same because they are determined by the number of interacting qubits in the different terms of the respective Hamiltonian. However, the measurement outcomes $E_i$ in Eq.~\eqref{eq:finalEtrans} deviate from the ones in Eq.~\eqref{eq:finalE} because $E_0$ in Eq.~\eqref{eq:Etrans} is not simply the sum of the $J$- and $h$-dependent parts of the ground-state energy as in Eq.~\eqref{eq:E0lI}.

The different measurement outcomes $E_i$ in Eq.~\eqref{eq:finalEtrans} can be derived in the following way. First, we know that $E_1=E_0$. Second, we know that $\En \tilde{E}$ vanishes for $|J|=|h|$ and $p=1$ because the two terms in the Hamiltonian~\eqref{eq:HtI} contribute equally to $E_0$ and thus cancel for $p=1$. This cancellation happens due to opposite signs of the non-interacting and interacting energy contributions in case of a total bit flip, as discussed above. In particular, any mixed terms, such as the mixed $Jh$-term in Eq.~\eqref{eq:Etrans}, vanish for $p=1$, as also discussed above. This fixes $E_3$. Third, we know that $\En \tilde{E}(p=0.5)=0$, so we can find $E_2$ by solving Eq.~\eqref{eq:finalEtrans} for $p=0.5$ and inserting the known expressions for $E_1$ and $E_3$. In total, we obtain
\begin{align}
  \begin{split}
  E_1&=E_{ZZ}+E_X,\\
	E_2&=-E_{ZZ},\\
  E_3&=E_{ZZ}-E_X,
  \end{split}
  \label{eq:EiTI}
\end{align}
which is similar to Eq.~\eqref{eq:finalE}, but with $E_{ZZ}$ and $E_X$ given by
\begin{align}
  \begin{split}
    E_{ZZ}=&-\frac{1}{2}\sum_k \gamma\left[4J^2-4Jh\cos\left(\frac{2\pi k}{N}\right)\right],\\
    E_X=&-\frac{1}{2}\sum_k \gamma\left[4h^2-4Jh\cos\left(\frac{2\pi k}{N}\right)\right].
  \end{split}
  \label{eq:E0JE0hTI}
\end{align}
Thus, the mean energy in Eq.~\eqref{eq:finalEtrans} can be brought into a similar form as the true ground-state energy in Eq.~\eqref{eq:Etrans},
\begin{align}
  \begin{split}
    \En \tilde{E}_0 =\;  & (1-2p)E_X+(1-2p)^2 E_{ZZ}\\
    =& - \frac{1}{2}\sum_k \gamma \left[(1-2p)4h^2 + (1-2p)^2 4J^2 \right.\\
      & - \left.(1-3p+2p^2)8Jh\cos\left(\frac{2\pi k}{N}\right)\right].
  \end{split}
  \label{eq:EtransGS}
\end{align}
The resulting parameter $\Delta \tilde{E}_0$ now has three contributions,
\begin{align}
  \Delta \tilde{E}_0 & = - \frac{1}{\nshots}\left(2E_X+4E_{ZZ}-4pE_{ZZ}\right).
  \label{eq:dELI}
\end{align}
We note that this expression is identical to the one for the LI model but with different $E_{ZZ}$ and $E_{Z/X}$. For the LI model, $\Delta \tilde{E}_0$ rises strictly linearly with $N$. For the TI model, the sum over $k$ yields $N$ contributions to each $E_i$ in Eq.~\eqref{eq:EiTI}, which are equal for $E_3$ but differ for $E_1$ and $E_2$ due to the $N$-dependence of the cosine in Eq.~\eqref{eq:EtransGS}. Thus, $\Delta \tilde{E}_0(N)$ only becomes approximately linear for large $N$, where these small differences average out.

In Fig.~\ref{fig:transIsing}, we plot the energy histograms for the ground state of $\mathcal{H}_{\rm TI}$ with different $N$, $J$, $h$, $\nshots$, and $p$, where we again measure the ground state 2048 times for each parameter combination. As before, the noise model with the mean energy from Eq.~\eqref{eq:EtransGS} and the variance from Eq.~\eqref{eq:bit-string-variance} agrees with the data for any choice of parameters we study. Note that the variance is larger compared to the longitudinal case in Fig.~\ref{fig:longIsing}, because the measurement $Z$-basis is not an eigenbasis of the $X_i$ operator. Thus, the histograms are wider for the transversal case.

\section{Prediction for the variances of noisy expectation values}\label{app:variances}

In this appendix, we derive the variances of the different noisy expectation values presented in Secs.~\ref{sec:mitenergy} and \ref{sec:mitmeasure}. To this end, we construct again random operators whose expectation value yields the variance with respect to the bit-flip probability.
We follow the structure of Sec.~\ref{sec:mitmeasure}: we first discuss a single $Z$ operator in Sec.~\ref{app:single_Z_section}, followed by the general case of $Z_Q\otimes\cdots\otimes Z_1$ operators in Sec.~\ref{app:multi_Z_section}. We then simplify our results to the case of equal bit-flip probabilities in Sec.~\ref{app:multi_Z_section_same_p}. Finally, we discuss the case of measuring general operators from bit-string distributions of $\ket{\psi}$ in Sec.~\ref{app:general_operators}. In this section, we will also discuss different measurement paradigms and their impact on the variance of means extracted from histogram data. We will eventually return to the TI model for an explicit illustration.

\subsection{Measurement of single $Z$ operator}\label{app:single_Z_section}

For computing the variance $\Vn \tilde{Z}_q$ of the noisy expectation in Eq.~\eqref{eq:expvalue},
\begin{align}
  \begin{aligned}
    \Vn \tilde{Z}_q & = \En (\tilde{Z}_q\otimes \tilde{Z}_q)- (\En \tilde{Z}_q)^2\\
              & = \Phi_{\tilde{Z}_q}'(0)\otimes\Phi_{\tilde{Z}_q}'(0)-\Phi_{\tilde{Z}_q}''(0),
  \end{aligned}
  \label{eq:var}
\end{align}
we need to evaluate the derivatives $\Phi_q'(0)=i\En \tilde{Z}_q$ and $\Phi_{\tilde{Z}_q}''(0)=-\En (\tilde{Z}_q)^2$ of the characteristic function
\begin{align}
  \Phi_{\tilde{Z}_q}(t):=\En\exp\left[i\, \textrm{Tr}\left(t^*{\tilde{Z}_q}\right)\right].
\end{align}
This yields
\begin{align}
  \begin{split}
    \Phi_{\tilde{Z}_q}'(0)=&\ i(1-p_{q,0}-p_{q,1})Z_q+i(p_{q,1}-p_{q,0})\Id_q\\
    \Phi_{\tilde{Z}_q}''(0)=& -(1-p_{q,0}-p_{q,1}+2p_{q,0}p_{q,1})Z_q\otimes Z_q\\
    &-(p_{q,0}+p_{q,1}-2p_{q,0}p_{q,1})\Id_q\otimes\Id_q.
  \end{split}
\end{align}
Thus, the variance operator in Eq.~\eqref{eq:var} reads
\begin{align}
  \begin{split}
    \Vn \tilde{Z}_q  =\, &[(p_{q,0}+p_{q,1})(1-p_{q,0}-p_{q,1})+2p_{q,0}p_{q,1}]\\
    &\times Z_q\otimes Z_q\\
    &-(1-p_{q,0}-p_{q,1})(p_{q,1}-p_{q,0})Z_q\otimes\Id_q\\
    &-(1-p_{q,0}-p_{q,1})(p_{q,1}-p_{q,0})\Id_q\otimes Z_q\\
    &+(p_{q,0}+p_{q,1}-p_{q,0}^2-p_{q,1}^2)\Id_q\otimes\Id_q.
  \end{split}
  \label{eq:varQ1}
\end{align}

\subsection{Measurement of $Z_Q\otimes\cdots\otimes Z_1$ operators}\label{app:multi_Z_section}

We now generalize the variance for $Q=1$ in Eq.~\eqref{eq:varQ1} to operators acting on multiple qubits, i.e., $Q>1$. According to Eq.~\eqref{eq:independence-operators-different-qubits}, operators $\tilde{O}_1$ and $\tilde{O}_2$ acting on different qubits are uncorrelated, i.e., the covariance vanishes,
\begin{align}
  \begin{split}
  \Cov_\otimes(\tilde{O}_1,\tilde{O}_2)&:=\En(\tilde{O}_1\otimes \tilde{O}_2)-\En(\tilde{O}_1)\otimes\En(\tilde{O}_2)\\
                          &\phantom{:}=0.
  \end{split}
\end{align}
Hence, we obtain the variance operator
\begin{align}
  \begin{split}
    \Vn & \left(\tilde{Z}_Q\otimes\cdots\otimes \tilde{Z}_1\right) \\
    =&\ \En\left(\tilde{Z}_Q\otimes\cdots\otimes \tilde{Z}_1\otimes \tilde{Z}_Q\otimes\cdots\otimes \tilde{Z}_1\right)\\
    &-\En\left(\tilde{Z}_Q\otimes\cdots\otimes  \tilde{Z}_1\right)\otimes\En\left(\tilde{Z}_Q\otimes\cdots\otimes \tilde{Z}_1\right)\\
    = &\ U^*\Biggl(\En\left(\tilde{Z}_Q\otimes \tilde{Z}_Q\right)\otimes\cdots\otimes\En\left(\tilde{Z}_1\otimes \tilde{Z}_1\right)\Biggr.\\
    &-\left.\bigotimes_{q=1}^Q\left(\En \tilde{Z}_q\otimes\En \tilde{Z}_q\right)\right)U\\
    =&\ U^*\Biggl(\left(\Vn \tilde{Z}_Q+\En \tilde{Z}_Q\otimes\En \tilde{Z}_Q\right)\otimes\cdots\Biggr.\\
    &\cdots\otimes\left(\Vn \tilde{Z}_1+\En \tilde{Z}_1\otimes\En \tilde{Z}_1\right)\\
		&\left.-\bigotimes_{q=1}^Q\left(\En \tilde{Z}_q\otimes\En \tilde{Z}_q\right)\right)U,
  \end{split}
  \label{eq:multivar}
\end{align}
where the unitary operation $U$ re-orders the tensor products from $\ket{\psi_Q}\otimes\cdots\otimes\ket{\psi_1}\otimes\ket{\psi_Q}\otimes\cdots\otimes\ket{\psi_1}$ to $(\ket{\psi_Q}\otimes\ket{\psi_Q})\otimes\cdots\otimes(\ket{\psi_1}\otimes\ket{\psi_1})$.
That is, for two qubits the re-ordering maps the basis state $\ket{b_3b_2b_1b_0}$ to $\ket{b_3b_1b_2b_0}$, and for three qubits the re-ordering maps $\ket{b_5b_4b_3b_2b_1b_0}$ to $\ket{b_5b_2b_4b_1b_3b_0}$, etc.

\subsection{Measurement of $Z_Q\otimes\cdots\otimes Z_1$ operators assuming equal bit-flip probabilities}\label{app:multi_Z_section_same_p}

For $Q=1$, the variance in Eq.~\eqref{eq:multivar} reduces to
\begin{align}
  \Vn \tilde{Z}_q=2p(1-p) (Z_q\otimes Z_q+\Id_q\otimes\Id_q).
\end{align}
For $Q=2$, the re-ordering of the tensor product $\ket{\psi}\otimes\ket{\psi}$ in Eq.~\eqref{eq:multivar} becomes important, which yields
\begin{align}
  \begin{split}
    \Vn &\left(\bra\psi \tilde{Z}_2\otimes \tilde{Z}_1\ket\psi\right) \\
    =&\ (\bra{\psi}\otimes\bra\psi) U^*\left(\Vn \tilde{Z}_2\otimes\Vn \tilde{Z}_1 \right)U(\ket{\psi}\otimes\ket\psi)\\
    +&(\bra{\psi}\otimes\bra\psi) U^*\left(\Vn \tilde{Z}_2\otimes\En \tilde{Z}_1\otimes\En \tilde{Z}_1 \right)U(\ket{\psi}\otimes\ket\psi)\\
    +&(\bra{\psi}\otimes\bra\psi) U^*\left(\En \tilde{Z}_2\otimes\En \tilde{Z}_2\otimes\Vn \tilde{Z}_1 \right)U(\ket{\psi}\otimes\ket\psi).
  \end{split}
\end{align}
For arbitrary $Q$, we can evaluate the variance operator in Eq.~\eqref{eq:multivar} for the ground state, $\ket\psi=\ket{0\ldots0}$, and obtain the expression
\begin{align}
  \begin{split}
    \Vn & \left(\bra\psi \tilde{Z}_Q\otimes\cdots\otimes \tilde{Z}_1\ket\psi \right) \\
    =&\ \prod_{q=1}^Q\bra{\psi} \Vn \tilde{Z}_q+\En \tilde{Z}_q\otimes\En \tilde{Z}_q\ket{\psi}\\    -&\prod_{q=1}^Q\bra{\psi} \En \tilde{Z}_q\otimes\En \tilde{Z}_q\ket{\psi}\\
    =&\ \prod_{q=1}^Q\left(4p(1-p)+(1-2p)^2\right)-(1-2p)^{2Q}\\
    =&\ 1-(1-2p)^{2Q}.
  \end{split}
\end{align}
This surprisingly simple result can be verified directly by noting that the measurement of $\bra{0\ldots0} \tilde{Z}_Q\otimes\cdots\otimes \tilde{Z}_1\ket{0\ldots0}$ yields the values $+1$ with probability $p_1$ and $-1$ with probability $p_{-1}$. Thus, we conclude
\begin{align}
  \begin{split}
    \Vn &\left(\bra{0\ldots0} \tilde{Z}_Q\otimes\cdots\otimes \tilde{Z}_1\ket{0\ldots0}\right)\\
    =&\ \En\left(\bra{0\ldots0} \tilde{Z}_Q\otimes\cdots\otimes \tilde{Z}_1\ket{0\ldots0}^2\right)\\
    &-\En\left(\bra{0\ldots0} \tilde{Z}_Q\otimes\cdots\otimes \tilde{Z}_1\ket{0\ldots0}\right)^2\\
    =&\ p_1+p_{-1}(-1)^2\\
    &-(1-2p)^{2Q}\bra{0\ldots0}Z_Q\otimes\cdots\otimes Z_1\ket{0\ldots0}^2\\
    =&\ 1-(1-2p)^{2Q}.
  \end{split}
\end{align}

\subsection{Measurement of general operators $\mathcal{H}$ from bit-string distributions of $\ket{\psi}$}\label{app:general_operators}

\subsubsection{Prediction for the variance of operators}

While measuring the entire Hamiltonian simultaneously makes no difference for the measured mean value, the variance on the other hand is affected by this change in measurement paradigm. If we consider $\mathcal{H}_{ZZ}$ with $N=2$ and $J=1$, i.e., $\mathcal{H}_{ZZ}=Z_2Z_1+Z_1Z_2$, then we would formally compute $\bra\psi Z_2\otimes Z_1\ket\psi$ twice independently using the approach considered so far, whereas the expectation from the bit-string distribution of $\ket\psi$ directly extracts $2\bra\psi Z_2\otimes Z_1\ket\psi$. Thus the variance using independent histograms for each summand is given by
\begin{align}
  \begin{split}
    &\Vn_{\mathrm{ind}}\bra\psi\tilde{\mathcal{H}}_{ZZ}\ket\psi
    =\\
    &=\Vn\bra\psi \tilde{Z}_2\otimes \tilde{Z}_1\ket\psi+\Vn\bra\psi \tilde{Z}_2\otimes \tilde{Z}_1\ket\psi\\
    &=2\Vn\bra\psi \tilde{Z}_2\otimes \tilde{Z}_1\ket\psi
  \end{split}
\end{align}
whereas the variance using the bit-string distribution of $\ket\psi$ is
\begin{align}
  \begin{split}
    \Vn_{\mathrm{bsd}}\bra\psi\tilde{\mathcal{H}}_{ZZ}\ket\psi=\,&\Vn(2\bra\psi \tilde{Z}_2\otimes \tilde{Z}_1\ket\psi)\\
    =\,&4\Vn\bra\psi \tilde{Z}_2\otimes \tilde{Z}_1\ket\psi\\
    =\,&2\Vn_{\mathrm{ind}}\bra\psi\tilde{\mathcal{H}}_{ZZ}\ket\psi.
  \end{split}
\end{align}
In general, if $\tilde{\mathcal{H}}=\sum_\alpha \lambda_\alpha U_\alpha^*\tilde{O}_\alpha U_\alpha$, we are still able to predict the variance $\Vn_{\mathrm{bsd}}\bra\psi\tilde{\mathcal{H}}\ket\psi$ using the same method as above albeit the covariance terms no longer vanish (each $\tilde{O}_\alpha$ is a tensor product $\tilde{O}_{\alpha,Q}\otimes\cdots\otimes \tilde{O}_{\alpha,1}$). For $\tilde{O}_{\alpha,q}=\tilde{Z}_q$, $\tilde{O}_{\alpha,q}$ takes one of the possible values $\{Z_q,-\Id_q,\Id_q,-Z_q\}$, as in Sec.~\ref{sec:single_Z_section}. For $\tilde{O}_{\alpha,q}=\tilde{\Id}_q$, $\tilde{O}_{\alpha,q}$ always takes the value $\Id_q$. Using these replacements for all summands in $\tilde{\mathcal{H}}$, we obtain that $\tilde{\mathcal{H}}$ takes finitely many (up to $2^N$) values $\mathcal{H}_\alpha$ with probability~$p_\alpha$. Hence, the characteristic function $\Phi_{\tilde{\mathcal{H}}}$ is given by
\begin{align}
  \begin{split}
    \Phi_{\tilde{\mathcal{H}}}(t):=&\,\En\exp\left(i\tr\left(t^*\tilde{\mathcal{H}}\right)\right)\\
    =&\sum_\alpha p_\alpha \exp\left(i\tr\left(t^*\mathcal{H}_\alpha\right)\right).
  \end{split}
\end{align}
As such, we can directly conclude
\begin{align}
  \Phi_{\tilde{\mathcal{H}}}'(0)=  & \sum_\alpha p_\alpha i\mathcal{H}_\alpha=i\En\tilde{\mathcal{H}},        \\
  \Phi_{\tilde{\mathcal{H}}}''(0)= & -\sum_\alpha p_\alpha \mathcal{H}_\alpha\otimes\mathcal{H}_\alpha,
\end{align}
and find the variance operator
\begin{align}
  \begin{split}
    &\Vn_{\mathrm{bsd}}\tilde{\mathcal{H}}=\Phi_{\tilde{\mathcal{H}}}'(0)\otimes \Phi_{\tilde{\mathcal{H}}}'(0)-\Phi_{\tilde{\mathcal{H}}}''(0)\\
    &=\left(\sum_\alpha p_\alpha \mathcal{H}_\alpha\otimes\mathcal{H}_\alpha\right)-\left(\En\tilde{\mathcal{H}}\right)\otimes\left(\En\tilde{\mathcal{H}}\right)\\
    &=\left(\sum_\alpha p_\alpha \mathcal{H}_\alpha\otimes\mathcal{H}_\alpha\right)-\left(\sum_{\alpha,\beta}p_\alpha p_\beta\mathcal{H}_\alpha\otimes\mathcal{H}_\beta\right).
  \end{split}
\end{align}
Similarly, we can measure the operator $\tilde{\mathcal{H}}$ on the state $\ket\psi$ and obtain the variance
\begin{align}
  \begin{split}
    \Vn_{\mathrm{bsd}}&\bra\psi\tilde{\mathcal{H}}\ket\psi
    =\left(\sum_\alpha p_\alpha \bra\psi\mathcal{H}_\alpha\ket\psi^2\right)\\
    &-\left(\sum_{\alpha,\beta}p_\alpha p_\beta\bra\psi\mathcal{H}_\alpha\ket\psi\bra\psi\mathcal{H}_\beta\ket\psi\right).
  \end{split}
  \label{eq:bit-string-variance}
\end{align}

\subsubsection{Prediction for the variance of histogram means}
Lastly, we can combine the bit-flip induced variances with quantum mechanically induced variances to obtain the full variances observed in measuring histogram means. In particular, we will construct the variances for the three methods discussed in Sec.~\ref{sec:variance-by-method}. There we measured the bit-flip corrected TI Hamiltonian $\tilde{\mathcal{H}}_{\rm TI,bfc}=J_p\sum_{j=1}^N \tilde{Z}_j\tilde{Z}_{j+1}+h_p\sum_{j=1}^N \tilde{X}_j$ in Eq.~\eqref{eq:HTIbfc} subject to bit flips on the ground state of the ``true'' TI Hamiltonian $\mathcal{H}_{\rm TI}=J\sum_{j=1}^N Z_jZ_{j+1}+h\sum_{j=1}^N X_j$. For simplicity, we assumed that all bit-flip probabilities~$p_{q,b}$ equal~$p$. The three methods are

\begin{itemize}
  \item \emph{Method 1:} measure each $\tilde{Z}_j\tilde{Z}_{j+1}$ and $\tilde{X}_j$ in Eq.~\eqref{eq:HTIbfc} independently
  \item \emph{Method 2:} measure the entire Hamiltonian $\tilde{\mathcal{H}}_{\rm TI,bfc}$ in Eq.~\eqref{eq:HTIbfc} from distributions of $\ket\psi$ measurements
  \item \emph{Method 3:} measure $\tilde{\mathcal{H}}_{ZZ}:=J_p\sum_{j=1}^N \tilde{Z}_j\tilde{Z}_{j+1}$ and $\tilde{\mathcal{H}}_X:=h_p\sum_{j=1}^N \tilde{X}_j$ independently from distributions of $\ket\psi$ measurements
\end{itemize}

\emph{Method 1:} Since each $\tilde{Z}_j\tilde{Z}_{j+1}$ and $\tilde{X}_j$ is measured independently, the bit-flip contributions $\Vn_{\rm bf}\bra\psi \tilde{Z}_j\tilde{Z}_{j+1}\ket\psi$ and $\Vn_{\rm bf}\bra\psi \tilde{X}_j\ket\psi$ to the variance can be directly obtained from Eq.~\eqref{eq:multivar} keeping in mind that $\tilde{X}_j=H_j\tilde{Z}_jH_j$ where $H_j$ is the Hadamard gate on qubit~$j$. But since $\ket\psi$, in general, will not be an eigenstate of all $Z_jZ_{j+1}$ and $X_j$ simultaneously, we also have a contribution from the quantum mechanical variances $\Vn_{\rm QM}\bra\psi Z_jZ_{j+1}\ket\psi=1-\bra\psi Z_jZ_{j+1}\ket\psi^2$ and $\Vn_{\rm QM}\bra\psi X_j\ket\psi=1-\bra\psi X_j\ket\psi^2$. We therefore obtain the variance of  histogram means
\begin{align}
  \begin{split}
    \Vn_{\rm M1}\bra\psi\tilde{\mathcal{H}}_{\rm TI,bfc}\ket\psi=\,&\frac{J_p^2}{\nshots}\sum_{j=1}^N\Vn_{\rm bf}\bra\psi \tilde{Z}_j\tilde{Z}_{j+1}\ket\psi\\
    +&\frac{J_p^2}{\nshots}\sum_{j=1}^N\Vn_{\rm QM}\bra\psi Z_jZ_{j+1}\ket\psi\\
    +&\frac{h_p^2}{\nshots}\sum_{j=1}^N\Vn_{\rm bf}\bra\psi \tilde{X}_j\ket\psi\\
    +&\frac{h_p^2}{\nshots}\sum_{j=1}^N\Vn_{\rm QM}\bra\psi X_j\ket\psi.
  \end{split}
\end{align}
In particular, if the state $\ket\psi$ is translationally invariant, such as the ground state of $\mathcal{H}_{\rm TI}$, then this further simplifies to
\begin{align}
  \begin{split}
    \Vn_{\rm M1}\bra\psi\tilde{\mathcal{H}}_{\rm TI,bfc}\ket\psi=\,&\frac{J_p^2N}{\nshots}\Vn_{\rm bf}\bra\psi \tilde{Z}_j\tilde{Z}_{j+1}\ket\psi\\
    &+\frac{J_p^2N}{\nshots}\Vn_{\rm QM}\bra\psi Z_jZ_{j+1}\ket\psi\\
    &+\frac{h_p^2N}{\nshots}\Vn_{\rm bf}\bra\psi \tilde{X}_j\ket\psi\\
    &+\frac{h_p^2N}{\nshots}\Vn_{\rm QM}\bra\psi X_j\ket\psi
  \end{split}\label{eq:var-m1}
\end{align}
for any choice of $j$.

\emph{Method 2:} In this case, the bit-flip contribution $\Vn_{\rm bf}\bra\psi\tilde{\mathcal{H}}_{\rm TI,bfc}\ket\psi$ is given by Eq.~\eqref{eq:bit-string-variance} and the quantum mechanical variance is given by
\begin{align}
  \begin{split}
    \Vn_{\rm QM}\bra\psi\tilde{\mathcal{H}}_{\rm TI,bfc}\ket\psi=\,&\bra\psi(\tilde{\mathcal{H}}_{\rm TI,bfc})^2\ket\psi\\
    &-\bra\psi\tilde{\mathcal{H}}_{\rm TI,bfc}\ket\psi^2.
  \end{split}
\end{align}
Hence, the variance of  histogram means is
\begin{align}
  \begin{split}
    \Vn_{\rm M2}\bra\psi\tilde{\mathcal{H}}_{\rm TI,bfc}\ket\psi =\, & \frac{1}{\nshots}\Vn_{\rm bf}\bra\psi\tilde{\mathcal{H}}_{\rm TI,bfc}\ket\psi\\
    &+\frac{1}{\nshots}\Vn_{\rm QM}\bra\psi\tilde{\mathcal{H}}_{\rm TI,bfc}\ket\psi.
  \end{split}\label{eq:var-m2}
\end{align}
While this expression appears simpler than its counterpart for Method 1, it is also important to note that $O(4^N)$ terms are required to compute $\Vn_{\rm M2}\bra\psi\tilde{\mathcal{H}}_{\rm TI,bfc}\ket\psi$ whereas the number of terms required to compute $\Vn_{\rm M1}\tilde{\mathcal{H}}_{\rm TI,bfc}$ is only $O(N)$ and can even be reduced to $O(1)$ for translationally invariant states $\ket\psi$.

\emph{Method 3:} Being a combination of Method 1 and Method 2, the variance can be constructed combining the results from Method 1 and 2. The bit-flip contributions $\Vn_{\rm bf}\bra\psi\tilde{\mathcal{H}}_{ZZ}\ket\psi$ and $\Vn_{\rm bf}\bra\psi\tilde{\mathcal{H}}_{X}\ket\psi$ follow from Eq.~\eqref{eq:bit-string-variance} again. Furthermore, the quantum mechanical variances contribute as
\begin{align}
  \begin{split}
    \Vn_{\rm QM}\bra\psi\tilde{\mathcal{H}}_{ZZ}\ket\psi=\, & \bra\psi(\tilde{\mathcal{H}}_{ZZ})^2\ket\psi\\
    &-\bra\psi\tilde{\mathcal{H}}_{ZZ}\ket\psi^2
  \end{split}
\end{align}
and
\begin{align}
  \Vn_{\rm QM}\bra\psi\tilde{\mathcal{H}}_{X}\ket\psi=\, & \bra\psi(\tilde{\mathcal{H}}_{X})^2\ket\psi    -\bra\psi\tilde{\mathcal{H}}_{X}\ket\psi^2.
\end{align}
The variance of histogram means is thus
\begin{align}
  \begin{split}
    \Vn_{\rm M3}\bra\psi\tilde{\mathcal{H}}_{\rm TI,bfc}\ket\psi =\, & \frac{1}{\nshots}\Vn_{\rm bf}\bra\psi\tilde{\mathcal{H}}_{ZZ}\ket\psi\\
    &+\frac{1}{\nshots}\Vn_{\rm bf}\bra\psi\tilde{\mathcal{H}}_{X}\ket\psi\\
    &+\frac{1}{\nshots}\Vn_{\rm QM}\bra\psi\tilde{\mathcal{H}}_{ZZ}\ket\psi\\
    &+\frac{1}{\nshots}\Vn_{\rm QM}\bra\psi\tilde{\mathcal{H}}_{X}\ket\psi.
  \end{split}\label{eq:var-m3}
\end{align}

Methods 1 and 2 are the two extreme cases, which we discussed in Secs.~\ref{sec:single_Z_section}--\ref{sec:multi_Z_section_same_p} and Sec.~\ref{sec:general_operators}, respectively. Method 3 is a reasonable compromise, which is closely related to implementations of quantum algorithms that are optimized for the number of calls to the quantum device. In such implementations, only parts of an operator can be measured simultaneously, such that both Methods 1 and 2 are impractical to various degrees.

\section{Technical details of the simulations}
Here we briefly summarize the details on how to determine the bit-flip probabilities, the simulations, and data evaluation procedure for the results shown in Sec.~\ref{sec:results}.

\subsection{Calibration of the bit-flip probabilities\label{app:calibration}}
Although the Qiskit SDK~\cite{Qiskit} provides values for the bit-flip probabilities for the different qubits on the different chips, we choose to calibrate $p_{q,0}$ and $p_{q,1}$ ourselves. To obtain $p_{q,0}$, we simply measure the initial state using $\nshots_\text{calibration}$ shots and record the number of 1 outcomes. Similarly, we determine $p_{q,1}$ by first applying an $X$ gate to the qubit $q$, thus preparing the state $\ket{1}$, and we measure the resulting state again $\nshots_\text{calibration}$ times and record the number of 0 outcomes. For all data shown in the main text, we use $\nshots_\text{calibration} = 8192$, which is the maximum number of repetitions possible on the real quantum hardware. Moreover, to acquire some statistics on how the obtained values for the bit-flip probabilities fluctuate, we repeat this procedure multiple times. Subsequently, we average all the data obtained for $p_{q,b}$. The resulting bit-flip probabilities are the ones used for correcting the data in Sec.~\ref{sec:results}.

\subsubsection{Single-qubit case}
\begin{figure}[t]
  \centering
  \includegraphics[width=1.0\columnwidth]{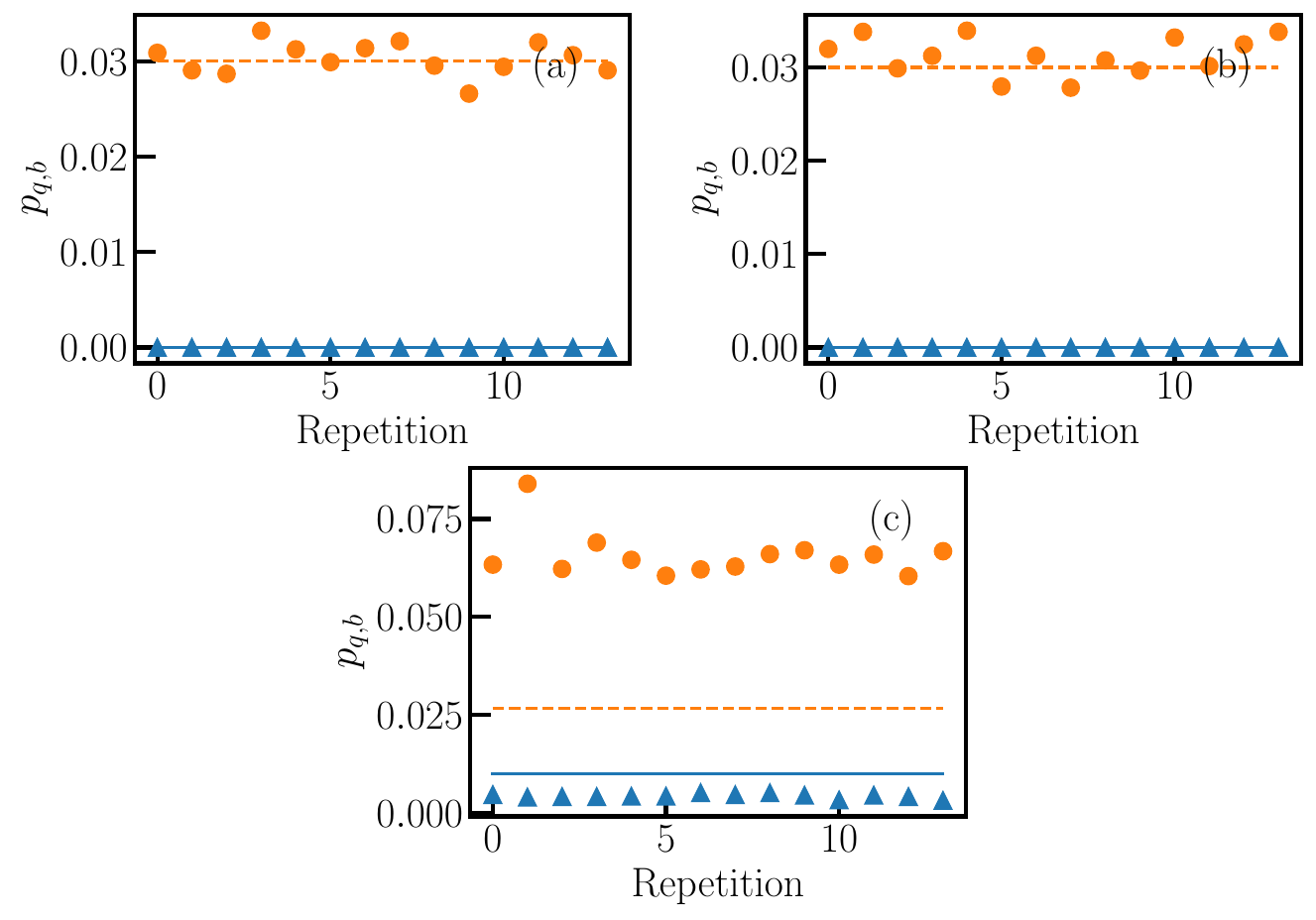}
  \caption{Bit-flip probabilities $p_{0,0}$ (blue triangles) and $p_{0,1}$ (orange dots) for the single qubit case measured with the calibration procedure as a function of the repetition for (a) classically simulating ibmq\_london with readout errors only, (b) the full noise model, and (c)  data obtained on the quantum hardware. The blue solid and the orange dashed line represent the corresponding data provided by the noise model.}
  \label{fig:ibmq_london_single_qubit_flip_probabilities}
\end{figure}

\begin{figure}[t]
  \centering
  \includegraphics[width=1.0\columnwidth]{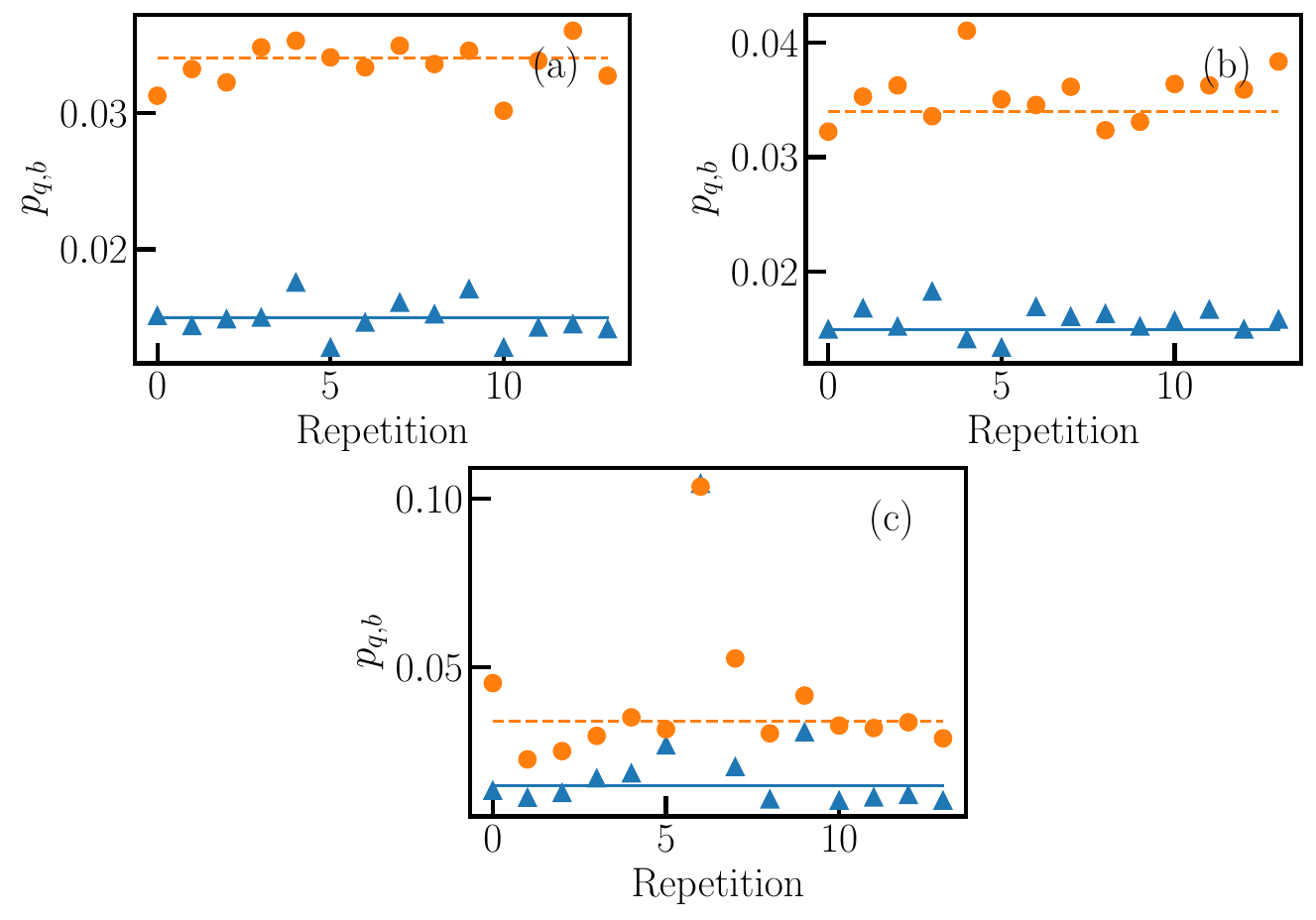}
  \caption{Bit-flip probabilities $p_{0,0}$ (blue triangles) and $p_{0,1}$ (orange dots) for the single qubit case measured with the calibration procedure as a function of the repetition for (a) classically simulating ibmq\_burlington with readout error only, (b) the full noise model, and (c)  data obtained on the hardware. The blue solid and the orange dashed line represent the corresponding data provided by the noise model.}
  \label{fig:ibmq_burlington_single_qubit_flip_probabilities}
\end{figure}

In Fig.~\ref{fig:ibmq_london_single_qubit_flip_probabilities} and Fig.~\ref{fig:ibmq_burlington_single_qubit_flip_probabilities}, we show the bit-flip probabilities we obtained for ibmq\_london and ibmq\_burlington.
Looking at the data resulting from simulating ibmq\_london classically with readout noise only in Fig.~\ref{fig:ibmq_london_single_qubit_flip_probabilities}(a), we observe that the bit-flip probabilities our calibration procedure yields scatter around the value provided by the noise model. Using the full noise model does not change the picture a lot; only the values for $p_{0,1}$ scatter slightly more around the value of the noise model, as Fig.~\ref{fig:ibmq_london_single_qubit_flip_probabilities}(b) reveals. The data generated on the actual ibmq\_london quantum hardware in Fig.~\ref{fig:ibmq_london_single_qubit_flip_probabilities}(c) do not agree very well with the values of the noise model. Even the values for $p_{0,0}$, which do not involve a single gate, are in general lower than the value provided by the noise model. In contrast, $p_{0,1}$ exceeds the value of the noise model.  Despite the fact that the values for the experimentally obtained bit-flip probabilities deviate from the noise model, they only fluctuate moderately and we can extract a reasonable bit-flip probability by averaging over all repetitions. Comparing the different panels of Fig.~\ref{fig:ibmq_london_single_qubit_flip_probabilities} closely, one can also observe that the values for the bit-flip probabilities provided by the noise model in panel (c) differ slightly from those in panel (a) and (b). The reason for that is that the data in the noise model are updated every day, and our classical simulations as well as our simulations on real quantum hardware were not carried out the same day.

The corresponding results for imbq\_burlington are shown in Fig.~\ref{fig:ibmq_burlington_single_qubit_flip_probabilities}. Again, the classical simulation of the chip using the noise model produces as expected bit-flip probabilities in agreement with the values provided. Looking at the data from the real chip in Fig.~\ref{fig:ibmq_burlington_single_qubit_flip_probabilities}(c), we see that these fluctuate over a wide range between different repetitions. Thus, in this case the bit-flip probabilities cannot be extracted as reliably as for imbq\_london. Since our correction procedure relies on being able to estimate the bit-flip probabilities precisely, this partially explains why the improvement in Sec.~\ref{sec:results_single_qubit} after applying the correction to our data for ibmq\_burlington is smaller.

\subsubsection{Two-qubit case}
\begin{figure}[t]
  \centering
  \includegraphics[width=1.0\columnwidth]{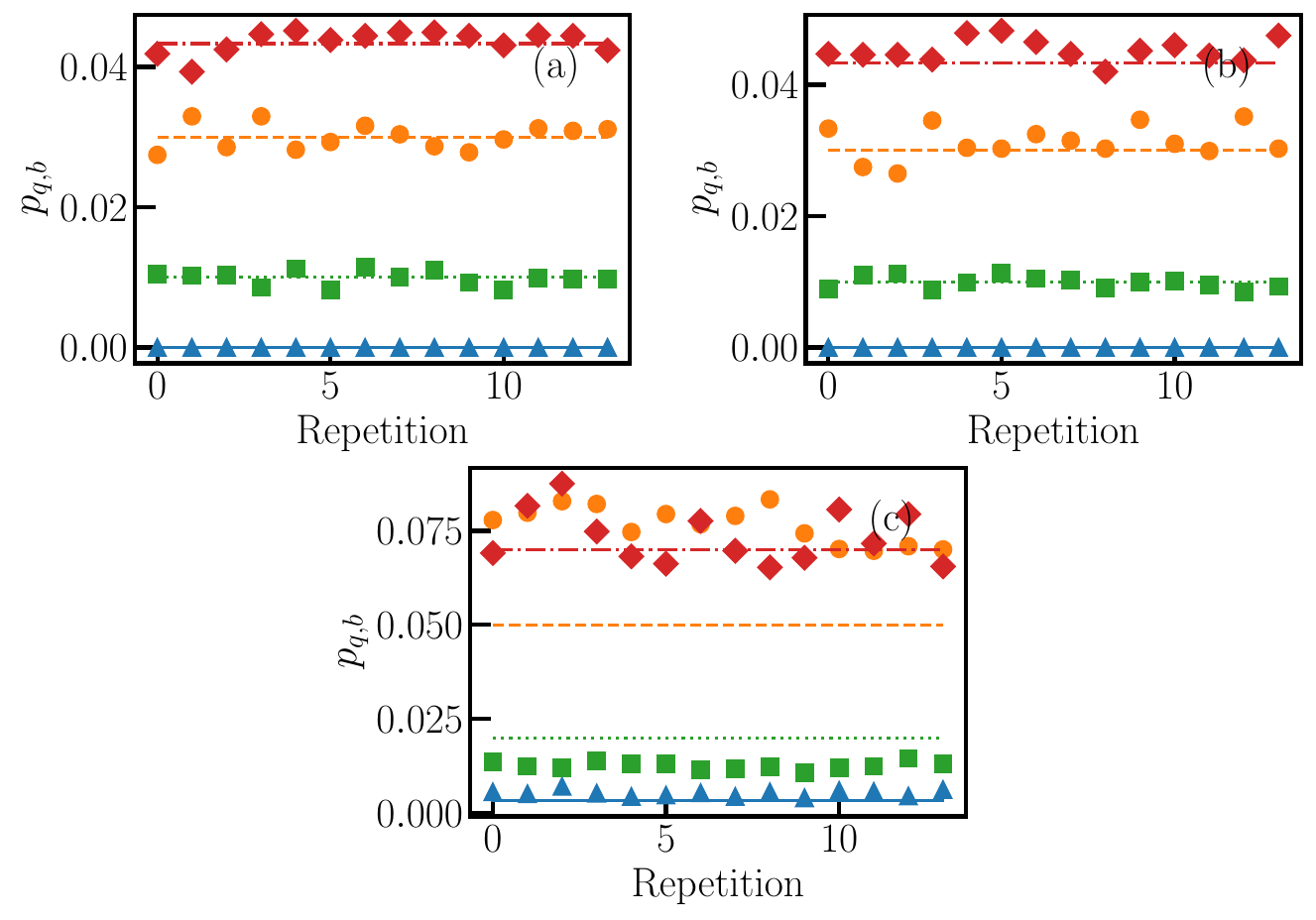}
  \caption{Bit-flip probabilities $p_{0,0}$ (blue triangles), $p_{0,1}$ (orange dots), $p_{1,0}$ (green squares), $p_{1,1}$ (red diamonds) for the two-qubit case measured with the calibration procedure as a function of the repetition for (a) classically simulating ibmq\_london with readout error only, (b) the full noise model, and (c)  data obtained on the hardware. The blue solid, the orange dashed, the green dotted, and the red dash-dotted line represent the corresponding data provided by the noise model.}
  \label{fig:ibmq_london_two_qubit_flip_probabilities}
\end{figure}

\begin{figure}[t]
  \centering
  \includegraphics[width=1.0\columnwidth]{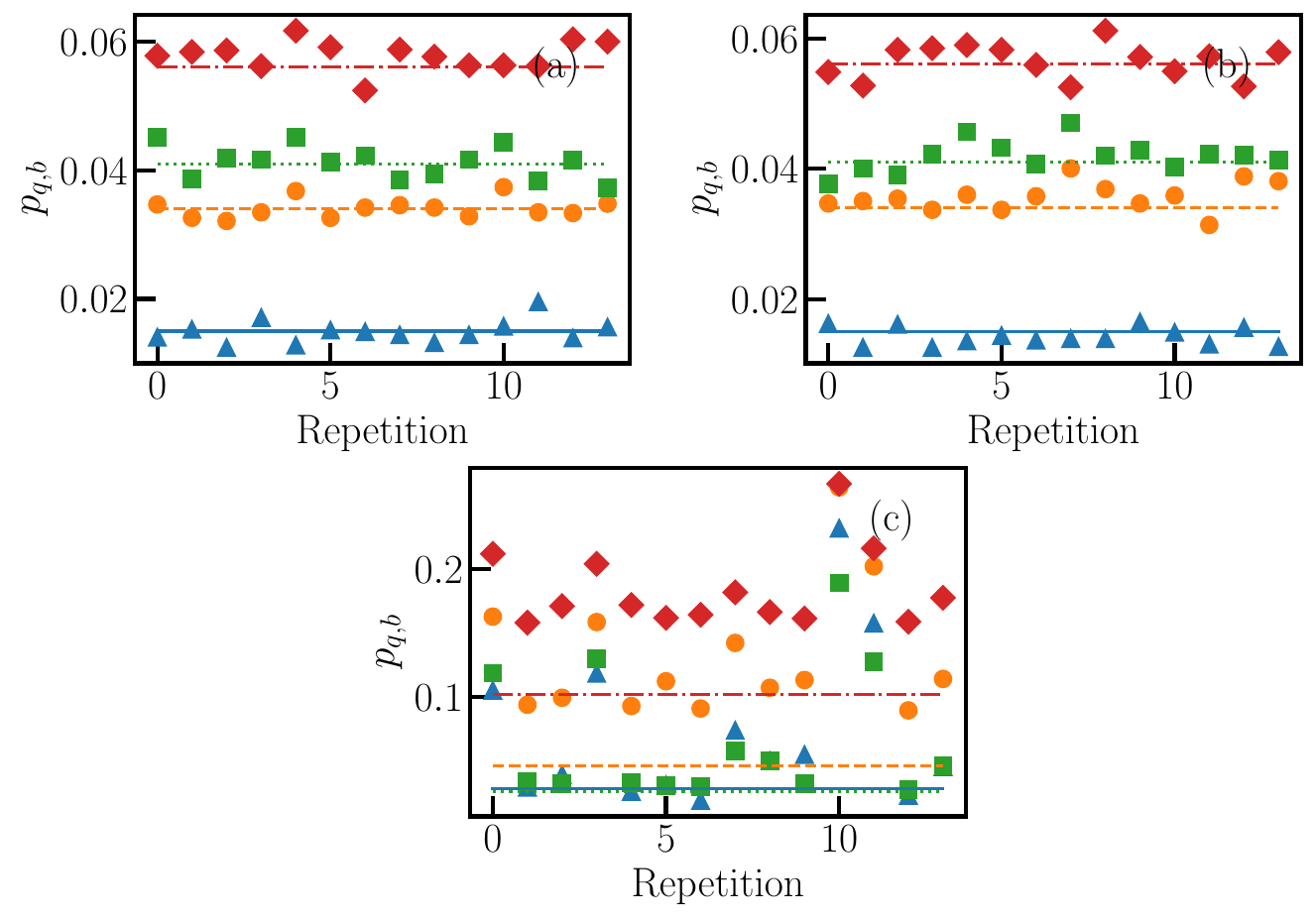}
  \caption{Bit-flip probabilities $p_{0,0}$ (blue triangles), $p_{0,1}$ (orange dots), $p_{1,0}$ (green squares), $p_{1,1}$ (red diamonds) for the two-qubit case measured with the calibration procedure as a function of the repetition for (a) classically simulating ibmq\_burlington with readout error only, (b) the full noise model, and (c)  data obtained on the hardware. The blue solid, the orange dashed, the green dotted, and the red dash-dotted line represent the corresponding data provided by the noise model.}
  \label{fig:ibmq_burlington_two_qubit_flip_probabilities}
\end{figure}

Analogously to the single-qubit case, Fig.~\ref{fig:ibmq_london_two_qubit_flip_probabilities} and Fig.~\ref{fig:ibmq_burlington_two_qubit_flip_probabilities} show the data for extracting the bit-flip probabilities for ibmq\_london and ibmq\_burlington obtained in our two-qubit simulations.

The results for the two-qubit case on ibmq\_london in Fig.~\ref{fig:ibmq_london_two_qubit_flip_probabilities} show a fairly similar behavior to the single-qubit case. The classical simulations in panels (a) and (b) yield as expected good agreement with the values provided in the noise model. In contrast, the data obtained on the real quantum device (Fig.~\ref{fig:ibmq_london_two_qubit_flip_probabilities}(c)) do not agree with the data in the noise model, in particular for $p_{0,1}$ and $p_{1,0}$. Nevertheless the experimental data is fairly consistent and allows us to reliably determine the bit-flip probabilities for ibmq\_london. Again, we see that the theoretical values differ noticeably between the panels in the upper row and the lower row. This is once more due to the fact that the hardware data was taken on a different day than the simulator data and the noise model was updated in between.

The bit-flip probabilities obtained from classically simulating ibmq\_burlington in in Fig.~\ref{fig:ibmq_burlington_two_qubit_flip_probabilities}(a) and Fig.~\ref{fig:ibmq_burlington_two_qubit_flip_probabilities}(b) show a similar picture to the previous cases, and they agree well with the values provided in the noise model. On the contrary, the data from the real quantum device again do not agree very well with the values provided in the noise model. Moreover, the values for $p_{0,0}$ and $p_{0,1}$ show large fluctuations. In this case as well, the theoretical values for the bit-flip probabilities differ between the simulator data and the hardware data. Most noticeably, the theoretical value for $p_{1,1}$ almost doubled during the time span between carrying out the classical simulations and the experiments on quantum hardware.

\subsection{Technical details for generating the experimental data on quantum hardware}
Each of the data points in Figs.~\ref{fig:ibmq_burlington_london_single_qubit_vs_shots_all}(c), \ref{fig:ibmq_burlington_london_single_qubit_vs_shots_all}(f) and Figs.~\ref{fig:ibmq_burlington_london_two_qubit_vs_shots_all}(c), \ref{fig:ibmq_burlington_london_two_qubit_vs_shots_all}(f) is obtained by preparing 1050 random wave functions $\ket{\psi}$ using the circuits shown in Sec.~\ref{sec:results}. While running the 1050 circuits is unproblematic for classical simulations, as of completion of this paper one can only submit 75 circuits per job to real quantum hardware. Thus, we divide them into 14 chunks of 75 circuits. This procedure is repeated for every value of $\nshots$. Since we have to run a considerable number of jobs, which might take some time depending on how busy the queue of the device is, we insert a job running the circuits for determining the bit-flip probabilities before every chunk. This way we can monitor the bit-flip probabilities over the duration of the run and detect potential outliers.

Moreover, before running our circuits, we use the transpiler to optimize them for the hardware we intend to use. To ensure that we have the same mapping between logical and physical qubits in every case, we inspect the transpiler results obtained for the circuits used to extract the bit-flip probabilities and to prepare the random wave function $\ket{\psi}$. For all the data reported in the main text we checked that the mapping between logical and physical qubits is indeed the same and the bit-flip probabilities we extract correspond to the qubits we use for generating our random wave functions.

\bibliography{Papers}

\begin{thebibliography}{62}%
\makeatletter
\providecommand \@ifxundefined [1]{%
 \@ifx{#1\undefined}
}%
\providecommand \@ifnum [1]{%
 \ifnum #1\expandafter \@firstoftwo
 \else \expandafter \@secondoftwo
 \fi
}%
\providecommand \@ifx [1]{%
 \ifx #1\expandafter \@firstoftwo
 \else \expandafter \@secondoftwo
 \fi
}%
\providecommand \natexlab [1]{#1}%
\providecommand \enquote  [1]{``#1''}%
\providecommand \bibnamefont  [1]{#1}%
\providecommand \bibfnamefont [1]{#1}%
\providecommand \citenamefont [1]{#1}%
\providecommand \href@noop [0]{\@secondoftwo}%
\providecommand \href [0]{\begingroup \@sanitize@url \@href}%
\providecommand \@href[1]{\@@startlink{#1}\@@href}%
\providecommand \@@href[1]{\endgroup#1\@@endlink}%
\providecommand \@sanitize@url [0]{\catcode `\\12\catcode `\$12\catcode
  `\&12\catcode `\#12\catcode `\^12\catcode `\_12\catcode `\%12\relax}%
\providecommand \@@startlink[1]{}%
\providecommand \@@endlink[0]{}%
\providecommand \url  [0]{\begingroup\@sanitize@url \@url }%
\providecommand \@url [1]{\endgroup\@href {#1}{\urlprefix }}%
\providecommand \urlprefix  [0]{URL }%
\providecommand \Eprint [0]{\href }%
\providecommand \doibase [0]{https://doi.org/}%
\providecommand \selectlanguage [0]{\@gobble}%
\providecommand \bibinfo  [0]{\@secondoftwo}%
\providecommand \bibfield  [0]{\@secondoftwo}%
\providecommand \translation [1]{[#1]}%
\providecommand \BibitemOpen [0]{}%
\providecommand \bibitemStop [0]{}%
\providecommand \bibitemNoStop [0]{.\EOS\space}%
\providecommand \EOS [0]{\spacefactor3000\relax}%
\providecommand \BibitemShut  [1]{\csname bibitem#1\endcsname}%
\let\auto@bib@innerbib\@empty
\bibitem [{\citenamefont {Montanaro}(2016)}]{Montanaro2016}%
  \BibitemOpen
  \bibfield  {author} {\bibinfo {author} {\bibfnamefont {A.}~\bibnamefont
  {Montanaro}},\ }\bibfield  {title} {\bibinfo {title} {Quantum algorithms: an
  overview},\ }\href {https://doi.org/10.1038/npjqi.2015.23} {\bibfield
  {journal} {\bibinfo  {journal} {npj Quantum Information}\ }\textbf {\bibinfo
  {volume} {2}},\ \bibinfo {pages} {15023} (\bibinfo {year}
  {2016})}\BibitemShut {NoStop}%
\bibitem [{\citenamefont {{Brandao}}\ and\ \citenamefont
  {{Svore}}(2017)}]{Brandao2017}%
  \BibitemOpen
  \bibfield  {author} {\bibinfo {author} {\bibfnamefont {F.~G. S.~L.}\
  \bibnamefont {{Brandao}}}\ and\ \bibinfo {author} {\bibfnamefont {K.~M.}\
  \bibnamefont {{Svore}}},\ }\bibfield  {title} {\bibinfo {title} {Quantum
  speed-ups for solving semidefinite programs},\ }in\ \href
  {https://doi.org/10.1109/FOCS.2017.45} {\emph {\bibinfo {booktitle} {2017
  IEEE 58th Annual Symposium on Foundations of Computer Science (FOCS)}}}\
  (\bibinfo {year} {2017})\ pp.\ \bibinfo {pages} {415--426}\BibitemShut
  {NoStop}%
\bibitem [{\citenamefont {Gisin}\ \emph {et~al.}(2002)\citenamefont {Gisin},
  \citenamefont {Ribordy}, \citenamefont {Tittel},\ and\ \citenamefont
  {Zbinden}}]{Gisin2002}%
  \BibitemOpen
  \bibfield  {author} {\bibinfo {author} {\bibfnamefont {N.}~\bibnamefont
  {Gisin}}, \bibinfo {author} {\bibfnamefont {G.}~\bibnamefont {Ribordy}},
  \bibinfo {author} {\bibfnamefont {W.}~\bibnamefont {Tittel}},\ and\ \bibinfo
  {author} {\bibfnamefont {H.}~\bibnamefont {Zbinden}},\ }\bibfield  {title}
  {\bibinfo {title} {Quantum cryptography},\ }\href
  {https://doi.org/10.1103/RevModPhys.74.145} {\bibfield  {journal} {\bibinfo
  {journal} {Rev. Mod. Phys.}\ }\textbf {\bibinfo {volume} {74}},\ \bibinfo
  {pages} {145} (\bibinfo {year} {2002})}\BibitemShut {NoStop}%
\bibitem [{\citenamefont {Pirandola}\ \emph {et~al.}(2020)\citenamefont
  {Pirandola}, \citenamefont {Andersen}, \citenamefont {Banchi}, \citenamefont
  {Berta}, \citenamefont {Bunandar}, \citenamefont {Colbeck}, \citenamefont
  {Englund}, \citenamefont {Gehring}, \citenamefont {Lupo}, \citenamefont
  {Ottaviani}, \citenamefont {Pereira}, \citenamefont {Razavi}, \citenamefont
  {Shaari}, \citenamefont {Tomamichel}, \citenamefont {Usenko}, \citenamefont
  {Vallone}, \citenamefont {Villoresi},\ and\ \citenamefont
  {Wallden}}]{Pirandola2020}%
  \BibitemOpen
  \bibfield  {author} {\bibinfo {author} {\bibfnamefont {S.}~\bibnamefont
  {Pirandola}}, \bibinfo {author} {\bibfnamefont {U.~L.}\ \bibnamefont
  {Andersen}}, \bibinfo {author} {\bibfnamefont {L.}~\bibnamefont {Banchi}},
  \bibinfo {author} {\bibfnamefont {M.}~\bibnamefont {Berta}}, \bibinfo
  {author} {\bibfnamefont {D.}~\bibnamefont {Bunandar}}, \bibinfo {author}
  {\bibfnamefont {R.}~\bibnamefont {Colbeck}}, \bibinfo {author} {\bibfnamefont
  {D.}~\bibnamefont {Englund}}, \bibinfo {author} {\bibfnamefont
  {T.}~\bibnamefont {Gehring}}, \bibinfo {author} {\bibfnamefont
  {C.}~\bibnamefont {Lupo}}, \bibinfo {author} {\bibfnamefont {C.}~\bibnamefont
  {Ottaviani}}, \bibinfo {author} {\bibfnamefont {J.~L.}\ \bibnamefont
  {Pereira}}, \bibinfo {author} {\bibfnamefont {M.}~\bibnamefont {Razavi}},
  \bibinfo {author} {\bibfnamefont {J.~S.}\ \bibnamefont {Shaari}}, \bibinfo
  {author} {\bibfnamefont {M.}~\bibnamefont {Tomamichel}}, \bibinfo {author}
  {\bibfnamefont {V.~C.}\ \bibnamefont {Usenko}}, \bibinfo {author}
  {\bibfnamefont {G.}~\bibnamefont {Vallone}}, \bibinfo {author} {\bibfnamefont
  {P.}~\bibnamefont {Villoresi}},\ and\ \bibinfo {author} {\bibfnamefont
  {P.}~\bibnamefont {Wallden}},\ }\bibfield  {title} {\bibinfo {title}
  {Advances in quantum cryptography},\ }\href
  {https://doi.org/10.1364/AOP.361502} {\bibfield  {journal} {\bibinfo
  {journal} {Adv. Opt. Photon.}\ }\textbf {\bibinfo {volume} {12}},\ \bibinfo
  {pages} {1012} (\bibinfo {year} {2020})}\BibitemShut {NoStop}%
\bibitem [{\citenamefont {Schuld}\ \emph {et~al.}(2015)\citenamefont {Schuld},
  \citenamefont {Sinayskiy},\ and\ \citenamefont {Petruccione}}]{Schuld2015}%
  \BibitemOpen
  \bibfield  {author} {\bibinfo {author} {\bibfnamefont {M.}~\bibnamefont
  {Schuld}}, \bibinfo {author} {\bibfnamefont {I.}~\bibnamefont {Sinayskiy}},\
  and\ \bibinfo {author} {\bibfnamefont {F.}~\bibnamefont {Petruccione}},\
  }\bibfield  {title} {\bibinfo {title} {An introduction to quantum machine
  learning},\ }\href {https://doi.org/10.1080/00107514.2014.964942} {\bibfield
  {journal} {\bibinfo  {journal} {Contemporary Physics}\ }\textbf {\bibinfo
  {volume} {56}},\ \bibinfo {pages} {172} (\bibinfo {year} {2015})}\BibitemShut
  {NoStop}%
\bibitem [{\citenamefont {Biamonte}\ \emph {et~al.}(2017)\citenamefont
  {Biamonte}, \citenamefont {Wittek}, \citenamefont {Pancotti}, \citenamefont
  {Rebentrost}, \citenamefont {Wiebe},\ and\ \citenamefont
  {Lloyd}}]{Biamonte2017}%
  \BibitemOpen
  \bibfield  {author} {\bibinfo {author} {\bibfnamefont {J.}~\bibnamefont
  {Biamonte}}, \bibinfo {author} {\bibfnamefont {P.}~\bibnamefont {Wittek}},
  \bibinfo {author} {\bibfnamefont {N.}~\bibnamefont {Pancotti}}, \bibinfo
  {author} {\bibfnamefont {P.}~\bibnamefont {Rebentrost}}, \bibinfo {author}
  {\bibfnamefont {N.}~\bibnamefont {Wiebe}},\ and\ \bibinfo {author}
  {\bibfnamefont {S.}~\bibnamefont {Lloyd}},\ }\bibfield  {title} {\bibinfo
  {title} {Quantum machine learning},\ }\href
  {https://doi.org/10.1038/nature23474} {\bibfield  {journal} {\bibinfo
  {journal} {Nature}\ }\textbf {\bibinfo {volume} {549}},\ \bibinfo {pages}
  {195} (\bibinfo {year} {2017})}\BibitemShut {NoStop}%
\bibitem [{\citenamefont {Preskill}(2018)}]{Preskill:2018}%
  \BibitemOpen
  \bibfield  {author} {\bibinfo {author} {\bibfnamefont {J.}~\bibnamefont
  {Preskill}},\ }\bibfield  {title} {\bibinfo {title} {Quantum {C}omputing in
  the {NISQ} era and beyond},\ }\href
  {https://doi.org/10.22331/q-2018-08-06-79} {\bibfield  {journal} {\bibinfo
  {journal} {{Quantum}}\ }\textbf {\bibinfo {volume} {2}},\ \bibinfo {pages}
  {79} (\bibinfo {year} {2018})}\BibitemShut {NoStop}%
\bibitem [{\citenamefont {Arute}\ \emph {et~al.}(2019)\citenamefont {Arute}
  \emph {et~al.}}]{Arute2019}%
  \BibitemOpen
  \bibfield  {author} {\bibinfo {author} {\bibfnamefont {F.}~\bibnamefont
  {Arute}} \emph {et~al.},\ }\bibfield  {title} {\bibinfo {title} {Quantum
  supremacy using a programmable superconducting processor},\ }\href
  {https://doi.org/10.1038/s41586-019-1666-5} {\bibfield  {journal} {\bibinfo
  {journal} {Nature}\ }\textbf {\bibinfo {volume} {574}},\ \bibinfo {pages}
  {505} (\bibinfo {year} {2019})}\BibitemShut {NoStop}%
\bibitem [{\citenamefont {Peruzzo}\ \emph {et~al.}(2014)\citenamefont
  {Peruzzo}, \citenamefont {McClean}, \citenamefont {Shadbolt}, \citenamefont
  {Yung}, \citenamefont {Zhou}, \citenamefont {Love}, \citenamefont
  {Aspuru-Guzik},\ and\ \citenamefont {O'Brien}}]{Peruzzo2014}%
  \BibitemOpen
  \bibfield  {author} {\bibinfo {author} {\bibfnamefont {A.}~\bibnamefont
  {Peruzzo}}, \bibinfo {author} {\bibfnamefont {J.}~\bibnamefont {McClean}},
  \bibinfo {author} {\bibfnamefont {P.}~\bibnamefont {Shadbolt}}, \bibinfo
  {author} {\bibfnamefont {M.}~\bibnamefont {Yung}}, \bibinfo {author}
  {\bibfnamefont {X.}~\bibnamefont {Zhou}}, \bibinfo {author} {\bibfnamefont
  {P.~J.}\ \bibnamefont {Love}}, \bibinfo {author} {\bibfnamefont
  {A.}~\bibnamefont {Aspuru-Guzik}},\ and\ \bibinfo {author} {\bibfnamefont
  {J.~L.}\ \bibnamefont {O'Brien}},\ }\bibfield  {title} {\bibinfo {title} {A
  variational eigenvalue solver on a photonic quantum processor},\ }\href
  {https://doi.org/10.1038/ncomms5213} {\bibfield  {journal} {\bibinfo
  {journal} {Nat. Comm.}\ }\textbf {\bibinfo {volume} {5}},\ \bibinfo {pages}
  {4213} (\bibinfo {year} {2014})}\BibitemShut {NoStop}%
\bibitem [{\citenamefont {McClean}\ \emph {et~al.}(2016)\citenamefont
  {McClean}, \citenamefont {Romero}, \citenamefont {Babbush},\ and\
  \citenamefont {Aspuru-Guzik}}]{McClean:2016}%
  \BibitemOpen
  \bibfield  {author} {\bibinfo {author} {\bibfnamefont {J.~R.}\ \bibnamefont
  {McClean}}, \bibinfo {author} {\bibfnamefont {J.}~\bibnamefont {Romero}},
  \bibinfo {author} {\bibfnamefont {R.}~\bibnamefont {Babbush}},\ and\ \bibinfo
  {author} {\bibfnamefont {A.}~\bibnamefont {Aspuru-Guzik}},\ }\bibfield
  {title} {\bibinfo {title} {The theory of variational hybrid quantum-classical
  algorithms},\ }\href {https://doi.org/10.1088/1367-2630/18/2/023023}
  {\bibfield  {journal} {\bibinfo  {journal} {New J. Phys.}\ }\textbf {\bibinfo
  {volume} {18}},\ \bibinfo {pages} {023023} (\bibinfo {year}
  {2016})}\BibitemShut {NoStop}%
\bibitem [{\citenamefont {O'Malley}\ \emph {et~al.}(2016)\citenamefont
  {O'Malley} \emph {et~al.}}]{OMalley:2016}%
  \BibitemOpen
  \bibfield  {author} {\bibinfo {author} {\bibfnamefont {P.~J.~J.}\
  \bibnamefont {O'Malley}} \emph {et~al.},\ }\bibfield  {title} {\bibinfo
  {title} {Scalable quantum simulation of molecular energies},\ }\href
  {https://doi.org/10.1103/PhysRevX.6.031007} {\bibfield  {journal} {\bibinfo
  {journal} {Phys. Rev. X}\ }\textbf {\bibinfo {volume} {6}},\ \bibinfo {pages}
  {031007} (\bibinfo {year} {2016})}\BibitemShut {NoStop}%
\bibitem [{\citenamefont {Kandala}\ \emph {et~al.}(2017)\citenamefont
  {Kandala}, \citenamefont {Mezzacapo}, \citenamefont {Temme} \emph
  {et~al.}}]{Kandala:2017}%
  \BibitemOpen
  \bibfield  {author} {\bibinfo {author} {\bibfnamefont {A.}~\bibnamefont
  {Kandala}}, \bibinfo {author} {\bibfnamefont {A.}~\bibnamefont {Mezzacapo}},
  \bibinfo {author} {\bibfnamefont {K.}~\bibnamefont {Temme}}, \emph {et~al.},\
  }\bibfield  {title} {\bibinfo {title} {Hardware-efficient variational quantum
  eigensolver for small molecules and quantum magnets},\ }\href
  {https://doi.org/10.1038/nature23879} {\bibfield  {journal} {\bibinfo
  {journal} {{Nat.}}\ }\textbf {\bibinfo {volume} {549}},\ \bibinfo {pages}
  {242} (\bibinfo {year} {2017})}\BibitemShut {NoStop}%
\bibitem [{\citenamefont {Shen}\ \emph {et~al.}(2017)\citenamefont {Shen},
  \citenamefont {Zhang}, \citenamefont {Zhang}, \citenamefont {Zhang},
  \citenamefont {Yung},\ and\ \citenamefont {Kim}}]{Shen:2017}%
  \BibitemOpen
  \bibfield  {author} {\bibinfo {author} {\bibfnamefont {Y.}~\bibnamefont
  {Shen}}, \bibinfo {author} {\bibfnamefont {X.}~\bibnamefont {Zhang}},
  \bibinfo {author} {\bibfnamefont {S.}~\bibnamefont {Zhang}}, \bibinfo
  {author} {\bibfnamefont {J.-N.}\ \bibnamefont {Zhang}}, \bibinfo {author}
  {\bibfnamefont {M.-H.}\ \bibnamefont {Yung}},\ and\ \bibinfo {author}
  {\bibfnamefont {K.}~\bibnamefont {Kim}},\ }\bibfield  {title} {\bibinfo
  {title} {Quantum implementation of the unitary coupled cluster for simulating
  molecular electronic structure},\ }\href
  {https://doi.org/10.1103/PhysRevA.95.020501} {\bibfield  {journal} {\bibinfo
  {journal} {Phys. Rev. A}\ }\textbf {\bibinfo {volume} {95}},\ \bibinfo
  {pages} {020501} (\bibinfo {year} {2017})}\BibitemShut {NoStop}%
\bibitem [{\citenamefont {Colless}\ \emph {et~al.}(2018)\citenamefont
  {Colless}, \citenamefont {Ramasesh}, \citenamefont {Dahlen}, \citenamefont
  {Blok}, \citenamefont {Kimchi-Schwartz}, \citenamefont {McClean},
  \citenamefont {Carter}, \citenamefont {de~Jong},\ and\ \citenamefont
  {Siddiqi}}]{Colless:2018}%
  \BibitemOpen
  \bibfield  {author} {\bibinfo {author} {\bibfnamefont {J.~I.}\ \bibnamefont
  {Colless}}, \bibinfo {author} {\bibfnamefont {V.~V.}\ \bibnamefont
  {Ramasesh}}, \bibinfo {author} {\bibfnamefont {D.}~\bibnamefont {Dahlen}},
  \bibinfo {author} {\bibfnamefont {M.~S.}\ \bibnamefont {Blok}}, \bibinfo
  {author} {\bibfnamefont {M.~E.}\ \bibnamefont {Kimchi-Schwartz}}, \bibinfo
  {author} {\bibfnamefont {J.~R.}\ \bibnamefont {McClean}}, \bibinfo {author}
  {\bibfnamefont {J.}~\bibnamefont {Carter}}, \bibinfo {author} {\bibfnamefont
  {W.~A.}\ \bibnamefont {de~Jong}},\ and\ \bibinfo {author} {\bibfnamefont
  {I.}~\bibnamefont {Siddiqi}},\ }\bibfield  {title} {\bibinfo {title}
  {Computation of molecular spectra on a quantum processor with an
  error-resilient algorithm},\ }\href
  {https://doi.org/10.1103/PhysRevX.8.011021} {\bibfield  {journal} {\bibinfo
  {journal} {Phys. Rev. X}\ }\textbf {\bibinfo {volume} {8}},\ \bibinfo {pages}
  {011021} (\bibinfo {year} {2018})}\BibitemShut {NoStop}%
\bibitem [{\citenamefont {Dumitrescu}\ \emph {et~al.}(2018)\citenamefont
  {Dumitrescu}, \citenamefont {McCaskey}, \citenamefont {Hagen}, \citenamefont
  {Jansen}, \citenamefont {Morris}, \citenamefont {Papenbrock}, \citenamefont
  {Pooser}, \citenamefont {Dean},\ and\ \citenamefont
  {Lougovski}}]{Dumitrescu:2018}%
  \BibitemOpen
  \bibfield  {author} {\bibinfo {author} {\bibfnamefont {E.~F.}\ \bibnamefont
  {Dumitrescu}}, \bibinfo {author} {\bibfnamefont {A.~J.}\ \bibnamefont
  {McCaskey}}, \bibinfo {author} {\bibfnamefont {G.}~\bibnamefont {Hagen}},
  \bibinfo {author} {\bibfnamefont {G.~R.}\ \bibnamefont {Jansen}}, \bibinfo
  {author} {\bibfnamefont {T.~D.}\ \bibnamefont {Morris}}, \bibinfo {author}
  {\bibfnamefont {T.}~\bibnamefont {Papenbrock}}, \bibinfo {author}
  {\bibfnamefont {R.~C.}\ \bibnamefont {Pooser}}, \bibinfo {author}
  {\bibfnamefont {D.~J.}\ \bibnamefont {Dean}},\ and\ \bibinfo {author}
  {\bibfnamefont {P.}~\bibnamefont {Lougovski}},\ }\bibfield  {title} {\bibinfo
  {title} {Cloud quantum computing of an atomic nucleus},\ }\href
  {https://doi.org/10.1103/PhysRevLett.120.210501} {\bibfield  {journal}
  {\bibinfo  {journal} {Phys. Rev. Lett.}\ }\textbf {\bibinfo {volume} {120}},\
  \bibinfo {pages} {210501} (\bibinfo {year} {2018})}\BibitemShut {NoStop}%
\bibitem [{\citenamefont {Hempel}\ \emph {et~al.}(2018)\citenamefont {Hempel}
  \emph {et~al.}}]{Hempel2018}%
  \BibitemOpen
  \bibfield  {author} {\bibinfo {author} {\bibfnamefont {C.}~\bibnamefont
  {Hempel}} \emph {et~al.},\ }\bibfield  {title} {\bibinfo {title} {Quantum
  chemistry calculations on a trapped-ion quantum simulator},\ }\href
  {https://doi.org/10.1103/PhysRevX.8.031022} {\bibfield  {journal} {\bibinfo
  {journal} {Phys. Rev. X}\ }\textbf {\bibinfo {volume} {8}},\ \bibinfo {pages}
  {031022} (\bibinfo {year} {2018})},\ \Eprint
  {https://arxiv.org/abs/1803.10238v1} {1803.10238v1} \BibitemShut {NoStop}%
\bibitem [{\citenamefont {Ganzhorn}\ \emph {et~al.}(2019)\citenamefont
  {Ganzhorn}, \citenamefont {Egger}, \citenamefont {Barkoutsos}, \citenamefont
  {Ollitrault}, \citenamefont {Salis}, \citenamefont {Moll}, \citenamefont
  {Roth}, \citenamefont {Fuhrer}, \citenamefont {Mueller}, \citenamefont
  {Woerner}, \citenamefont {Tavernelli},\ and\ \citenamefont
  {Filipp}}]{Ganzhorn:2019}%
  \BibitemOpen
  \bibfield  {author} {\bibinfo {author} {\bibfnamefont {M.}~\bibnamefont
  {Ganzhorn}}, \bibinfo {author} {\bibfnamefont {D.}~\bibnamefont {Egger}},
  \bibinfo {author} {\bibfnamefont {P.}~\bibnamefont {Barkoutsos}}, \bibinfo
  {author} {\bibfnamefont {P.}~\bibnamefont {Ollitrault}}, \bibinfo {author}
  {\bibfnamefont {G.}~\bibnamefont {Salis}}, \bibinfo {author} {\bibfnamefont
  {N.}~\bibnamefont {Moll}}, \bibinfo {author} {\bibfnamefont {M.}~\bibnamefont
  {Roth}}, \bibinfo {author} {\bibfnamefont {A.}~\bibnamefont {Fuhrer}},
  \bibinfo {author} {\bibfnamefont {P.}~\bibnamefont {Mueller}}, \bibinfo
  {author} {\bibfnamefont {S.}~\bibnamefont {Woerner}}, \bibinfo {author}
  {\bibfnamefont {I.}~\bibnamefont {Tavernelli}},\ and\ \bibinfo {author}
  {\bibfnamefont {S.}~\bibnamefont {Filipp}},\ }\bibfield  {title} {\bibinfo
  {title} {Gate-efficient simulation of molecular eigenstates on a quantum
  computer},\ }\href {https://doi.org/10.1103/PhysRevApplied.11.044092}
  {\bibfield  {journal} {\bibinfo  {journal} {Phys. Rev. Applied}\ }\textbf
  {\bibinfo {volume} {11}},\ \bibinfo {pages} {044092} (\bibinfo {year}
  {2019})}\BibitemShut {NoStop}%
\bibitem [{\citenamefont {Kokail}\ \emph {et~al.}(2019)\citenamefont {Kokail},
  \citenamefont {Maier}, \citenamefont {van Bijnen} \emph
  {et~al.}}]{Kokail:2019}%
  \BibitemOpen
  \bibfield  {author} {\bibinfo {author} {\bibfnamefont {C.}~\bibnamefont
  {Kokail}}, \bibinfo {author} {\bibfnamefont {C.}~\bibnamefont {Maier}},
  \bibinfo {author} {\bibfnamefont {R.}~\bibnamefont {van Bijnen}}, \emph
  {et~al.},\ }\bibfield  {title} {\bibinfo {title} {Self-verifying variational
  quantum simulation of lattice models},\ }\href
  {https://doi.org/10.1038/s41586-019-1177-4} {\bibfield  {journal} {\bibinfo
  {journal} {{Nat.}}\ }\textbf {\bibinfo {volume} {569}},\ \bibinfo {pages}
  {355} (\bibinfo {year} {2019})}\BibitemShut {NoStop}%
\bibitem [{\citenamefont {Hartung}\ and\ \citenamefont
  {Jansen}(2019)}]{Hartung2019}%
  \BibitemOpen
  \bibfield  {author} {\bibinfo {author} {\bibfnamefont {T.}~\bibnamefont
  {Hartung}}\ and\ \bibinfo {author} {\bibfnamefont {K.}~\bibnamefont
  {Jansen}},\ }\bibfield  {title} {\bibinfo {title} {Zeta-regularized vacuum
  expectation values},\ }\href {https://doi.org/10.1063/1.5085866} {\bibfield
  {journal} {\bibinfo  {journal} {J. Math. Phys.}\ }\textbf {\bibinfo {volume}
  {60}},\ \bibinfo {pages} {093504} (\bibinfo {year} {2019})}\BibitemShut
  {NoStop}%
\bibitem [{\citenamefont {Jansen}\ and\ \citenamefont
  {Hartung}(2020)}]{Jansen2020}%
  \BibitemOpen
  \bibfield  {author} {\bibinfo {author} {\bibfnamefont {K.}~\bibnamefont
  {Jansen}}\ and\ \bibinfo {author} {\bibfnamefont {T.}~\bibnamefont
  {Hartung}},\ }\bibfield  {title} {\bibinfo {title} {Zeta-regularized vacuum
  expectation values fromquantum computing simulations},\ }\href
  {https://doi.org/10.22323/1.363.0153} {\bibfield  {journal} {\bibinfo
  {journal} {Proc. Sci. LATTICE2019}\ }\textbf {\bibinfo {volume} {363}},\
  \bibinfo {pages} {153} (\bibinfo {year} {2020})}\BibitemShut {NoStop}%
\bibitem [{\citenamefont {Li}\ and\ \citenamefont {Benjamin}(2017)}]{Li:2017}%
  \BibitemOpen
  \bibfield  {author} {\bibinfo {author} {\bibfnamefont {Y.}~\bibnamefont
  {Li}}\ and\ \bibinfo {author} {\bibfnamefont {S.~C.}\ \bibnamefont
  {Benjamin}},\ }\bibfield  {title} {\bibinfo {title} {Efficient variational
  quantum simulator incorporating active error minimization},\ }\href
  {https://doi.org/10.1103/PhysRevX.7.021050} {\bibfield  {journal} {\bibinfo
  {journal} {Phys. Rev. X}\ }\textbf {\bibinfo {volume} {7}},\ \bibinfo {pages}
  {021050} (\bibinfo {year} {2017})}\BibitemShut {NoStop}%
\bibitem [{\citenamefont {Temme}\ \emph {et~al.}(2017)\citenamefont {Temme},
  \citenamefont {Bravyi},\ and\ \citenamefont {Gambetta}}]{Temme:2017}%
  \BibitemOpen
  \bibfield  {author} {\bibinfo {author} {\bibfnamefont {K.}~\bibnamefont
  {Temme}}, \bibinfo {author} {\bibfnamefont {S.}~\bibnamefont {Bravyi}},\ and\
  \bibinfo {author} {\bibfnamefont {J.~M.}\ \bibnamefont {Gambetta}},\
  }\bibfield  {title} {\bibinfo {title} {Error mitigation for short-depth
  quantum circuits},\ }\href {https://doi.org/10.1103/PhysRevLett.119.180509}
  {\bibfield  {journal} {\bibinfo  {journal} {Phys. Rev. Lett.}\ }\textbf
  {\bibinfo {volume} {119}},\ \bibinfo {pages} {180509} (\bibinfo {year}
  {2017})}\BibitemShut {NoStop}%
\bibitem [{\citenamefont {McClean}\ \emph {et~al.}(2017)\citenamefont
  {McClean}, \citenamefont {Kimchi-Schwartz}, \citenamefont {Carter},\ and\
  \citenamefont {de~Jong}}]{McClean:2017}%
  \BibitemOpen
  \bibfield  {author} {\bibinfo {author} {\bibfnamefont {J.~R.}\ \bibnamefont
  {McClean}}, \bibinfo {author} {\bibfnamefont {M.~E.}\ \bibnamefont
  {Kimchi-Schwartz}}, \bibinfo {author} {\bibfnamefont {J.}~\bibnamefont
  {Carter}},\ and\ \bibinfo {author} {\bibfnamefont {W.~A.}\ \bibnamefont
  {de~Jong}},\ }\bibfield  {title} {\bibinfo {title} {Hybrid quantum-classical
  hierarchy for mitigation of decoherence and determination of excited
  states},\ }\href {https://doi.org/10.1103/PhysRevA.95.042308} {\bibfield
  {journal} {\bibinfo  {journal} {Phys. Rev. A}\ }\textbf {\bibinfo {volume}
  {95}},\ \bibinfo {pages} {042308} (\bibinfo {year} {2017})}\BibitemShut
  {NoStop}%
\bibitem [{\citenamefont {Bonet-Monroig}\ \emph {et~al.}(2018)\citenamefont
  {Bonet-Monroig}, \citenamefont {Sagastizabal}, \citenamefont {Singh},\ and\
  \citenamefont {O'Brien}}]{BonetMonroig:2018}%
  \BibitemOpen
  \bibfield  {author} {\bibinfo {author} {\bibfnamefont {X.}~\bibnamefont
  {Bonet-Monroig}}, \bibinfo {author} {\bibfnamefont {R.}~\bibnamefont
  {Sagastizabal}}, \bibinfo {author} {\bibfnamefont {M.}~\bibnamefont
  {Singh}},\ and\ \bibinfo {author} {\bibfnamefont {T.~E.}\ \bibnamefont
  {O'Brien}},\ }\bibfield  {title} {\bibinfo {title} {Low-cost error mitigation
  by symmetry verification},\ }\href
  {https://doi.org/10.1103/PhysRevA.98.062339} {\bibfield  {journal} {\bibinfo
  {journal} {Phys. Rev. A}\ }\textbf {\bibinfo {volume} {98}},\ \bibinfo
  {pages} {062339} (\bibinfo {year} {2018})}\BibitemShut {NoStop}%
\bibitem [{\citenamefont {Endo}\ \emph {et~al.}(2018)\citenamefont {Endo},
  \citenamefont {Benjamin},\ and\ \citenamefont {Li}}]{Endo:2018}%
  \BibitemOpen
  \bibfield  {author} {\bibinfo {author} {\bibfnamefont {S.}~\bibnamefont
  {Endo}}, \bibinfo {author} {\bibfnamefont {S.~C.}\ \bibnamefont {Benjamin}},\
  and\ \bibinfo {author} {\bibfnamefont {Y.}~\bibnamefont {Li}},\ }\bibfield
  {title} {\bibinfo {title} {Practical quantum error mitigation for near-future
  applications},\ }\href {https://doi.org/10.1103/PhysRevX.8.03102} {\bibfield
  {journal} {\bibinfo  {journal} {Phys. Rev. X}\ }\textbf {\bibinfo {volume}
  {8}},\ \bibinfo {pages} {031027} (\bibinfo {year} {2018})}\BibitemShut
  {NoStop}%
\bibitem [{\citenamefont {McArdle}\ \emph {et~al.}(2019)\citenamefont
  {McArdle}, \citenamefont {Yuan},\ and\ \citenamefont
  {Benjamin}}]{McArdle:2019}%
  \BibitemOpen
  \bibfield  {author} {\bibinfo {author} {\bibfnamefont {S.}~\bibnamefont
  {McArdle}}, \bibinfo {author} {\bibfnamefont {X.}~\bibnamefont {Yuan}},\ and\
  \bibinfo {author} {\bibfnamefont {S.}~\bibnamefont {Benjamin}},\ }\bibfield
  {title} {\bibinfo {title} {Error-mitigated digital quantum simulation},\
  }\href {https://doi.org/10.1103/PhysRevLett.122.180501} {\bibfield  {journal}
  {\bibinfo  {journal} {Phys. Rev. Lett.}\ }\textbf {\bibinfo {volume} {122}},\
  \bibinfo {pages} {180501} (\bibinfo {year} {2019})}\BibitemShut {NoStop}%
\bibitem [{\citenamefont {Endo}\ \emph {et~al.}(2019)\citenamefont {Endo},
  \citenamefont {Zhao}, \citenamefont {Li}, \citenamefont {Benjamin},\ and\
  \citenamefont {Yuan}}]{Endo:2019}%
  \BibitemOpen
  \bibfield  {author} {\bibinfo {author} {\bibfnamefont {S.}~\bibnamefont
  {Endo}}, \bibinfo {author} {\bibfnamefont {Q.}~\bibnamefont {Zhao}}, \bibinfo
  {author} {\bibfnamefont {Y.}~\bibnamefont {Li}}, \bibinfo {author}
  {\bibfnamefont {S.}~\bibnamefont {Benjamin}},\ and\ \bibinfo {author}
  {\bibfnamefont {X.}~\bibnamefont {Yuan}},\ }\bibfield  {title} {\bibinfo
  {title} {Mitigating algorithmic errors in a hamiltonian simulation},\ }\href
  {https://doi.org/10.1103/PhysRevA.99.012334} {\bibfield  {journal} {\bibinfo
  {journal} {Phys. Rev. A}\ }\textbf {\bibinfo {volume} {99}},\ \bibinfo
  {pages} {012334} (\bibinfo {year} {2019})}\BibitemShut {NoStop}%
\bibitem [{\citenamefont {Kandala}\ \emph {et~al.}(2019)\citenamefont
  {Kandala}, \citenamefont {Temme}, \citenamefont {Córcoles} \emph
  {et~al.}}]{Kandala:2019}%
  \BibitemOpen
  \bibfield  {author} {\bibinfo {author} {\bibfnamefont {A.}~\bibnamefont
  {Kandala}}, \bibinfo {author} {\bibfnamefont {K.}~\bibnamefont {Temme}},
  \bibinfo {author} {\bibfnamefont {A.~D.}\ \bibnamefont {Córcoles}}, \emph
  {et~al.},\ }\bibfield  {title} {\bibinfo {title} {Error mitigation extends
  the computational reach of a noisy quantum processor},\ }\href
  {https://doi.org/10.1038/s41586-019-1040-7} {\bibfield  {journal} {\bibinfo
  {journal} {{Nat.}}\ }\textbf {\bibinfo {volume} {567}},\ \bibinfo {pages}
  {491} (\bibinfo {year} {2019})}\BibitemShut {NoStop}%
\bibitem [{\citenamefont {McClean}\ \emph {et~al.}(2020)\citenamefont
  {McClean}, \citenamefont {Jiang}, \citenamefont {Rubin}, \citenamefont
  {Babbush},\ and\ \citenamefont {Neven}}]{McClean:2020}%
  \BibitemOpen
  \bibfield  {author} {\bibinfo {author} {\bibfnamefont {J.~R.}\ \bibnamefont
  {McClean}}, \bibinfo {author} {\bibfnamefont {Z.}~\bibnamefont {Jiang}},
  \bibinfo {author} {\bibfnamefont {N.~C.}\ \bibnamefont {Rubin}}, \bibinfo
  {author} {\bibfnamefont {R.}~\bibnamefont {Babbush}},\ and\ \bibinfo {author}
  {\bibfnamefont {H.}~\bibnamefont {Neven}},\ }\bibfield  {title} {\bibinfo
  {title} {Decoding quantum errors with subspace expansions},\ }\bibfield
  {journal} {\bibinfo  {journal} {Nat. Comm.}\ }\textbf {\bibinfo {volume}
  {11}},\ \href {https://doi.org/10.1038/s41467-020-14341-w}
  {10.1038/s41467-020-14341-w} (\bibinfo {year} {2020})\BibitemShut {NoStop}%
\bibitem [{\citenamefont {Otten}\ and\ \citenamefont
  {Gray}(2019{\natexlab{a}})}]{Otten:2019a}%
  \BibitemOpen
  \bibfield  {author} {\bibinfo {author} {\bibfnamefont {M.}~\bibnamefont
  {Otten}}\ and\ \bibinfo {author} {\bibfnamefont {S.~K.}\ \bibnamefont
  {Gray}},\ }\bibfield  {title} {\bibinfo {title} {Recovering noise-free
  quantum observables},\ }\href {https://doi.org/10.1103/PhysRevA.99.012338}
  {\bibfield  {journal} {\bibinfo  {journal} {Phys. Rev. A}\ }\textbf {\bibinfo
  {volume} {99}},\ \bibinfo {pages} {012338} (\bibinfo {year}
  {2019}{\natexlab{a}})}\BibitemShut {NoStop}%
\bibitem [{\citenamefont {Otten}\ and\ \citenamefont
  {Gray}(2019{\natexlab{b}})}]{Otten:2019b}%
  \BibitemOpen
  \bibfield  {author} {\bibinfo {author} {\bibfnamefont {M.}~\bibnamefont
  {Otten}}\ and\ \bibinfo {author} {\bibfnamefont {S.~K.}\ \bibnamefont
  {Gray}},\ }\bibfield  {title} {\bibinfo {title} {Accounting for errors in
  quantum algorithms via individual error reduction},\ }\href
  {https://doi.org/10.1038/s41534-019-0125-3} {\bibfield  {journal} {\bibinfo
  {journal} {npj Quantum Inf.}\ }\textbf {\bibinfo {volume} {5}},\ \bibinfo
  {pages} {11} (\bibinfo {year} {2019}{\natexlab{b}})}\BibitemShut {NoStop}%
\bibitem [{\citenamefont {Sagastizabal}\ \emph {et~al.}(2019)\citenamefont
  {Sagastizabal} \emph {et~al.}}]{Sagastizabal:2019}%
  \BibitemOpen
  \bibfield  {author} {\bibinfo {author} {\bibfnamefont {R.}~\bibnamefont
  {Sagastizabal}} \emph {et~al.},\ }\bibfield  {title} {\bibinfo {title}
  {Experimental error mitigation via symmetry verification in a variational
  quantum eigensolver},\ }\href {https://doi.org/10.1103/PhysRevA.100.010302}
  {\bibfield  {journal} {\bibinfo  {journal} {Phys. Rev. A}\ }\textbf {\bibinfo
  {volume} {100}},\ \bibinfo {pages} {010302} (\bibinfo {year}
  {2019})}\BibitemShut {NoStop}%
\bibitem [{\citenamefont {Urbanek}\ \emph {et~al.}(2019)\citenamefont
  {Urbanek}, \citenamefont {Nachman},\ and\ \citenamefont
  {de~Jong}}]{Urbanek:2019}%
  \BibitemOpen
  \bibfield  {author} {\bibinfo {author} {\bibfnamefont {M.}~\bibnamefont
  {Urbanek}}, \bibinfo {author} {\bibfnamefont {B.}~\bibnamefont {Nachman}},\
  and\ \bibinfo {author} {\bibfnamefont {W.~A.}\ \bibnamefont {de~Jong}},\
  }\bibfield  {title} {\bibinfo {title} {Quantum error detection improves
  accuracy of chemical calculations on a quantum computer},\ }\href
  {https://arxiv.org/abs/1910.00129} {\bibfield  {journal} {\bibinfo  {journal}
  {arXiv:1910.00129}\ } (\bibinfo {year} {2019})},\ \Eprint
  {https://arxiv.org/abs/1910.00129v1} {1910.00129v1} \BibitemShut {NoStop}%
\bibitem [{\citenamefont {Crawford}\ \emph {et~al.}(2019)\citenamefont
  {Crawford}, \citenamefont {van Straaten}, \citenamefont {Wang}, \citenamefont
  {Parks}, \citenamefont {Campbell},\ and\ \citenamefont
  {Brierley}}]{Crawford:2019}%
  \BibitemOpen
  \bibfield  {author} {\bibinfo {author} {\bibfnamefont {O.}~\bibnamefont
  {Crawford}}, \bibinfo {author} {\bibfnamefont {B.}~\bibnamefont {van
  Straaten}}, \bibinfo {author} {\bibfnamefont {D.}~\bibnamefont {Wang}},
  \bibinfo {author} {\bibfnamefont {T.}~\bibnamefont {Parks}}, \bibinfo
  {author} {\bibfnamefont {E.}~\bibnamefont {Campbell}},\ and\ \bibinfo
  {author} {\bibfnamefont {S.}~\bibnamefont {Brierley}},\ }\bibfield  {title}
  {\bibinfo {title} {Efficient quantum measurement of pauli operators in the
  presence of finite sampling error},\ }\href
  {https://arxiv.org/abs/1908.06942} {\bibfield  {journal} {\bibinfo  {journal}
  {arXiv:1908.06942}\ } (\bibinfo {year} {2019})},\ \Eprint
  {https://arxiv.org/abs/1908.06942v2} {1908.06942v2} \BibitemShut {NoStop}%
\bibitem [{\citenamefont {Chungheon}\ \emph {et~al.}(2019)\citenamefont
  {Chungheon}, \citenamefont {Tomohiro}, \citenamefont {Seigo},\ and\
  \citenamefont {Byung-Soo}}]{Baek:2019}%
  \BibitemOpen
  \bibfield  {author} {\bibinfo {author} {\bibfnamefont {B.}~\bibnamefont
  {Chungheon}}, \bibinfo {author} {\bibfnamefont {O.}~\bibnamefont {Tomohiro}},
  \bibinfo {author} {\bibfnamefont {T.}~\bibnamefont {Seigo}},\ and\ \bibinfo
  {author} {\bibfnamefont {C.}~\bibnamefont {Byung-Soo}},\ }\bibfield  {title}
  {\bibinfo {title} {Density matrix simulation of quantum error correction
  codes for near-term quantum devices},\ }\href
  {https://doi.org/10.1088/2058-9565/ab5887} {\bibfield  {journal} {\bibinfo
  {journal} {Quantum Sci. Technol.}\ }\textbf {\bibinfo {volume} {5}},\
  \bibinfo {pages} {015002} (\bibinfo {year} {2019})}\BibitemShut {NoStop}%
\bibitem [{\citenamefont {Córcoles}\ \emph {et~al.}(2015)\citenamefont
  {Córcoles}, \citenamefont {Magesan}, \citenamefont {Srinivasan},
  \citenamefont {Cross}, \citenamefont {Steffen}, \citenamefont {Gambetta},\
  and\ \citenamefont {Chow}}]{Corcoles:2015}%
  \BibitemOpen
  \bibfield  {author} {\bibinfo {author} {\bibfnamefont {A.~D.}\ \bibnamefont
  {Córcoles}}, \bibinfo {author} {\bibfnamefont {E.}~\bibnamefont {Magesan}},
  \bibinfo {author} {\bibfnamefont {S.~J.}\ \bibnamefont {Srinivasan}},
  \bibinfo {author} {\bibfnamefont {A.~W.}\ \bibnamefont {Cross}}, \bibinfo
  {author} {\bibfnamefont {M.}~\bibnamefont {Steffen}}, \bibinfo {author}
  {\bibfnamefont {J.~M.}\ \bibnamefont {Gambetta}},\ and\ \bibinfo {author}
  {\bibfnamefont {J.~M.}\ \bibnamefont {Chow}},\ }\bibfield  {title} {\bibinfo
  {title} {Demonstration of a quantum error detection code using a square
  lattice of four superconducting qubits},\ }\href
  {https://doi.org/10.1038/ncomms7979} {\bibfield  {journal} {\bibinfo
  {journal} {Nat. Commun.}\ }\textbf {\bibinfo {volume} {6}},\ \bibinfo {pages}
  {6979} (\bibinfo {year} {2015})}\BibitemShut {NoStop}%
\bibitem [{\citenamefont {Sheldon}\ \emph {et~al.}(2016)\citenamefont
  {Sheldon}, \citenamefont {Magesan}, \citenamefont {Chow},\ and\ \citenamefont
  {Gambetta}}]{Sheldon:2016}%
  \BibitemOpen
  \bibfield  {author} {\bibinfo {author} {\bibfnamefont {S.}~\bibnamefont
  {Sheldon}}, \bibinfo {author} {\bibfnamefont {E.}~\bibnamefont {Magesan}},
  \bibinfo {author} {\bibfnamefont {J.~M.}\ \bibnamefont {Chow}},\ and\
  \bibinfo {author} {\bibfnamefont {J.~M.}\ \bibnamefont {Gambetta}},\
  }\bibfield  {title} {\bibinfo {title} {Procedure for systematically tuning up
  cross-talk in the cross-resonance gate},\ }\href
  {https://doi.org/10.1103/PhysRevA.93.060302} {\bibfield  {journal} {\bibinfo
  {journal} {Phys. Rev. A}\ }\textbf {\bibinfo {volume} {93}},\ \bibinfo
  {pages} {060302} (\bibinfo {year} {2016})}\BibitemShut {NoStop}%
\bibitem [{\citenamefont {Tannu}\ and\ \citenamefont
  {Qureshi}(2019)}]{Tannu:2019}%
  \BibitemOpen
  \bibfield  {author} {\bibinfo {author} {\bibfnamefont {S.~S.}\ \bibnamefont
  {Tannu}}\ and\ \bibinfo {author} {\bibfnamefont {M.~K.}\ \bibnamefont
  {Qureshi}},\ }\bibfield  {title} {\bibinfo {title} {Mitigating measurement
  errors in quantum computers by exploiting state-dependent bias},\ }in\ \href
  {https://doi.org/10.1145/3352460.3358265} {\emph {\bibinfo {booktitle}
  {Proceedings of the 52nd Annual IEEE/ACM International Symposium on
  Microarchitecture}}},\ \bibinfo {series and number} {MICRO ’52}\ (\bibinfo
  {publisher} {Association for Computing Machinery},\ \bibinfo {address} {New
  York, NY, USA},\ \bibinfo {year} {2019})\ p.\ \bibinfo {pages}
  {279}\BibitemShut {NoStop}%
\bibitem [{\citenamefont {Yeter-Aydeniz}\ \emph {et~al.}(2019)\citenamefont
  {Yeter-Aydeniz}, \citenamefont {Dumitrescu}, \citenamefont {McCaskey},
  \citenamefont {Bennink}, \citenamefont {Pooser},\ and\ \citenamefont
  {Siopsis}}]{YeterAydeniz2019}%
  \BibitemOpen
  \bibfield  {author} {\bibinfo {author} {\bibfnamefont {K.}~\bibnamefont
  {Yeter-Aydeniz}}, \bibinfo {author} {\bibfnamefont {E.~F.}\ \bibnamefont
  {Dumitrescu}}, \bibinfo {author} {\bibfnamefont {A.~J.}\ \bibnamefont
  {McCaskey}}, \bibinfo {author} {\bibfnamefont {R.~S.}\ \bibnamefont
  {Bennink}}, \bibinfo {author} {\bibfnamefont {R.~C.}\ \bibnamefont
  {Pooser}},\ and\ \bibinfo {author} {\bibfnamefont {G.}~\bibnamefont
  {Siopsis}},\ }\bibfield  {title} {\bibinfo {title} {Scalar quantum field
  theories as a benchmark for near-term quantum computers},\ }\href
  {https://doi.org/10.1103/PhysRevA.99.032306} {\bibfield  {journal} {\bibinfo
  {journal} {Phys. Rev. A}\ }\textbf {\bibinfo {volume} {99}},\ \bibinfo
  {pages} {032306} (\bibinfo {year} {2019})}\BibitemShut {NoStop}%
\bibitem [{\citenamefont {Yeter-Aydeniz}\ \emph {et~al.}(2020)\citenamefont
  {Yeter-Aydeniz}, \citenamefont {Pooser},\ and\ \citenamefont
  {Siopsis}}]{YeterAydeniz2020}%
  \BibitemOpen
  \bibfield  {author} {\bibinfo {author} {\bibfnamefont {K.}~\bibnamefont
  {Yeter-Aydeniz}}, \bibinfo {author} {\bibfnamefont {R.~C.}\ \bibnamefont
  {Pooser}},\ and\ \bibinfo {author} {\bibfnamefont {G.}~\bibnamefont
  {Siopsis}},\ }\bibfield  {title} {\bibinfo {title} {Practical quantum
  computation of chemical and nuclear energy levels using quantum imaginary
  time evolution and lanczos algorithms},\ }\bibfield  {journal} {\bibinfo
  {journal} {npj Quantum Information}\ }\textbf {\bibinfo {volume} {6}},\ \href
  {https://doi.org/10.1038/s41534-020-00290-1} {10.1038/s41534-020-00290-1}
  (\bibinfo {year} {2020})\BibitemShut {NoStop}%
\bibitem [{\citenamefont {van~den Berg}\ \emph {et~al.}(2021)\citenamefont
  {van~den Berg}, \citenamefont {Minev},\ and\ \citenamefont
  {Temme}}]{berg2021modelfree}%
  \BibitemOpen
  \bibfield  {author} {\bibinfo {author} {\bibfnamefont {E.}~\bibnamefont
  {van~den Berg}}, \bibinfo {author} {\bibfnamefont {Z.~K.}\ \bibnamefont
  {Minev}},\ and\ \bibinfo {author} {\bibfnamefont {K.}~\bibnamefont {Temme}},\
  }\href@noop {} {\bibinfo {title} {Model-free readout-error mitigation for
  quantum expectation values}} (\bibinfo {year} {2021}),\ \Eprint
  {https://arxiv.org/abs/2012.09738} {arXiv:2012.09738 [quant-ph]} \BibitemShut
  {NoStop}%
\bibitem [{\citenamefont {Kaufmann}\ \emph {et~al.}(2017)\citenamefont
  {Kaufmann}, \citenamefont {Ruster}, \citenamefont {Schmiegelow},
  \citenamefont {Luda}, \citenamefont {Kaushal}, \citenamefont {Schulz},
  \citenamefont {von Lindenfels}, \citenamefont {Schmidt-Kaler},\ and\
  \citenamefont {Poschinger}}]{Kaufmann2017}%
  \BibitemOpen
  \bibfield  {author} {\bibinfo {author} {\bibfnamefont {H.}~\bibnamefont
  {Kaufmann}}, \bibinfo {author} {\bibfnamefont {T.}~\bibnamefont {Ruster}},
  \bibinfo {author} {\bibfnamefont {C.~T.}\ \bibnamefont {Schmiegelow}},
  \bibinfo {author} {\bibfnamefont {M.~A.}\ \bibnamefont {Luda}}, \bibinfo
  {author} {\bibfnamefont {V.}~\bibnamefont {Kaushal}}, \bibinfo {author}
  {\bibfnamefont {J.}~\bibnamefont {Schulz}}, \bibinfo {author} {\bibfnamefont
  {D.}~\bibnamefont {von Lindenfels}}, \bibinfo {author} {\bibfnamefont
  {F.}~\bibnamefont {Schmidt-Kaler}},\ and\ \bibinfo {author} {\bibfnamefont
  {U.~G.}\ \bibnamefont {Poschinger}},\ }\bibfield  {title} {\bibinfo {title}
  {Fast ion swapping for quantum-information processing},\ }\href
  {https://doi.org/10.1103/PhysRevA.95.052319} {\bibfield  {journal} {\bibinfo
  {journal} {Phys. Rev. A}\ }\textbf {\bibinfo {volume} {95}},\ \bibinfo
  {pages} {052319} (\bibinfo {year} {2017})}\BibitemShut {NoStop}%
\bibitem [{\citenamefont {{Qiskit Aer API documentation and source
  code}}()}]{Qiskit:2020}%
  \BibitemOpen
  \bibfield  {author} {\bibinfo {author} {\bibnamefont {{Qiskit Aer API
  documentation and source code}}},\ }\href@noop {} {}\bibinfo {note}
  {{accessed on 2020/07/03}}\BibitemShut {NoStop}%
\bibitem [{\citenamefont {Mooney}\ \emph {et~al.}(2021)\citenamefont {Mooney},
  \citenamefont {White}, \citenamefont {Hill},\ and\ \citenamefont
  {Hollenberg}}]{Mooney2021}%
  \BibitemOpen
  \bibfield  {author} {\bibinfo {author} {\bibfnamefont {G.~J.}\ \bibnamefont
  {Mooney}}, \bibinfo {author} {\bibfnamefont {G.~A.~L.}\ \bibnamefont
  {White}}, \bibinfo {author} {\bibfnamefont {C.~D.}\ \bibnamefont {Hill}},\
  and\ \bibinfo {author} {\bibfnamefont {L.~C.~L.}\ \bibnamefont
  {Hollenberg}},\ }\bibfield  {title} {\bibinfo {title} {Generation and
  verification of 27-qubit greenberger-horne-zeilinger states in a
  superconducting quantum computer},\ }\href
  {https://doi.org/10.1088/2399-6528/ac1df7} {\bibfield  {journal} {\bibinfo
  {journal} {J. Phys. Commun.}\ }\textbf {\bibinfo {volume} {5}},\ \bibinfo
  {pages} {095004} (\bibinfo {year} {2021})}\BibitemShut {NoStop}%
\bibitem [{\citenamefont {Nielsen}\ and\ \citenamefont
  {Chuang}(2000)}]{Nielsen2000}%
  \BibitemOpen
  \bibfield  {author} {\bibinfo {author} {\bibfnamefont {M.~A.}\ \bibnamefont
  {Nielsen}}\ and\ \bibinfo {author} {\bibfnamefont {I.~L.}\ \bibnamefont
  {Chuang}},\ }\href@noop {} {\emph {\bibinfo {title} {Quantum Computation and
  Quantum Information}}}\ (\bibinfo  {publisher} {Cambridge University Press},\
  \bibinfo {address} {Cambridge},\ \bibinfo {year} {2000})\ p.\ \bibinfo
  {pages} {375}\BibitemShut {NoStop}%
\bibitem [{\citenamefont {Abraham}\ \emph {et~al.}(2019)\citenamefont {Abraham}
  \emph {et~al.}}]{Qiskit}%
  \BibitemOpen
  \bibfield  {author} {\bibinfo {author} {\bibfnamefont {H.}~\bibnamefont
  {Abraham}} \emph {et~al.},\ }\bibfield  {title} {\bibinfo {title} {Qiskit: An
  open-source framework for quantum computing},\ }\bibfield  {journal}
  {\bibinfo  {journal} {Zenodo}\ }\href
  {https://doi.org/10.5281/zenodo.2562110} {10.5281/zenodo.2562110} (\bibinfo
  {year} {2019})\BibitemShut {NoStop}%
\bibitem [{Lon()}]{London2020}%
  \BibitemOpen
  \href {https://quantum-computing.ibm.com} {\bibinfo {title}
  {{\textit{ibmq\_london} v1.1.0, IBM Quantum team}}},\ \bibinfo {note}
  {{Retrieved from https://quantum-computing.ibm.com (2020)}}\BibitemShut
  {NoStop}%
\bibitem [{Bur()}]{Burlington2020}%
  \BibitemOpen
  \href {https://quantum-computing.ibm.com} {\bibinfo {title}
  {{\textit{ibmq\_burlington} v1.1.4, IBM Quantum team}}},\ \bibinfo {note}
  {{Retrieved from https://quantum-computing.ibm.com (2020)}}\BibitemShut
  {NoStop}%
\bibitem [{\citenamefont {Geller}\ and\ \citenamefont
  {Sun}(2020)}]{Geller2020}%
  \BibitemOpen
  \bibfield  {author} {\bibinfo {author} {\bibfnamefont {M.~R.}\ \bibnamefont
  {Geller}}\ and\ \bibinfo {author} {\bibfnamefont {M.}~\bibnamefont {Sun}},\
  }\bibfield  {title} {\bibinfo {title} {Efficient correction of multiqubit
  measurement errors},\ }\href {https://arxiv.org/abs/2001.09980} {\bibfield
  {journal} {\bibinfo  {journal} {arXiv:2001.09980}\ } (\bibinfo {year}
  {2020})},\ \Eprint {https://arxiv.org/abs/2001.09980v2} {2001.09980v2}
  \BibitemShut {NoStop}%
\bibitem [{\citenamefont {Geller}(2020)}]{Geller2020a}%
  \BibitemOpen
  \bibfield  {author} {\bibinfo {author} {\bibfnamefont {M.~R.}\ \bibnamefont
  {Geller}},\ }\bibfield  {title} {\bibinfo {title} {Rigorous measurement error
  correction},\ }\href {https://doi.org/10.1088/2058-9565/ab9591} {\bibfield
  {journal} {\bibinfo  {journal} {Quantum Science and Technology}\ }\textbf
  {\bibinfo {volume} {5}},\ \bibinfo {pages} {03LT01} (\bibinfo {year}
  {2020})}\BibitemShut {NoStop}%
\bibitem [{Note1()}]{Note1}%
  \BibitemOpen
  \bibinfo {note} {We thank Tom Weber for pointing this out to us.}\BibitemShut
  {Stop}%
\bibitem [{\citenamefont {Ba\~{n}uls}\ \emph {et~al.}(2013)\citenamefont
  {Ba\~{n}uls}, \citenamefont {Cichy}, \citenamefont {Jansen},\ and\
  \citenamefont {Cirac}}]{Banuls2013}%
  \BibitemOpen
  \bibfield  {author} {\bibinfo {author} {\bibfnamefont {M.~C.}\ \bibnamefont
  {Ba\~{n}uls}}, \bibinfo {author} {\bibfnamefont {K.}~\bibnamefont {Cichy}},
  \bibinfo {author} {\bibfnamefont {K.}~\bibnamefont {Jansen}},\ and\ \bibinfo
  {author} {\bibfnamefont {J.~I.}\ \bibnamefont {Cirac}},\ }\bibfield  {title}
  {\bibinfo {title} {The mass spectrum of the schwinger model with matrix
  product states},\ }\href {https://doi.org/10.1007/JHEP11(2013)158} {\bibfield
   {journal} {\bibinfo  {journal} {J. High Energy Phys.}\ }\textbf {\bibinfo
  {volume} {2013}}\bibinfo  {number} { (11)},\ \bibinfo {pages}
  {158}}\BibitemShut {NoStop}%
\bibitem [{\citenamefont {Ba\~nuls}\ \emph {et~al.}(2017)\citenamefont
  {Ba\~nuls}, \citenamefont {Cichy}, \citenamefont {Cirac}, \citenamefont
  {Jansen},\ and\ \citenamefont {K\"uhn}}]{Banuls2016a}%
  \BibitemOpen
\bibfield  {number} {  }\bibfield  {author} {\bibinfo {author} {\bibfnamefont
  {M.~C.}\ \bibnamefont {Ba\~nuls}}, \bibinfo {author} {\bibfnamefont
  {K.}~\bibnamefont {Cichy}}, \bibinfo {author} {\bibfnamefont {J.~I.}\
  \bibnamefont {Cirac}}, \bibinfo {author} {\bibfnamefont {K.}~\bibnamefont
  {Jansen}},\ and\ \bibinfo {author} {\bibfnamefont {S.}~\bibnamefont
  {K\"uhn}},\ }\bibfield  {title} {\bibinfo {title} {Density induced phase
  transitions in the schwinger model: A study with matrix product states},\
  }\href {https://doi.org/10.1103/PhysRevLett.118.071601} {\bibfield  {journal}
  {\bibinfo  {journal} {Phys. Rev. Lett.}\ }\textbf {\bibinfo {volume} {118}},\
  \bibinfo {pages} {071601} (\bibinfo {year} {2017})}\BibitemShut {NoStop}%
\bibitem [{\citenamefont {Ba\~{n}uls}\ \emph {et~al.}(2016)\citenamefont
  {Ba\~{n}uls}, \citenamefont {Cichy}, \citenamefont {Cirac}, \citenamefont
  {Jansen}, \citenamefont {K\"uhn},\ and\ \citenamefont {Saito}}]{Banuls2016b}%
  \BibitemOpen
  \bibfield  {author} {\bibinfo {author} {\bibfnamefont {M.~C.}\ \bibnamefont
  {Ba\~{n}uls}}, \bibinfo {author} {\bibfnamefont {K.}~\bibnamefont {Cichy}},
  \bibinfo {author} {\bibfnamefont {J.~I.}\ \bibnamefont {Cirac}}, \bibinfo
  {author} {\bibfnamefont {K.}~\bibnamefont {Jansen}}, \bibinfo {author}
  {\bibfnamefont {S.}~\bibnamefont {K\"uhn}},\ and\ \bibinfo {author}
  {\bibfnamefont {H.}~\bibnamefont {Saito}},\ }\bibfield  {title} {\bibinfo
  {title} {The multi-flavor schwinger model with chemical potential -
  overcoming the sign problem with matrix product states},\ }\bibfield
  {journal} {\bibinfo  {journal} {PoS(LATTICE 2016)316}\ }\href
  {https://doi.org/10.22323/1.256.0316} {10.22323/1.256.0316} (\bibinfo {year}
  {2016})\BibitemShut {NoStop}%
\bibitem [{\citenamefont {Funcke}\ \emph {et~al.}(2020)\citenamefont {Funcke},
  \citenamefont {Jansen},\ and\ \citenamefont {K\"uhn}}]{Funcke:2019zna}%
  \BibitemOpen
  \bibfield  {author} {\bibinfo {author} {\bibfnamefont {L.}~\bibnamefont
  {Funcke}}, \bibinfo {author} {\bibfnamefont {K.}~\bibnamefont {Jansen}},\
  and\ \bibinfo {author} {\bibfnamefont {S.}~\bibnamefont {K\"uhn}},\
  }\bibfield  {title} {\bibinfo {title} {{T}opological vacuum structure of the
  {S}chwinger model with matrix product states},\ }\href
  {https://doi.org/10.1103/PhysRevD.101.054507} {\bibfield  {journal} {\bibinfo
   {journal} {Phys. Rev. D}\ }\textbf {\bibinfo {volume} {101}},\ \bibinfo
  {pages} {054507} (\bibinfo {year} {2020})},\ \Eprint
  {https://arxiv.org/abs/1908.00551} {arXiv:1908.00551 [hep-lat]} \BibitemShut
  {NoStop}%
\bibitem [{\citenamefont {Farhi}\ \emph {et~al.}(2014)\citenamefont {Farhi},
  \citenamefont {Goldstone},\ and\ \citenamefont {Gutmann}}]{Farhi2014}%
  \BibitemOpen
  \bibfield  {author} {\bibinfo {author} {\bibfnamefont {E.}~\bibnamefont
  {Farhi}}, \bibinfo {author} {\bibfnamefont {J.}~\bibnamefont {Goldstone}},\
  and\ \bibinfo {author} {\bibfnamefont {S.}~\bibnamefont {Gutmann}},\
  }\bibfield  {title} {\bibinfo {title} {A quantum approximate optimization
  algorithm},\ }\href {https://arxiv.org/abs/1411.4028} {\bibfield  {journal}
  {\bibinfo  {journal} {arXiv:1411.4028}\ } (\bibinfo {year}
  {2014})}\BibitemShut {NoStop}%
\bibitem [{\citenamefont {Nation}\ \emph {et~al.}(2021)\citenamefont {Nation},
  \citenamefont {Kang}, \citenamefont {Sundaresan},\ and\ \citenamefont
  {Gambetta}}]{Nation:2021kye}%
  \BibitemOpen
  \bibfield  {author} {\bibinfo {author} {\bibfnamefont {P.~D.}\ \bibnamefont
  {Nation}}, \bibinfo {author} {\bibfnamefont {H.}~\bibnamefont {Kang}},
  \bibinfo {author} {\bibfnamefont {N.}~\bibnamefont {Sundaresan}},\ and\
  \bibinfo {author} {\bibfnamefont {J.~M.}\ \bibnamefont {Gambetta}},\
  }\bibfield  {title} {\bibinfo {title} {{Scalable mitigation of measurement
  errors on quantum computers}},\ }\href
  {https://doi.org/10.1103/PRXQuantum.2.040326} {\bibfield  {journal} {\bibinfo
   {journal} {PRX Quantum}\ }\textbf {\bibinfo {volume} {2}},\ \bibinfo {pages}
  {040326} (\bibinfo {year} {2021})},\ \Eprint
  {https://arxiv.org/abs/2108.12518} {arXiv:2108.12518 [quant-ph]} \BibitemShut
  {NoStop}%
\bibitem [{\citenamefont {Lieb}\ \emph {et~al.}(1961)\citenamefont {Lieb},
  \citenamefont {Schultz},\ and\ \citenamefont {Mattis}}]{Lieb1961}%
  \BibitemOpen
  \bibfield  {author} {\bibinfo {author} {\bibfnamefont {E.}~\bibnamefont
  {Lieb}}, \bibinfo {author} {\bibfnamefont {T.}~\bibnamefont {Schultz}},\ and\
  \bibinfo {author} {\bibfnamefont {D.}~\bibnamefont {Mattis}},\ }\bibfield
  {title} {\bibinfo {title} {Two soluble models of an antiferromagnetic
  chain},\ }\href {https://doi.org/10.1016/0003-4916(61)90115-4} {\bibfield
  {journal} {\bibinfo  {journal} {Ann. Phys.}\ }\textbf {\bibinfo {volume}
  {16}},\ \bibinfo {pages} {407} (\bibinfo {year} {1961})}\BibitemShut
  {NoStop}%
\bibitem [{\citenamefont {Pfeuty}(1970)}]{Pfeuty1970}%
  \BibitemOpen
  \bibfield  {author} {\bibinfo {author} {\bibfnamefont {P.}~\bibnamefont
  {Pfeuty}},\ }\bibfield  {title} {\bibinfo {title} {The one-dimensional
  {I}sing model with a transverse field},\ }\href
  {https://doi.org/10.1016/0003-4916(70)90270-8} {\bibfield  {journal}
  {\bibinfo  {journal} {Ann. Phys.}\ }\textbf {\bibinfo {volume} {57}},\
  \bibinfo {pages} {79} (\bibinfo {year} {1970})}\BibitemShut {NoStop}%
\bibitem [{\citenamefont {Kitaev}(2001)}]{Kitaev2001}%
  \BibitemOpen
  \bibfield  {author} {\bibinfo {author} {\bibfnamefont {A.}~\bibnamefont
  {Kitaev}},\ }\bibfield  {title} {\bibinfo {title} {Unpaired majorana fermions
  in quantum wires},\ }\href {https://doi.org/10.1070/1063-7869/44/10s/s29}
  {\bibfield  {journal} {\bibinfo  {journal} {Phys. Usp.}\ }\textbf {\bibinfo
  {volume} {44}},\ \bibinfo {pages} {131} (\bibinfo {year} {2001})}\BibitemShut
  {NoStop}%
\bibitem [{\citenamefont {Vidal}\ \emph {et~al.}(2003)\citenamefont {Vidal},
  \citenamefont {Latorre}, \citenamefont {Rico},\ and\ \citenamefont
  {Kitaev}}]{Vidal2003a}%
  \BibitemOpen
  \bibfield  {author} {\bibinfo {author} {\bibfnamefont {G.}~\bibnamefont
  {Vidal}}, \bibinfo {author} {\bibfnamefont {J.~I.}\ \bibnamefont {Latorre}},
  \bibinfo {author} {\bibfnamefont {E.}~\bibnamefont {Rico}},\ and\ \bibinfo
  {author} {\bibfnamefont {A.}~\bibnamefont {Kitaev}},\ }\bibfield  {title}
  {\bibinfo {title} {Entanglement in quantum critical phenomena},\ }\href
  {https://doi.org/10.1103/PhysRevLett.90.227902} {\bibfield  {journal}
  {\bibinfo  {journal} {Phys. Rev. Lett.}\ }\textbf {\bibinfo {volume} {90}},\
  \bibinfo {pages} {227902} (\bibinfo {year} {2003})}\BibitemShut {NoStop}%
\bibitem [{\citenamefont {Latorre}\ \emph {et~al.}(2004)\citenamefont
  {Latorre}, \citenamefont {Rico},\ and\ \citenamefont {Vidal}}]{Latorre2004}%
  \BibitemOpen
  \bibfield  {author} {\bibinfo {author} {\bibfnamefont {J.~I.}\ \bibnamefont
  {Latorre}}, \bibinfo {author} {\bibfnamefont {E.}~\bibnamefont {Rico}},\ and\
  \bibinfo {author} {\bibfnamefont {G.}~\bibnamefont {Vidal}},\ }\bibfield
  {title} {\bibinfo {title} {Ground state entanglement in quantum spin
  chains},\ }\href {http://dl.acm.org/citation.cfm?id=2011572.2011576}
  {\bibfield  {journal} {\bibinfo  {journal} {Quantum Info. Comput.}\ }\textbf
  {\bibinfo {volume} {4}},\ \bibinfo {pages} {48} (\bibinfo {year}
  {2004})}\BibitemShut {NoStop}%
\end{thebibliography}%
\end{document}